\documentclass[aps,floats]{revtex4-2}
\usepackage{amsmath,amssymb}
\usepackage{graphicx,epsfig}
\usepackage[greek,english]{babel}
\usepackage{bbold}
\usepackage{amsfonts}

\usepackage{mathtools}
\usepackage{makecell,tabularx}
\setcellgapes{1pt}
\makegapedcells

\makeatletter\AtBeginDocument{\let\@elt\relax}\makeatother
 
\begin{document}
\bibliographystyle {plain}

\pdfoutput=1
\def\oppropto{\mathop{\propto}} 
\def\opsimeq{\mathop{\simeq}}
\def\opoverderline{\mathop{\overline}}
\def\operarrow{\mathop{\longrightarrow}}
\def\opsim{\mathop{\sim}}

\def\opmin{\mathop{\min}} 
\def\opmax{\mathop{\max}} 
\def\oplim{\mathop{\lim}}

\title{ On the Kemeny time for continuous-time reversible and irreversible Markov processes  
with applications to stochastic resetting and to conditioning towards forever-survival
} 


\author{Alain Mazzolo}
\affiliation{Universit\'e Paris-Saclay, CEA, Service d'\'Etudes des R\'eacteurs et de Math\'ematiques Appliqu\'ees, 91191, Gif-sur-Yvette, France}

\author{C\'ecile Monthus}
\affiliation{Universit\'e Paris-Saclay, CNRS, CEA, Institut de Physique Th\'eorique, 91191 Gif-sur-Yvette, France}

\begin{abstract}
For continuous-time ergodic Markov processes, the Kemeny time $\tau_*$ is the characteristic time needed to converge towards the steady state $P_*(x)$: in real-space, the Kemeny time $\tau_*$ corresponds to the average of the Mean-First-Passage-Time $\tau(x,x_0) $ over the final configuration $x$ drawn with the steady state $P_*(x)$, which turns out to be independent of the initial configuration $x_0$; in the spectral domain, the Kemeny time $\tau_*$ corresponds to the sum of the inverses of all the non-vanishing eigenvalues $\lambda_n \ne 0 $ of the opposite generator. We describe many illustrative examples involving jumps and/or diffusion in one dimension, where the Kemeny time can be explicitly computed as a function of the system-size, via its real-space definition and/or via its spectral definition: we consider both reversible processes satisfying detailed-balance where the eigenvalues are real, and irreversible processes characterized by non-vanishing steady currents where the eigenvalues can be complex. In particular, we study the specific properties of the Kemeny times for Markov processes with stochastic resetting, and for absorbing Markov processes conditioned to survive forever.

\end{abstract}

\maketitle

\section{  Introduction }

For discrete-time and continuous-time ergodic Markov processes, 
a very important characteristic time $\tau_*$
is known in the mathematical literature under the two main names 'Kemeny constant' \cite{Kemeny1960,Kemeny1976,Kemeny_surfer,Kemeny_doyle,Kemeny_catral,Kemeny_kirkland2010,Kemeny_palacios1,Kemeny_palacios2,Kemeny_palacios3,Kemeny_palaciosSumrules,kemeny_optimal,Kemeny_hunter,Hunter2014,Kemeny_kirkland2014,Kemeny_robotic,Kemeny_gustafson,Kemeny_kirkland2016,Kemeny_resistance,Kemeny_bini,Kemeny_pitman,Kemeny_InfiniteDiscreteTime,Kemeny_graphWWW,Kemeny_laplacian,Pinsky_kemeny,choi_resistance,Kemeny_paradox,Kemeny_china,Kemeny_realworld,Kemeny_2021,Kemeny_tree,Kemeny2022}
and 'Eigentime Identity' \cite{aldous,Mao2004,Mao2006Transient,Mao2010Irreversible,Mao2012hitting,MicloAbsorbing2015,MaoDiffusion2015,EigentimeTree,choi_diff,choi_stein,choi_diff_JacobiTaboo},
while many other names have been also introduced by different authors
rediscovering this essential observable again and again over the years since 1960.
In the physics literature, this Kemeny time $\tau_*$ does not seem to have been much considered in its own right,
 except recently in the contexts of networks \cite{kemeny_undirectednetwork,kemeny_fractalnetwork,Kemeny_corre,Kemeny_ComplexNetwork},
 of fractional laplacian \cite{Kemeny_fractional2015,Kemeny_fractional2017},
 and of lifted Monte-Carlo Markov Chains \cite{c_SkewDB,c_GW}.
 One goal of the present paper is thus to give a self-contained unified description of the Kemeny time
  in a broad perspective for all physicists working in the field of Markov processes:
 we will emphasize its various interesting physical meanings as well as
 its nice mathematical properties.
 To simplify the discussions and the notations, 
 we will consider only Markov processes in continuous-time, using the same notations 
 as in the previous study \cite{c_SkewDB},
 while the case of discrete-time Markov Chains is discussed in detail in the recent work \cite{c_GW}.
 Since we will consider both Markov chains with discrete configurations 
 and diffusion processes with continuous configurations, it will be more convenient to first explain
 the properties of the Kemeny time for an ergodic Markov Chain with a finite number $N$ of configurations.
 
 
 \subsection{ Motivation: physical meanings of the Kemeny time for an ergodic Markov Chain on $N$ configurations   }

 For an ergodic continuous-time Markov Chain in a space of $N$ configurations converging towards the steady state $P_*(x)$, the Kemeny time $\tau_*$ has two definitions,
 one in real-space and one in the spectral domain
  \begin{eqnarray}
\tau_{*} = \tau^{Space}_{*} =\tau^{Spectral}_{*}
\label{kemeny}
\end{eqnarray}
that we recall below in Eqs \ref{taukemeny} and \ref{tauspectral} respectively.
 
 
  \subsubsection{ $\tau^{Space}_{*} $ as the Mean-First-Passage-Time $\tau(x,x_0) $ 
  averaged over the final configuration $x$ drawn with the steady state $P_*(x)$  }

 When the Mean-First-Passage-Time $\tau(x,x_0)$ at configuration $x$ when starting at configuration $x_0$
is averaged over the final configuration $x$ drawn with the steady state $ P_*(x)$,
the result $\tau_{*}(x_0) $ is actually independent of the initial configuration $x_0$
and is the real-space Kemeny time $\tau_*^{Space}$
 \begin{eqnarray}
\tau_{*}(x_0) \equiv  \sum_{x=1}^N P_{*}(x) \tau(x,x_0) \equiv \tau^{Space}_{*} \ \ \ \ { \rm independent \ \ of } \ \ x_0
\label{taukemeny}
\end{eqnarray}
This independence with respect to the initial configuration $x_0$ 
is a very striking property at first and explains the name 'Kemeny constant' 
used the mathematical literature. 
From the physical point of view,
 Eq. \ref{taukemeny} is a very appealing definition of the characteristic time $\tau_*^{Space}$ 
 needed to converge towards the steady state $P_*(x)$ independently of the initial configuration $x_0$.
 One of the most important issue is then to compute this characteristic time $\tau_*^{Space}$ 
 as a function of the system size $N$, in particular for large $N$.
 As a consequence, we prefer to avoid the mathematical terminology 'Kemeny constant' 
 that could be misleading and to use the name 'Kemeny time' in the present paper,
  where our goal will be to compute it in many different examples.


  \subsubsection{ $\tau_*^{Spectral}$ as the sum of the inverses of the non-vanishing eigenvalues $\lambda_n \ne 0 $ of the opposite generator $[-w]$ }

Let us first consider the case where the opposite generator $[-w]$ of the Markov chain is diagonalizable in terms of its $N$ eigenvalues 
$\lambda_n$. The vanishing eigenvalue $\lambda_0=0$ is associated to the conservation of probability and to
the steady state $P_*(x)$,
while the other $(N-1)$ eigenvalues have positive real parts 
 \begin{eqnarray}
{\rm Re}(\lambda_n) >0 \ \ \ { \rm for } \ \ \ n=1,..,N-1
\label{Relambdanpos}
\end{eqnarray}
and are associated to the $(N-1)$ relaxation modes towards the steady state.
The corresponding $(N-1)$ relaxation times $\tau_n$ can be defined from the  
absolute values of the exponential factors $e^{- t \lambda_n} $ that appear in the propagator 
 \begin{eqnarray}
  \vert e^{- t \lambda_n} \vert = e^{- t {\rm Re}(\lambda_n) } \equiv e^{- \frac{t}{\tau_n} }
  \ \ \ {\rm with } \ \ \ \tau_n = \frac{1}{{\rm Re}(\lambda_n)} \ \ {\rm for } \ \ \ n=1,..,N-1
\label{taurelax}
\end{eqnarray}
As will be explained in detail in section \ref{sec_green}, 
the real-space Kemeny time $\tau^{Space}_{*} $ of Eq. \ref{taukemeny}
actually coincides with the spectral Kemeny time $\tau^{Spectral}_{*}  $ 
defined as the sum of the inverses of the $(N-1)$ non-vanishing eigenvalues $\lambda_n \ne 0$ with $n=1,..,N-1$
 \begin{eqnarray}
\tau^{Spectral}_{*} 
\equiv  \sum_{n=1}^{N-1} \frac{1}{\lambda_n} 
\label{tauspectral}
\end{eqnarray}
When the matrix $[-w]$ is not diagonalizable,
its rewriting in the Jordan canonical form with the diagonal elements $\lambda_n$,
and with $1$ on the upper diagonal in the Jordan blocks,
yields that the spectral expression of Eq. \ref{tauspectral} for the Kemeny time remains valid.

From the physical point of view, the spectral expression Eq. \ref{tauspectral} is very nice because
it is a global characteristic time $\tau^{Spectral}_*$ that involves 
all the non-vanishing eigenvalues $\lambda_n \ne 0$,
while the $(N-1)$ relaxation-times $\tau_n$ of Eq. \ref{taurelax} are associated to specific relaxation eigenmodes.
 In particular, the maximal value among the $(N-1)$ relaxation-times $\tau_n$
 \begin{eqnarray}
\tau_{max} = \opmax_{1 \leq n \leq N-1} \tau_n = \frac{1}{ \displaystyle \opmin_{1 \leq n \leq N-1} {\rm Re}(\lambda_n) }  
\label{taurelaxmax}
\end{eqnarray}
plays a major role as the relaxation-time of the slowest eigenmode,
 that governs the asymptotic convergence towards the steady state in $e^{- \frac{t}{\tau_{max}} } $ 
 for large time $t \to + \infty$.
 In conclusion, the spectral Kemeny time $\tau^{Spectral}_{*}  $ of Eq. \ref{tauspectral}
 and the maximal relaxation time $\tau_{max}$ of Eq.  \ref{taurelaxmax} 
 are two different characteristic times that are both essential to understand the convergence properties
 towards the steady state. Let us stress some of their similarities and differences in the next subsection.
 

  \subsubsection{ Examples of comparison between the Kemeny time $\tau_{*}  $ 
 and the maximal relaxation time $\tau_{max}$ }
 
 The Kemeny time $\tau_{*}  $ 
 and the maximal relaxation time $\tau_{max}$
 can display the same scaling with the system size $L$,
 either with two different prefactors or with the same prefactor,
 but that they can also be completely different.
 Let us illustrate these three possibilities with the following simple examples 
 that will be discussed in more details later:
 
 (i) In subsection \ref{subsec_DiffpureL} concerning the pure diffusion on the interval $[0,L]$
 with reflecting boundary conditions,
 the maximal relaxation time $\tau_{max}=\frac{1}{\lambda_1}$ of Eq. \ref{eigenexciteduni}
and the Kemeny time $\tau_*$ of Eq. \ref{tauspectraluni}
 \begin{eqnarray}
\tau_{max}^{Diffusion} && = \frac{1}{\lambda_1} =\frac{L^2}{\pi^2}
\nonumber \\
\tau_*^{Diffusion} && =   \sum_{n=1}^{+\infty} \frac{1}{\lambda_n} = \frac{L^2}{ 6 } 
\nonumber 
\end{eqnarray}
both display the same expected diffusive scaling in $L^2$ but with different prefactors.
 
 (ii) In subsection \ref{subsec_sawBarrier} containing the case of a symmetric saw-tooth potential barrier 
 of slopes $\pm \mu$ of between two valleys minima at $x=0$ and $x=L$,
 the Kemeny time $\tau_{*}  $ of Eq. \ref{kemeny1dsawdemi}
 and the maximal relaxation time $\tau_{max}=\frac{1}{\lambda_1 }$ 
associated to the lowest non-vanishing eigenvalue $\lambda_1$
of Eq. \ref{group2hgapinverse} 
become asymptotically equal for large system-size $L$
\begin{eqnarray}
\tau_1^{SawTooth} && = \frac{1}{\lambda_1 } \opsimeq_{L \to + \infty} \frac{1}{\mu^2}   e^{ \frac{\mu}{2} L}
\nonumber \\
\tau_*^{SawTooth}  && =   \sum_{n=1}^{+\infty} \frac{1}{\lambda_n}  \opsimeq_{L \to + \infty} \frac{1}{\mu^2} e^{\frac{\mu}{2} L} 
\nonumber 
\end{eqnarray}
with the exponential growth with respect to the barrier $B(L)=\frac{\mu}{2} L $ that diverges with $L$,
and with exactly the same prefactor $\frac{1}{\mu^2} $.

 (iii) In Appendix \ref{app_infinite}, we will discuss why the Kemeny time $\tau_*$ can diverge
 for some diffusion processes when the system-size is infinite $L=+\infty$. The simplest example
 is the well-known Ornstein-Uhlenbeck on the whole line $]-\infty,+\infty[$ 
 corresponding to a quadratic confining potential:
its linear spectrum $\lambda_n=n \omega $ in $n$ of Eq. \ref{eigenOU}
 yields that the maximal relaxation time $\tau_{max}$ is finite
 while the corresponding spectral Kemeny time $\tau_*$ is infinite
 \begin{eqnarray}
\tau_{max}^{OU} && =  \frac{1}{\lambda_1 }   =\frac{1}{\omega}
\nonumber \\
\tau_{*}^{OU} && =  \sum_{n=1}^{+\infty} \frac{1}{\lambda_n} = \frac{1}{\omega} \sum_{n=1}^{+\infty} \frac{1}{n} =+\infty
\nonumber
\end{eqnarray}
This example shows that these two characteristic times can be completely different.

Besides the scaling with respect to the system-size $L$ that we have just discussed,
it is also interesting to compare the Kemeny time $\tau^{Spectral}_{*}  $ 
 and the maximal relaxation time $\tau_{max}$
from the practical point of view of their computations in given Markov models:

(a) on one hand,
it is clear that the spectral Kemeny time $\tau^{Spectral}_*$ is a priori more complicated 
that the maximal relaxation time $\tau_{max}$, since one needs to know all the eigenvalues
instead of only the dominant non-vanishing eigenvalue. 
And indeed from the numerical point of view (that will not be discussed in the present paper), 
the efforts required to compute the whole spectrum
or only the dominant eigenvalue are completely different.
Note however that from the analytical point of view that we will also consider in the present paper, 
the efforts are actually not different since it is the same
calculation that leads to the whole list of explicit eigenvalues.

(b) on the other hand, the Kemeny time can be alternatively computed
 via its real-space definition $\tau_*^{Space} $ of Eq. \ref{taukemeny},
 while we are not aware 
 of any real-space expression for the maximal relaxation time $\tau_{max}$.
 As a consequence, it is possible in some cases to obtain 
 an explicit real-space expression for the Kemeny time,
 even if one is not able to compute any individual eigenvalue, 
 i.e. even if one is not able to compute
 the maximal relaxation time $\tau_{max}$.

 
 \subsection{ Goals and organization of the paper  }
 
 The goal of the present paper is to describe in a unified framework
 the general properties of the Kemeny time for continuous-time Markov processes,
 and to give explicit results in simple illustrative examples,
 for both reversible or irreversible dynamics,
 for both discrete configurations or continuous configurations,
 with a special attention to stochastic resetting,
 and to absorbing Markov processes conditioned to survive forever.
 In particular, we will stress the respective advantages 
 of the real-space and of the spectral definitions of the Kemeny time
 that can be summarized as follows.


 \subsubsection{ Main advantage of the real-space definition $\tau_*^{Space} $ of the Kemeny time  }
 
 The main advantage of the real-space definition $\tau_*^{Space} $
 is that one can write explicit expressions for general Markov models
 in simple geometries, in particular in dimension $d=1$ on which we focus in the present paper:
 
 (i) for Markov chains with arbitrary nearest-neighbor transitions rates $w(x \pm 1,x)$,
 the real-space Kemeny time $\tau_*^{Space} $ is explicit 
 for one-dimensional intervals $x=1,2,...,N$ with reflecting boundary conditions (Eq. \ref{taueqDBM})
 and for one-dimensional rings with periodic boundary conditions
  (with the special directed case in Eq. \ref{kemenyDir}),
  as well as for some generalizations with one absorbing boundary condition (Eq. \ref{AbsN}).
 
 (ii) for diffusion processes with an arbitrary force $F(x)$ and an arbitrary diffusion coefficient $D(x)$,
 the real-space Kemeny time $\tau_*^{Space} $ is explicit 
 for one-dimensional intervals $x \in ]0,L[$ with reflecting boundary conditions (Eq. \ref{kemeny1d})
 and for one-dimensional rings with periodic boundary conditions (Eq. \ref{kemenyring}),
 as well as for some generalizations for one absorbing boundary condition (Eq. \ref{mftpAbs}).

 
  \subsubsection{ Main advantage of the spectral definition $\tau^{Spectral}_* $ of the Kemeny time  }
  
  Whenever all the eigenvalues $\lambda_n$ of a Markov process are explicitly known,
  it is very easy to analyze the addition of stochastic resetting at rate $\gamma$:
  all the excited eigenvalues are simply shifted $ \lambda_n^{[Reset]} = \lambda_n+ \gamma $ (Eq. \ref{lambdanreset})
  and one directly obtains the corresponding Kemeny time $\tau^{[Reset]Spectral}_* $ via Eq. \ref{KemenyReset}.
  Simple examples that will be discussed later in the present paper are given in the three Tables 
  \ref{tableSpectralChains},
  \ref{tableSpectralDiffusion} and
  \ref{tableSpectralRing}.
  
  Similarly, whenever all the eigenvalues $\lambda_n^{[Abs]}$ of an absorbing Markov process are explicitly known,
  it is very easy to take into account the conditioning towards forever-survival:
  the new excited eigenvalues are simply obtained via the differences $\lambda_m^{[Cond]}  = \lambda_{1+m}^{[Abs]} -\lambda^{[Abs]}_1 $ of Eq. \ref{lambdaCond},
  and one directly obtains the corresponding Kemeny time via Eq. \ref{KemenyCond}.
  Simple examples that will be discussed later in the present paper can be found in Table \ref{tableSpectralDiffusion}.
  

\begin{table}[!h]
\setcellgapes{4pt}
\begin{tabular}{|p{3.5cm}||c|c|}
\hline
Markov chains $ \ \ \ \ \ \ \ $ on the interval 
 of $N$ sites $x=1,2,...,N$  
& $ \begin{aligned} & \mathrm{Excited ~Eigenvalues~} \lambda_n \\ 
& \mathrm{with~} n=1,2,...,N-1 \end{aligned}$
& Spectral Kemeny time $\tau_{*} \equiv  \displaystyle \sum_{n=1}^{N-1} \frac{1}{\lambda_n}$  \\
\hline \hline
Symmetric random walk $w(x \pm 1,x)=1$  
& $\lambda_n =  2 \left[ 1 - \cos\left( n  \frac{ \pi}{N}\right) \right]$ 
& $\tau_{*} = \frac{N^2-1}{6}$  \\
\hline
Symmetric random walk $w(x \pm 1,x)=1  \ \ \ \ \ \ \ \ \ $ 
with resetting rate $\gamma$  
& $\lambda^{[Reset]}_n = \gamma+ 2 \left[ 1 - \cos\left( n  \frac{ \pi}{N}\right) \right] $ 
& $\tau_{*}  =  \frac{1}{  \sqrt{ \gamma ( 4 + \gamma ) } }
 \left[ N
   \frac{ \left( \sqrt{ \frac{\gamma}{4} }+ \sqrt{ \frac{\gamma}{4} +1}\right)^{4N} -1   }
   { \left( \sqrt{ \frac{\gamma}{4} }+ \sqrt{ \frac{\gamma}{4} +1}\right)^{4N} +1}
-    \frac{  \gamma +2 }{ \sqrt{ \gamma ( 4 + \gamma ) }} \right] $  \\
\hline
Asymmetric random walk $w(x \pm 1,x)=e^{\pm \frac{\mu}{2} }$ 
& $\lambda_n =  2 \cosh \left(  \frac{\mu}{2} \right)  - 2 \cos\left( n  \frac{ \pi}{N}\right)$ 
& $\tau_{*} 
=\frac{ 1}
{ 2 \sinh \left( \frac{\mu}{2} \right) }  
 \left[  \frac{ N }{\tanh \left( N \frac{\mu}{2} \right)}
  - \frac{ 1 }{ \tanh \left( \frac{\mu}{2} \right) }
  \right]$  \\
\hline
Asymmetric random walk $w(x \pm 1,x)=e^{\pm \frac{\mu}{2} }$ 
with resetting rate $\gamma$
& $\lambda_n^{[Reset]} = \gamma+ 2 \cosh \left(  \frac{\mu}{2} \right)  - 2 \cos\left( n  \frac{ \pi}{N}\right)$ 
& $\tau_{*} 
=\frac{\left[ N
   \frac{ \left(  \frac{\sqrt{\gamma +4 \sinh^2 \left(  \frac{\mu}{4} \right)}+\sqrt{\gamma +4 \cosh^2 \left(  \frac{\mu}{4} \right)} }{2} \right)^{4N} -1   }
   { \left(  \frac{\sqrt{\gamma +4 \sinh^2 \left(  \frac{\mu}{4} \right)}+\sqrt{\gamma +4 \cosh^2 \left(  \frac{\mu}{4} \right)} }{2} \right)^{4N} +1}
-    \frac{  \gamma + 2 \cosh \left(  \frac{\mu}{2} \right) }
{ \sqrt{ \left[ \gamma +4 \sinh^2 \left(  \frac{\mu}{4} \right)\right] 
\left[ \gamma +4 \cosh^2 \left(  \frac{\mu}{4} \right)\right] }} \right]}
 {  \sqrt{ \left[ \gamma +4 \sinh^2 \left(  \frac{\mu}{4} \right)\right] 
\left[ \gamma +4 \cosh^2 \left(  \frac{\mu}{4} \right)\right] } }$  \\
\hline
\end{tabular}
\caption{Examples of spectral Kemeny times $\tau^{Spectral}_{*}$ for some Markov chains
with nearest-neighbor transition rates $w(x \pm 1,x)$ on the interval 
 of $N$ sites $x=1,2,...,N$.
} 
\label{tableSpectralChains}
\end{table}


\begin{table}[!h]
\setcellgapes{6pt}
\begin{tabular}{|p{6cm}||c|c|}
\hline
Diffusion processes on $]0,L[$ 
& $ \begin{aligned} & \mathrm{Excited ~Eigenvalues~} \lambda_n \\ 
& \mathrm{with~} n=1,2,...,+\infty \end{aligned}$ 
& Spectral Kemeny time $\tau_{*} \equiv  \displaystyle \sum_{n=1}^{+\infty} \frac{1}{\lambda_n}$  \\
\hline \hline
Pure diffusion $D(x)=1$ 
and $F(x)=0$ 
& $\lambda_n =    \frac{ \pi^2 n^2}{L^2}$ 
& $\tau_{*} = \frac{L^2}{ 6 }  $  \\
\hline
Pure diffusion $D(x)=1$ 
and $F(x)=0 \ \ \ \ \ $
 with resetting rate $\gamma$
& $ \lambda_n^{[Reset]} = \gamma+   \frac{ \pi^2 n^2}{L^2}$ 
& $\tau_{*} 
    = \frac{1}{ 2 \gamma } \left[ \frac{ \sqrt{\gamma} L }{\tanh \left(\sqrt{\gamma} L \right)}  -1\right] $  \\
\hline
Biased diffusion 
$D(x)=1$ and $F(x)=\mu $  
& $\lambda_n =   \frac{ \mu^2 }{4  } + \frac{ \pi^2 n^2}{L^2}$ 
& $\tau_*
  = \frac{2}{ \mu^2 } \left[ \frac{ \frac{\mu L}{2} }{\tanh \left(\frac{\mu L}{2} \right)}  -1\right] $  \\
\hline
Biased diffusion 
$D(x)=1$ and $F(x)=\mu $
with resetting rate $\gamma$
& $\lambda_n^{[Reset]} = \gamma+  \frac{ \mu^2 }{4  } + \frac{ \pi^2 n^2}{L^2}$ 
& $\tau_*  = \frac{1}{ 2 \left(\gamma+  \frac{ \mu^2 }{4  } \right) } \left[ \frac{ \sqrt{\gamma+  \frac{ \mu^2 }{4  }} L }{\tanh \left(\sqrt{\gamma+  \frac{ \mu^2 }{4  }} L \right)}  -1\right]  $  \\  
\hline
Pure diffusion $D(x)=1$ 
and $F(x)=0 \ \ \ $ with two absorbing boundaries  
$ \ \ \ \ \ \ \ \ $
conditioned to survive forever  
& $\lambda_n^{[Cond]} = \frac{\pi^2}{L^2} ( n^2+2n )$ 
& $\tau_*  =  \frac{3 L^2}{4 \pi^2 } $  \\  
\hline
Pure diffusion $D(x)=1$ 
and $F(x)=0 \ \ \ \ $ with one absorbing boundary  
$ \ \ \ \ \ \ \ \ \ \ \ \ $
conditioned to survive forever  
& $\lambda_n^{[Cond]} = \frac{ \pi^2   }{ L^2}  (n^2+n ) $ 
& $\tau_*  =  \frac{L^2}{\pi^2 } $  \\  
\hline
 All the diffusion processes with the Jacobi spectrum
 of parameters $A>0$ and $B> -1$
& $\lambda_n =  A (n^2 +B n)$ 
& $\tau_*  =  \frac{1}{A B} \left[ \gamma_{Euler}+ \frac{ \Gamma’(B+1) }{\Gamma(B+1) }\right] $ \\
  \hline
\end{tabular}
\caption{Examples of spectral Kemeny times $\tau^{Spectral}_{*}$ for some diffusion processes
with diffusion coefficient $D(x)$ and force $F(x)$
on the interval $]0,L[$.
} 
\label{tableSpectralDiffusion}
\end{table}


\begin{table}[!h]
\setcellgapes{3pt}
\begin{tabular}{|p{5cm}||c|c|}
\hline
Irreversible Markov processes $ \ \ \ \ \ \ $ on periodic Rings  
& $ \begin{aligned} & \mathrm{Excited ~Eigenvalues~} \lambda_n \\ 
& \mathrm{with~imaginary~parts}  \end{aligned}$
& Spectral Kemeny time $\tau_{*} \equiv  \displaystyle \sum_{n \ne 0} \frac{1}{\lambda_n}$  \\
\hline \hline 
Biased diffusion $D(x)=1$ and $F(x)=\mu$ on the ring of length $L$  
& $ \begin{aligned} &  \lambda_n = \frac{4 \pi^2}{L^2}  \left[     n^2 +  i n \frac{\mu L}{2 \pi}  \right]\\ 
& \mathrm{with~} n=\pm 1,\pm 2,...\pm \infty \end{aligned}$
& $\tau_{*} =   \frac{1}{\mu^2 } \left[ \frac{ \frac{\mu L}{2 } }{ \tanh \left( \frac{\mu L}{2 } \right)} -1 \right]
$  \\
\hline 
Biased diffusion $D(x)=1$ and $F(x)=\mu$ on the ring of length $L$  
with resetting rate $\gamma$
& $ \begin{aligned} &  \lambda_n^{[Reset]} =\gamma+ \frac{4 \pi^2}{L^2}  \left[     n^2 +  i n \frac{\mu L}{2 \pi}  \right]\\ 
& \mathrm{with~} n=\pm 1,\pm 2,...\pm \infty \end{aligned}$
& $\tau_{*} =    - \frac{1}{\gamma} 
+ \frac{L \sinh \left( \frac{L }{2} \sqrt{\mu^2+4 \gamma}\right)}{2 \sqrt{\mu^2+4 \gamma}  
\sinh \left( L \frac{\sqrt{\mu^2+4 \gamma}+ \mu}{4}\right) 
\sinh \left( L \frac{\sqrt{\mu^2+4 \gamma}- \mu}{4}\right)}
$  \\
\hline
Directed Pure Trap model $\ \ \ \ \  \ \ \ \ \ $
$w(x+1,x)=2$ on the ring of $N$ sites  
& $ \begin{aligned} & \lambda_n  =   2 \left( 1- e^{- i 2 \pi  \frac{n}{N} } \right)\\ 
& \mathrm{with~} n=n=1,..,N-1 \end{aligned}$
& $\tau_{*}  = \frac{N-1}{4} $  \\
\hline
Directed Pure Trap model $\ \ \ \ \  \ \ \ \ \ $
$w(x+1,x)=2$ on the ring of $N$ sites  
with resetting rate $\gamma$
& $ \begin{aligned} & \lambda_n^{[Reset]}  = \gamma+    2 \left( 1- e^{- i 2 \pi  \frac{n}{N} } \right)\\ 
& \mathrm{with~} n=n=1,..,N-1 \end{aligned}$
& $\tau_{*}  =  \frac{ N   }{ \left( \gamma +2\right)
 \left[ 1-\left( \frac{\gamma}{2}+1\right)^{-N} \right] } - \frac{ 1 }{  \gamma } $  \\
\hline
\end{tabular}
\caption{Examples of spectral Kemeny time $\tau^{Spectral}_{*}$ for some irreversible processes 
on rings with complex eigenvalues.
} 
\label{tableSpectralRing}
\end{table}


 
 \subsubsection{ Organization of the paper  }

  The paper is organized as follows.
  In section \ref{sec_green}, the general properties of the Kemeny time $\tau_*$ 
are explained in detail for the case of an ergodic continuous-time Markov Chain
 in a space of $N$ configurations converging towards the steady state $P_*(x)$. We also describe the specific properties of the Kemeny times for 
 Markov processes in the presence of stochastic resetting at constant rate,
 and for absorbing Markov processes conditioned to survive forever. 
In all the other sections, this general framework is applied to various types of Markov processes as follows.

We first consider the simple geometry of one-dimensional finite intervals, 
where the zero-current condition at the two boundaries yields that the steady current has to vanish also 
everywhere in the bulk of the one-dimensional interval, so that detailed-balance is always satisfied:
we give the explicit expressions of the real-space 
Kemeny time for Markov chains with 
arbitrary transitions rates $w(x \pm 1,x)$
in section \ref{sec_ChainInterval}
and for diffusion processes 
with an arbitrary force $F(x)$ and an arbitrary diffusion coefficient $D(x)$
in section \ref{sec_DiffInterval}. We also consider simple examples where the eigenvalues are explicitly known
in order to analyze the effect of stochastic resetting and of conditioning towards forever-survival.

We then focus on the periodic ring which is the simplest geometry
where non-equilibrium steady states can exist 
with a non-vanishing steady current along the ring:
we give the explicit expression of the real-space 
Kemeny time for diffusion processes 
with an arbitrary force $F(x)$ and an arbitrary diffusion coefficient $D(x)$ in section \ref{sec_DiffRing}, 
while in section \ref{sec_DirectedTrap}, we focus on
the Directed Markov chains on the ring.
Again, we also consider simple examples where the eigenvalues are explicitly known and possibly complex
in order to analyze the effect of stochastic resetting.

Finally in section \ref{sec_Absorbing}, 
we discuss the generalizations of the Kemeny time for absorbing Markov processes.
Our conclusions are summarized in section \ref{sec_conclusions}.
The various appendices contain either complementary discussions or more technical computations.
 

\section{ Properties of the Kemeny time $\tau_*$ explained via the Green function $G$   }

\label{sec_green}

In this section, we describe the general properties of the Kemeny time $\tau_*$ of Eq. \ref{kemeny}
for an ergodic continuous-time Markov chain in a space of $N$ configurations 
converging towards the steady state $P_*(x)$.
The generator $w$ is then an $N \times N$ matrix with the following properties: 
the positive off-diagonal $x \ne y$ matrix element $w(x,y) \geq 0$   
represents the transition rate from configuration $y$ to configuration $x $,
while the diagonal elements are negative $w(x,x) \leq 0 $ 
and are fixed by the conservation of probability to be
\begin{eqnarray}
w(x,x) \equiv  - \sum_{y \ne x} w(y,x) 
\label{wdiag}
\end{eqnarray}


\subsection{ Properties of the propagator $P_t( x \vert x_0 ) $ converging towards the steady state $P_*(x)$}

The propagator $P_t( x \vert x_0 ) $, i.e. 
the probability to be in configuration $x$ at time $t$ when starting in configuration $x_0$ at time $0$
\begin{eqnarray}
P_t( x \vert x_0 )   \equiv \langle x \vert e^{w t} \vert x_0 \rangle
\label{propagatordef}
\end{eqnarray}
satisfies the forward master equation with respect to the final configuration $x$
\begin{eqnarray}
 \partial_t P_t( x \vert x_0 )    =   \langle x \vert w \ e^{w t} \vert x_0 \rangle 
 = \sum_{y }  w(x,y) P_t( y \vert x_0 ) 
\label{forward}
\end{eqnarray}
and the backward master equation with respect to the initial configuration $x_0$
\begin{eqnarray}
  \partial_{t} P_t( x \vert x_0 )    =    \langle x \vert  e^{w t} \ w \vert x_0 \rangle
 =  \sum_{y_0 } P_t( x \vert y_0 )    w(y_0,x_0)
\label{backward}
\end{eqnarray}

The steady state $P_*(x) $
satisfies the steady version of the forward master Eq. \ref{forward}
\begin{eqnarray}
 0   =   \sum_{y }  w(x,y) P_*(y)
\label{mastersteady}
\end{eqnarray}
Eqs \ref{wdiag} and \ref{mastersteady} mean that 
 the highest eigenvalue of the Markov matrix $w(.,.)$ is $\lambda_0=0$ 
 \begin{eqnarray}
  0  && = \langle l_0 \vert w 
 \nonumber \\
  0   && =  w \vert r_0 \rangle 
\label{mastereigen0}
\end{eqnarray}
where the positive left eigenvector is trivial 
\begin{eqnarray}
 l_0(x)=1
\label{markovleft}
\end{eqnarray}
while the positive right eigenvector is given by the steady state
\begin{eqnarray}
 r_0(y)=P_*(y) 
\label{markovright}
\end{eqnarray}


\subsection{ Properties of the Green function $G(x,x_0)$ inherited from the propagator 
$P_t( x \vert x_0 ) $  }

To characterize the convergence of the propagator $P_t( x \vert x_0 )$ towards the steady state $P_*(x) $ 
for large time $t \to + \infty$
\begin{eqnarray}
P_t( x \vert x_0 )    \opsimeq_{ t \to + \infty} P_*(x)
\label{propagatorCV}
\end{eqnarray}
it is useful to introduce the Green function corresponding to the integration over time $t \in [0,+\infty[$
of the difference $\left[ P_t( x \vert x_0 )  - P_{*}(x)\right] $ 
\begin{eqnarray}
G(x,x_0) \equiv \int_0^{+\infty} dt \left[ P_t( x \vert x_0 )  - P_{*}(x)\right]
= \int_0^{+\infty} dt \left[ \langle x \vert e^{wt} \vert x_0 \rangle - P_{*}(x)\right]
\label{defgreen}
\end{eqnarray}
while the many different names used in the mathematical literature include
'the fundamental matrix' \cite{Kemeny1976},
'the ergodic potential' \cite{syski1978},
'the deviation matrix' \cite{Kemeny_bini,deviationMatrix},
'the deviation kernel' \cite{Mao2002},
'the centered resolvent' \cite{MicloCenteredResolvent}.

The forward and backward equations of Eqs \ref{forward}
and \ref{backward}
for the propagator $P_t( x \vert x_0 ) $ translate into the following 
forward and backwards equations
for the Green function
\begin{eqnarray}
\sum_y w(x,y)  G(y,x_0) && = \int_0^{+\infty} dt \partial_t P_t(x \vert x_0)= \left[  P_t(x \vert x_0) \right]_{t=0}^{t=+\infty}= P_{*}(x) - \delta_{ x,x_0}  
 \nonumber \\
\sum_{y_0}  G(x,y_0)  w(y_0,x_0) && =\int_0^{+\infty} dt \partial_t P_t(x \vert x_0)
= \left[  P_t(x \vert x_0) \right]_{t=0}^{t=+\infty} =P_{*}(x) - \delta_{ x,x_0}  
 \label{eqG}
\end{eqnarray}
In addition, the orthogonality relations $\langle l_0 \vert  G =0 $ and $G \vert r_0 \rangle =0 $ with respect to the left and right eigenvectors associated to
the vanishing eigenvalues $\lambda_0=0$ yield
 \begin{eqnarray}
&& {\rm for \ any \ x_0 } : \ \ \ \ 
 0=  \langle l_0 \vert  G \vert x_0 \rangle  = \sum_x  \langle l_0 \vert x \rangle G(x,x_0) =  \sum_x  G(x,x_0)  
 \nonumber \\
&& {\rm for \ any \ x } : \ \ \ \ 
 0=  \langle x \vert   G \vert r_0 \rangle  =  \sum_{x_0}   G(x,x_0) \langle x_0 \vert r_0 \rangle
 =  \sum_{x_0}   G(x,x_0) P_*(x_0) 
\label{Gorthog}
\end{eqnarray}

All these properties can be summarized at the matrix level as follows:
 the Green function $G$ 
is the inverse of the matrix $[-w]$ 
 within the subspace $\left( \mathbb{1} - \vert r_0 \rangle \langle l_0 \vert   \right) $ orthogonal to 
the subspace $\left(  \vert r_0 \rangle \langle l_0 \vert   \right) $ 
associated to the zero eigenvalue of the generator $w$
\begin{eqnarray}
   G  = \left( \mathbb{1} - \vert r_0 \rangle \langle l_0 \vert  \right) \frac{1}{[ -w ] }  \left( \mathbb{1} - \vert r_0 \rangle \langle l_0 \vert  \right)
 \label{Gasinverse}
\end{eqnarray}
This form is reminiscent of the second-order perturbation theory of energy levels in quantum mechanics,
and indeed, the Green function $G$ of Eq. \ref{Gasinverse}
governs the perturbation theory of the lowest eigenvalue $\lambda_0$ 
when the unperturbed Markov matrix $w(x,y)$ is changed into the new matrix $[w(x,y)+\epsilon w_{per}(x,y)]$
involving some perturbation $\epsilon w_{per} (x,y)$ of small amplitude $\epsilon \to 0$
that allows to construct the perturbation theory in powers of $\epsilon$
(see Appendix A of \cite{c_ruelle} for details).


 \subsection{ Link with the averaged time $t_{[0,T]}(x,x_0)$ spent in configuration $x$ during $[0,T]$
when starting at $x_0$  }

The averaged time $t_{[0,T]}(x,x_0)$ spent in configuration $x$ during the time-window $[0,T]$
when starting at configuration $x_0$ can be computed from the propagator
of Eq. \ref{propagatordef}
\begin{eqnarray}
t_{[0,T]}(x,x_0) \equiv \int_0^{T} dt  P_t( x \vert x_0 )
\label{defavlocaltime}
\end{eqnarray}
Its large-time behavior is extensive in $T$ and involves the steady-state $P_*(x)$ 
of Eq. \ref{propagatorCV}
\begin{eqnarray}
t_{[0,T]}(x,x_0) \opsimeq_{T \to +\infty}  T P_*(x) + o(T)
\label{avlocaltimeExtensive}
\end{eqnarray}
So the Green function of Eq. \ref{defgreen}
represents the remaining finite contribution to $t_{[0,T]}(x,x_0) $ 
when the leading extensive contribution
of Eq. \ref{avlocaltimeExtensive} is subtracted from Eq. \ref{defavlocaltime}
\begin{eqnarray}
G(x,x_0) = \oplim_{T \to +\infty}  \left[ \int_0^{T} dt P_t( x \vert x_0 ) -  T P_{*}(x)\right]
= \oplim_{T \to +\infty}  \left[ t_{[0,T]}(x,x_0)  -  T P_{*}(x)\right]
\label{greenlocaltime}
\end{eqnarray}
As a consequence, the averaged time $t_{[0,T]}(x,x_0)$ 
spent at configuration $x$ for large $T$ of Eq. \ref{avlocaltimeExtensive} 
displays the extensive leading term $T P_*(x) $ and the subleading finite term $G(x,x_0) $
\begin{eqnarray}
t_{[0,T]}(x,x_0) \opsimeq_{T \to +\infty}  T P_*(x) +G(x,x_0) + o(T^0)
\label{avlocaltimeExtensivefinite}
\end{eqnarray}


 \subsection{ Link with the Mean-First-Passage-Time $\tau(x,x_0)$ at configuration $x$ 
when starting at configuration $x_0$  }

To analyze further the averaged time $t_{[0,T]}(x,x_0)$ spent in configuration $x$ during $[0,T]$
when starting at $x_0$ discussed in the previous subsection,
it is useful to introduce the Mean-First-Passage-Time $\tau(x,x_0)$
at configuration $x$ when starting at $x_0$
and to decompose the large time-window $[0,T]$ into two intervals:

(i) the initial interval $[0,\tau(x,x_0)]$ during which the particle is not yet able to visit the configuration $x$;

(ii) the remaining interval $[\tau(x,x_0),T]$, during which the time spent at configuration $x$
can be evaluated with $x$ as the initial configuration instead of $x_0$.

This decomposition thus yields the very simple relation
\begin{eqnarray}
t_{[0,T]}(x,x_0) = 0 +   t_{[0,T-  \tau(x,x_0)]}(x,x)
\label{avlocaltimetau}
\end{eqnarray}
Since $\tau(x,x_0) $ is a finite Mean-First-Passage-Time independent of the large time-window $[0,T]$,
one can plug the large-time behavior of Eq. \ref{avlocaltimeExtensivefinite} on the two sides of Eq. \ref{avlocaltimetau}
to obain
\begin{eqnarray}
 T P_*(x) +G(x,x_0) + o(T^0) =  [T -  \tau(x,x_0) ]P_*(x) +G(x,x) + o(T^0)
\label{avlocaltimetauT}
\end{eqnarray}
so that the Green function $G(x,x_0)$ and the Mean-First-Passage-Time $\tau(x,x_0)$  
are directly related via
\begin{eqnarray}
G(x,x_0)  =  G(x,x)  - P_{*}(x) \tau(x,x_0)  
\label{greenandfirst}
\end{eqnarray}
This relation can be alternatively obtained by considering the whole normalized probability distribution $F(t_1 ; x,x_0)$
of the first-passage-time $t_1 \in [0,+\infty[$ at $x$ when starting at $x_0$ in order to decompose the propagator 
via the convolution $P_t( x \vert x_0 ) = \int_0^{+\infty} dt_1 P_{t-t_1}( x \vert x )F(t_1 ; x,x_0)$ 
 (see \cite{c_SkewDB} for more details on this alternative derivation).

Plugging Eq. \ref{greenandfirst}
into the backward Eq. \ref{eqG} for the Green function
yields 
\begin{eqnarray}
-1 + \frac{ \delta_{ x,x_0}  }{ P_*(x)}  =  \sum_{y_0 }  \tau(x,y_0)   w(y_0,x_0) 
=  \sum_{y_0 \ne x_0}  \left[   \tau(x,y_0)  - \tau(x,x_0) \right] w(y_0,x_0) 
 \label{Ginvlefttau}
\end{eqnarray}
This backward equation for the MFPT $\tau(x,x_0)$ for $x_0 \ne x$
 is indeed a standard method to compute $\tau(x,x_0)$ with the boundary condition $\tau(x,x) =0 $
  (see for instance the textbooks \cite{gardiner,vankampen,risken,redner}).

Plugging Eq. \ref{greenandfirst} into the two orthogonality conditions of Eq. \ref{Gorthog} for the Green function
yield
 \begin{eqnarray}
&& {\rm for \ any \ x_0 } : \ \ \ \ 
     \sum_x  G(x,x)  =  \sum_x P_{*}(x) \tau(x,x_0)  
 \nonumber \\
&& {\rm for \ any \ x } : \ \ \ \ 
 G(x,x)= P_{*}(x)  \sum_{x_0}  \tau(x,x_0)   P_*(x_0)  
\label{Gorthogtau}
\end{eqnarray}
The second equation valid for any $x$ provides an interesting expression for the diagonal element $G(x,x)>0$
that is found to be proportional to the steady state $P_*(x)$ and to the average of the 
MFTP $\tau(x,x_0)$ at position $x$ when the initial condition $x_0$ is drawn with the steady state $P_*(x_0)$.
As a consequence, the Green function $G(x,x_0) $ of Eq. \ref{greenandfirst}
becomes
\begin{eqnarray}
G(x,x_0)  =  P_{*}(x) \left[ \sum_{y_0}  \tau(x,y_0)   P_*(y_0)    -  \tau(x,x_0)  \right]
\label{greenandfirstalone}
\end{eqnarray}
In particular, the sign of the Green function $G(x,x_0)$ is determined by the sign of the parenthesis, 
i.e. whether the MFTP $\tau(x,x_0)  $ is bigger or smaller than the average $\left[ \sum_{y_0}  \tau(x,y_0)   P_*(y_0) \right] $
over the initial point $y_0$ drawn with the steady state $P_*(y_0)$.


 \subsection{ The real-space Kemeny time $\tau^{Space}_*$ as the trace of the Green function  }

The first equation of Eq. \ref{Gorthogtau}
valid for any $x_0$ corresponds to the fact that 
the real-space Kemeny time of Eq. \ref{taukemeny} is independent of the initial point $x_0$ 
and coincides with the sum of the $N$ diagonal elements $G(x,x)$,
i.e. with the trace of Green function $G$
 \begin{eqnarray}
\tau^{Space}_{*}  \equiv  \sum_x P_{*}(x) \tau(x,x_0)  = \sum_{x=1}^N G(x,x) \equiv  {\rm Trace}(G) 
\label{tautraceG}
\end{eqnarray}
This trace can be also rewritten using Eq. \ref{defgreen} 
 \begin{eqnarray}
\tau^{Space}_{*} = {\rm Trace}(G) =\sum_{x=1}^N G(x,x) 
=   \int_0^{+\infty} dt \left[ \sum_{x=1}^N \langle x \vert e^{wt} \vert x \rangle - \sum_{x=1}^N P_{*}(x)\right]
= \int_0^{+\infty} dt \left[{\rm Trace} ( e^{wt} ) -1 \right]
\label{tautraceGbis}
\end{eqnarray}
in terms of the trace of the evolution operator ${\rm Trace} ( e^{wt} ) $.


\subsection{Link with the spectral Kemeny time $\tau^{Spectral}_*$ involving the eigenvalues $\lambda_n$ of the opposite generator $[-w]$ }

\subsubsection{ When the generator $w$ can be diagonalized in terms of right and left eigenvectors}

When the Markov matrix $w$ is diagonalizable, its spectral decomposition 
\begin{eqnarray}
 w  = -  \sum_{n=1}^{N-1} \lambda_n \vert r_n \rangle \langle l_n  \vert 
\label{jumpspectral}
\end{eqnarray}
involves the eigenvalues $(-\lambda_n) $ with $Re(\lambda_n)>0$ labeled by $n=1,..,N-1$ 
(while the vanishing highest eigenvalue $\lambda_0=0$ has been discussed in Eq. \ref{mastereigen0})
with their right eigenvectors $\vert r_n \rangle $ and their left eigenvectors $ \langle l_n  \vert  $
\begin{eqnarray}
 w \vert r_n \rangle && = -   \lambda_n \vert r_n \rangle 
 \nonumber \\
 \langle l_n  \vert  w  && = -   \lambda_n  \langle l_n  \vert 
\label{defeigenlambdan}
\end{eqnarray}
satisfying the othonormalization and closure relations 
\begin{eqnarray}
 \langle l_n  \vert r_m \rangle && = \delta_{nm}
 \nonumber \\
 \mathbb{1} && = \vert r_0 \rangle \langle l_0 \vert + \sum_{n=1}^{N-1}  \vert r_n \rangle \langle l_n  \vert
\label{fermeturej}
\end{eqnarray}
The spectral decomposition of Eq. \ref{jumpspectral} 
allows to write the decomposition into eigenmodes of the propagator of Eq. \ref{propagatordef}
\begin{eqnarray}
P_t(x \vert x_0) = \langle x \vert e^{wt} \vert x_0 \rangle 
 = P_*(x) +   \sum_{n=1}^{N-1} e^{-\lambda_n t} \langle x \vert r_n \rangle \langle l_n  \vert x_0 \rangle
\label{propagatorspectral}
\end{eqnarray}
and of the Green function $G(x,x_0)$ of Eq. \ref{defgreen} 
\begin{eqnarray}
G(x,x_0) = \int_0^{+\infty} dt \left[     \sum_{n=1}^{N-1} e^{- t\lambda_n } \langle x \vert r_n \rangle \langle l_n  \vert x_0 \rangle
\right]
 =   \sum_{n=1}^{N-1} \frac{  \langle x \vert r_n \rangle \langle l_n  \vert x_0 \rangle }{\lambda_n }
\label{Gspectral}
\end{eqnarray}
Its trace computed using Eq. \ref{fermeturej}
allows to rewrite the real-space Kemeny time 
 \begin{eqnarray}
\tau^{Space}_{*} =  \sum_{x=1}^{N} G(x,x) 
=  \sum_{n=1}^{N-1} \frac{ \sum_x \langle l_n  \vert x \rangle \langle x \vert r_n \rangle  }{\lambda_n }
=   \sum_{n=1}^{N-1} \frac{\langle l_n  \vert  r_n \rangle}{\lambda_n}
=   \sum_{n=1}^{N-1} \frac{1}{\lambda_n} \equiv \tau^{Spectral}_{*}
\label{tauspectraln}
\end{eqnarray}
as the sum of the inverses of the $(N-1)$ non-vanishing eigenvalues $\lambda_n \ne 0$
that corresponds to the spectral-Kemeny time introduced in Eq. \ref{tauspectral}.


\subsubsection{ When the generator is not diagonalizable and involves Jordan blocks}

 As already mentioned in the introduction
around Eq. \ref{tauspectral}, the expression of Eq. \ref{tauspectraln} remains valid even if
the generator $w$ is not diagonalizable and involves Jordan blocks,
as can be seen from the evaluation of the trace ${\rm Trace} ( e^{wt} ) $ of Eq. \ref{tautraceGbis}
when the generator is written in the canonical Jordan form with $\lambda_n$ on the diagonal
 \begin{eqnarray}
\tau^{Space}_{*} =  \int_0^{+\infty} dt \left[{\rm Trace} ( e^{wt} ) -1 \right]
= \int_0^{+\infty} dt \left[  \sum_{n=1}^{N-1} e^{- \lambda_n t} \right]
= \sum_{n=1}^{N-1} \frac{1}{\lambda_n} \equiv \tau^{Spectral}_{*}
\label{tautraceGbisjordan}
\end{eqnarray}
It should be stressed however that the presence of Jordan blocks will explicitly change 
other observables, in particular:

(i) the propagator of Eq. \ref{propagatorspectral} 
will contain powers of the time $t$ in front of the exponential functions 
associated to Jordan blocks;

(ii) the Green function of Eq. \ref{Gspectral}
will contain higher-inverse-powers of the eigenvalues associated to Jordan blocks.


\subsection{ Complementary discussions on the general properties of the Kemeny time $\tau_*$
 in Appendices }

Further discussions on the general properties of the Kemeny time $\tau_*$ 
can be found in the three Appendices:

$\bullet$ Appendix \ref{app_zeta} mentions the relations with the spectral zeta-function and the spectral partition function.

$\bullet$ Appendix \ref{app_DB} describes the specific spectral properties of reversible Markov processes satisfying detailed-balance.

$\bullet$ Appendix \ref{app_additive} recalls how the Green function allows to analyze the statistical properties of 
  all the time-additives observables of the Markov process.

 
\subsection{ Kemeny time for Markov processes with killing or absorption conditioned to survive forever}

\label{subsec_forever}
  
For a Markov jump process with $x=1,2,..N$ living configurations and 
one absorbing 'dead' configuration that will be denoted by $0$, the spectral analysis \cite{doorn,dickman,monasson,meleard,pollett,biblio,meleard_book,collet,c_population}
can be summarized as follows.
In the propagator of Eq. \ref{propagatorspectral}, the steady state is replaced by a delta function on the dead
configuration $x=0 $ 
\begin{eqnarray}
P^{[Abs]}_t(x \vert x_0) \equiv \langle x \vert e^{t w^{[Abs]} } \vert x_0 \rangle 
 = \delta_{x,0} +   \sum_{n=1}^{N} e^{- t \lambda_n^{[Abs]} } \langle x \vert r_n \rangle \langle l_n  \vert x_0 \rangle
\label{propagatorAbs}
\end{eqnarray}
It is then interesting to consider the process conditioned to survive forever, i.e. that remains forever among the $N$
living configurations $x=1,..,N$.
The corresponding conditioned Markov matrix $w^{[Cond]} $ can be computed from the initial Markov matrix $w^{[Abs]} $
with absorption 
and from the lowest relaxation rate $\lambda_{n=1}^{[Abs]} $ with its left eigenvector $\langle l_{n=1}  \vert $
(see \cite{c_population} for more details with the same notations)
\begin{eqnarray}
w^{[Cond]}(x , y) = \frac{\langle l_1  \vert x \rangle}{\langle l_1  \vert y \rangle} w^{[Abs]}(x , y) 
+ \lambda_{1}^{[Abs]} \delta_{x,y}
\label{wCond}
\end{eqnarray}
The propagator for this process conditioned to survive forever on the living configurations $x=1,..,N$
\begin{eqnarray}
P^{[Cond]}_t(x \vert x_0) && = \frac{\langle l_1  \vert x \rangle}{\langle l_1  \vert x_0 \rangle} 
e^{t \lambda_1^{[Abs]} } P^{[Abs]}_t(x \vert x_0) 
\nonumber \\
&&  =  \langle l_1  \vert x \rangle  \langle x \vert r_1 \rangle  
  + \sum_{n=2}^{N}  
   e^{- t \left[ \lambda_n^{[Abs]} - \lambda_1^{[Abs]}\right] } 
 \left[ \langle l_1  \vert x \rangle \langle x \vert r_n \rangle \right] \left[ 
  \frac{\langle l_n  \vert x_0 \rangle}{\langle l_1  \vert x_0 \rangle} \right]
\label{propagatorCond}
\end{eqnarray}
will converge towards the steady state 
 \begin{eqnarray}
P^{[Cond]}_*(x ) && =  \langle l_1  \vert x \rangle  \langle x \vert r_1 \rangle  
\label{steadyCond}
\end{eqnarray}
that is indeed normalized on the living configurations 
(again, see \cite{c_population} for more details with the same notations).
The $(N-1)$ non-vanishing eigenvalues that govern the convergence of the
propagator of Eq. \ref{propagatorCond}
towards its steady state
are simply given by the following differences of the initial eigenvalues $\lambda_n^{[Abs]} $
 \begin{eqnarray}
\lambda_m^{[Cond]}  = \lambda_{1+m}^{[Abs]} -\lambda^{[Abs]}_1 \ \ {\rm for } \ m=1,..,N-1
\label{lambdaCond}
\end{eqnarray}
For this process conditioned to survive forever,
the spectral Kemeny time of Eq. \ref{tauspectraln}
 \begin{eqnarray}
\tau^{[Cond]Spectral}_{*} =     \sum_{m=1}^{N-1} \frac{1}{\lambda^{[Cond]}_m} 
=  \sum_{m=1}^{N-1} \frac{1}{\lambda_{1+m}^{[Abs]} -\lambda^{[Abs]}_1}
\label{KemenyCond}
\end{eqnarray}
can be thus computed from the absorbing eigenvalues $\lambda_n^{[Abs]} $
of the initial absorbing Markov process, without even constructing the conditioned process.
Examples of Eq. \ref{lambdaCond} will be given in Eqs \ref{lambdaCondTwoAbs}
and \ref{lambdaCondOneAbs} with their corresponding Kemeny times.
  
 Besides the Kemeny time of Eq. \ref{KemenyCond} for the process conditioned to survive forever,
 its is also interesting to consider the analogs of the Kemeny time for
the initial unconditioned Markov process that gets absorbed in the dead configuration $0$,
but this discussion will be postponed to the very last section \ref{sec_Absorbing}
of the main text, in order to avoid interferences with all the other sections of the present paper
that concern ergodic Markov processes.
 
 
 
\subsection{ Kemeny time for Markov processes with stochastic resetting at rate $\gamma$ towards the distribution $\Pi$}

\label{subsec_reset}

Since stochastic resetting has been much studied recently 
(see the review \cite{review_reset} and references therein),
in particular from the point of view of their spectral properties \cite{garrahan_reset},
it is interesting to stress here the consequences for the spectral Kemeny time.

When the initial Markov jump process described by the Markov matrix $w$
is modified by the stochastic resetting at rate $\gamma$ towards some normalized distribution $\Pi$,
the process with resetting is described by the new Markov matrix
\begin{eqnarray}
w^{[Reset]}(x,y) \equiv w(x,y) + \gamma \left[ \Pi(x) - \delta_{x,y} \right]
\label{wresetgamma}
\end{eqnarray}
The propagator $P^{[Reset]}_t(x \vert x_0) $
 in the presence of resetting can be written in terms of the propagator $P_t(x \vert x_0)= \langle x \vert e^{t w } \vert x_0 \rangle $
 without resetting as
\begin{eqnarray}
P^{[Reset]}_t(x \vert x_0)  \equiv \langle x \vert e^{t w^{[Reset]} } \vert x_0 \rangle
= e^{- \gamma t} P_t(x \vert x_0) + \gamma \int_0^t d \tau e^{-\gamma \tau } 
\sum_y P_{\tau}(x \vert y) \Pi(y)
\label{propareset}
\end{eqnarray}
with the following interpretation.
The first contribution involves the probability $e^{- \gamma t} $ of no resetting during the time-window
$[0,t]$ and the initial propagator $P_t(x \vert x_0) $ from $x_0$ to $x$ in time $t$. 
The second contribution takes into account the last resetting event during the time-window
$[0,t]$, that occurs at some time $(t-\tau)$ and towards the position $y$ drawn with $\Pi(y)$,
so that it involves the initial propagator $P_{\tau}(x \vert y) $ from $y$ to $x$ in time $\tau$.
Plugging the spectral decomposition of Eq. \ref{propagatorspectral} for $P_t(x \vert x_0) $ and for $P_{\tau}(x \vert y) $
into Eq. \ref{propareset} 
\begin{eqnarray}
&& P^{[Reset]}_t(x \vert x_0) 
 = e^{- \gamma t} \left[ P_*(x) +   \sum_{n=1}^{N-1} e^{-\lambda_n t} \langle x \vert r_n \rangle \langle l_n  \vert x_0 \rangle \right]
+ \gamma \int_0^t d \tau e^{-\gamma \tau } 
\sum_y  \left[ P_*(x) +   \sum_{n=1}^{N-1} e^{-\lambda_n \tau} \langle x \vert r_n \rangle \langle l_n  \vert y \rangle \right] \Pi(y)
\nonumber \\
&& = e^{- \gamma t}  P_*(x) +   \sum_{n=1}^{N-1} e^{-(\gamma+ \lambda_n) t} \langle x \vert r_n \rangle \langle l_n  \vert x_0 \rangle 
+ [ 1- e^{-\gamma t } ]  P_*(x)  
  + \gamma 
    \sum_{n=1}^{N-1} \frac{ 1 - e^{-(\gamma+\lambda_n) \tau} }{ \gamma+\lambda_n } 
    \langle x \vert r_n \rangle \sum_y \langle l_n  \vert y \rangle \Pi(y) 
    \nonumber \\
&& = \left(  P_*(x) 
 + \gamma 
    \sum_{n=1}^{N-1} \frac{ \langle x \vert r_n \rangle  }{ \gamma+\lambda_n } 
    \left[  \sum_y \langle l_n  \vert y \rangle \Pi(y) \right] \right)
+   \sum_{n=1}^{N-1} e^{-(\gamma+ \lambda_n) t} \langle x \vert r_n \rangle 
\left( \langle l_n  \vert x_0 \rangle  
  -      \frac{  \gamma }{ \gamma+\lambda_n } 
     \left[ \sum_y \langle l_n  \vert y \rangle \Pi(y) \right] \right)
\label{proparesetspectral}
\end{eqnarray}
allows to identify the spectral properties of the process with resetting as follows.

(i) The vanishing eigenvalue $\lambda_0^{[Reset]} =0 $ is associated to the left trivial eigenvector
$\langle l_0^{[Reset]}  \vert x_0 \rangle =1$ as it should,
and to the following right eigenvector $\langle x \vert r_0^{[Reset]} \rangle $
corresponding to the new steady state
\begin{eqnarray}
P^{[Reset]}_*(x) \equiv \langle x \vert r_0^{[Reset]} \rangle
=P_*(x)
 + \gamma 
    \sum_{n=1}^{N-1} \frac{ \langle x \vert r_n \rangle  }{ \gamma+\lambda_n } 
    \left[  \sum_y \langle l_n  \vert y \rangle \Pi(y) \right]
\label{steadyreset}
\end{eqnarray}
Its physical meaning can be understood from the limit $t \to + \infty$ of Eq. \ref{propareset}
\begin{eqnarray}
P^{[Reset]}_*(x ) = \gamma \int_0^{+\infty} d \tau e^{-\gamma \tau } 
\sum_y P_{\tau}(x \vert y) \Pi(y)
\label{proparesetlimit}
\end{eqnarray}
where the initial propagator $P_{\tau}(x \vert y) $
is averaged over the time $\tau$ drawn with the exponential distribution $\left( \gamma  e^{-\gamma \tau } \right)$
and over the initial point $y$ drawn with the resetting probability $\Pi(y)$.

(ii) All the non-vanishing eigenvalues $\lambda_n^{[Reset]} \ne 0 $ 
are obtained from the initial non-vanishing eigenvalues $\lambda_n \ne 0 $
that are all increased by the same amount given by the resetting rate $\gamma$
 \begin{eqnarray}
\lambda_n^{[Reset]} = \lambda_n+ \gamma \ \ \ {\rm for } \ \ n=1,2,..,N-1
\label{lambdanreset}
\end{eqnarray}
The corresponding right eigenvectors are unchanged
 \begin{eqnarray}
\langle x \vert r_n^{[Reset]} \rangle = \langle x \vert r_n \rangle \ \ \ {\rm for } \ \ n=1,2,..,N-1
\label{rnreset}
\end{eqnarray}
while the new left eigenvectors read
 \begin{eqnarray}
\langle l_n^{[Reset]}  \vert x_0 \rangle = \langle l_n  \vert x_0 \rangle  
  -      \frac{  \gamma }{ \gamma+\lambda_n } 
     \left[ \sum_y \langle l_n  \vert y \rangle \Pi(y) \right] \ \ \ {\rm for } \ \ n=1,2,..,N-1
\label{lnreset}
\end{eqnarray}

Equivalently, all these spectral properties can be obtained from the corresponding eigenvalues equations
for the Markov matrix $w^{[Reset]}$ of Eq. \ref{wresetgamma} using Eqs \ref{defeigenlambdan} \ref{fermeturej},
as described in \cite{garrahan_reset}.

The spectral Kemeny time of Eq. \ref{tauspectraln} that involves the non-vanishing eigenvalues of 
Eq. \ref{lambdanreset}
 \begin{eqnarray}
\tau^{[Reset]Spectral}_{*} =     \sum_{n=1}^{N-1} \frac{1}{\lambda^{[Reset]}_n} 
=  \sum_{n=1}^{N-1} \frac{1}{\lambda_n+ \gamma}
\label{KemenyReset}
\end{eqnarray}
can be thus computed from the initial eigenvalues $\lambda_n \ne 0 $
and from the resetting rate $\gamma$.
Note that the resetting probability $\Pi(y)$ does not appear in the Kemeny time of Eq. \ref{KemenyReset}:
this property might seem somewhat surprising at first, since the resetting probability $\Pi(y)$
is essential to determine the new steady state $P^{[Reset]}_*(x ) $ of Eqs \ref{steadyreset} \ref{proparesetlimit},
towards which the convergence time is measured.
Examples of the formula of Eq. \ref{KemenyReset} 
are given in the three Tables 
  \ref{tableSpectralChains}
  \ref{tableSpectralDiffusion}
  \ref{tableSpectralRing}
and will be discussed in more details around Eqs 
\ref{KemenyResetcos}
\ref{lambdakcosnmuResetkemeny}
\ref{KemenyResetdiffuni}
\ref{eigenexcitedrhozeroreset}
\ref{KemenyResetDiff}
\ref{KemenyResetDiffmu0}
\ref{KemenyResetDirTrap}.


\section{ Markov chains on the interval of $N$ sites $x=1,2,..,N$ (always reversible)}

\label{sec_ChainInterval}

In this section, we focus on Markov chains on the interval of $N$ sites $x=1,2,..,N$,
where the dynamics always satisfy the detailed-balance condition 
as a consequence of the vanishing steady current at the two boundaries.

\subsection{ Forward master equation in terms of the flows and the currents of the links}

On each of the $(N-1)$ links $(x,x+1)$ with $x=1,..,N-1$,
it is useful to introduce the two flows ${\cal F}^{\pm}_t(x+1/2)$ in the two directions
produced by the two rates $w(x+1,x) $ and $w(x,x+1)$, 
and by the probability $P_t(y)$ to be at position $y$ at $t$
\begin{eqnarray}
{\cal F}_t^+ \left(x + \frac{1}{2} \right) && \equiv w(x+1,x) P_t(x)
\nonumber \\
{\cal F}_t^- \left(x + \frac{1}{2} \right) && \equiv w(x,x+1) P_t(x+1)
\label{flowslink}
\end{eqnarray}
so that their difference represents the current on this link
\begin{eqnarray}
J_t \left(x + \frac{1}{2} \right) && \equiv {\cal F}_t^+ \left(x + \frac{1}{2} \right) - {\cal F}_t^- \left(x + \frac{1}{2} \right) = w(x+1,x) P_t(x) - w(x,x+1) P_t(x+1)
\label{currentlink}
\end{eqnarray}

The forward master Eq. \ref{forward} for the probability $P_t(x)$ to be as position $x$ at time $t$
can be rewritten as continuity equations involving the currents of Eq. \ref{currentlink}
\begin{small}
\begin{eqnarray}
 x=1 :\ \ \  \ \ \partial_t P_t(1) && = - J_t \left( 1+\frac{1}{2} \right)
 = w(1,2) P_t(2) - w(2,1) P_t(1)  
\nonumber \\
x=2,..,N-1:   \partial_t P_t(x) && = J_t \left(x - \frac{1}{2} \right) - J_t \left(x + \frac{1}{2} \right)
= w(x,x-1) P_t(x-1) + w(x,x+1) P_t(x+1)- \left[ w(x-1,x) + w(x+1,x)\right] P_t(x)
\nonumber \\
x=N :\ \ \  \ \ \partial_t P_t(N) && =J_t \left(N - \frac{1}{2} \right)
= w(N,N-1) P_t(N-1) - w(N-1,N) P_t(N) 
\label{continuity}
\end{eqnarray}
\end{small}


\subsection{ Steady-state $P_*(x)$ with its vanishing steady current $J_*=0$ and its steady flows ${\cal F}_*\left(x + \frac{1}{2} \right) $}

For the steady state $P_*(x)$, 
Eq. \ref{continuity} yields that the steady current $J_* \left(x + \frac{1}{2} \right) $
should vanish on all the links $(x,x+1)$ 
\begin{eqnarray}
{\rm for } \ \ x=1,2,..,N-1 \ \ : \ \  0   && =   J_* \left(x + \frac{1}{2} \right) = 
 {\cal F}_*^+ \left(x + \frac{1}{2} \right) - {\cal F}_*^- \left(x + \frac{1}{2} \right)
\nonumber \\
&& =w(x+1,x) P_*(x) - w(x,x+1) P_*(x+1)
\label{zerosteadycurrent}
\end{eqnarray}
i.e.  the detailed-balance condition of Eq. \ref{DBdef} is satisfied with the many consequences summarized in Appendix \ref{app_DB}.
The steady state $P_*(x)$ can be obtained from the recurrence of Eq. \ref{zerosteadycurrent}
for arbitrary transition rates $w(.,.)$
\begin{eqnarray}
{\rm for } \ x=2,..,N : \ \ \ P_*(x) = P_*(1) \prod_{y=1}^{x-1} \frac{w(y+1,y)}{w(y,y+1) } \equiv \frac{ e^{- U(x) }}{Z}
\label{steady}
\end{eqnarray}
where the potential $U(x)$ can be computed with respect 
to the reference $U(x=1)=0$ to obtain for $x=2,..,N$
\begin{eqnarray}
U(x) =  \sum_{y=1}^{x-1} \ln \left( \frac{w(y,y+1) }{w(y+1,y)} \right) 
\label{steadyU}
\end{eqnarray}
The partition function $Z$ ensuring the normalization of the steady state of Eq. \ref{steady}
reads
\begin{eqnarray}
Z=\sum_{x=1}^N e^{-U(x)} =  1+ \sum_{x=2}^N \prod_{y=1}^{x-1} \frac{w(y+1,y)}{w(y,y+1) } 
\label{steadynorma}
\end{eqnarray}

As explained around Eq. \ref{DBdiff}, it is useful to introduce  
the diffusion coefficients associated to the $(N-1)$ links $(x,x+1)$
 \begin{eqnarray}
 D\left(x + \frac{1}{2} \right) \equiv \sqrt{w(x,x+1) w(x+1,x) } 
\label{DBdiffchain}
\end{eqnarray}
that will determine the steady flows of Eq. \ref{DBflow}
associated to the zero-current of Eq. \ref{zerosteadycurrent}
 \begin{eqnarray}
{\cal F}_*\left(x + \frac{1}{2} \right) && = w(x,x+1) P_*(x+1) =w(x+1,x) P_*(x)  
\nonumber \\
&& = D\left(x + \frac{1}{2} \right) \sqrt{P_*(x) P_*(x+1)  }
= D\left(x + \frac{1}{2} \right) \frac{ e^{- \frac{U(x)+U(x+1)}{2}} }{Z}
\label{DBflowchain}
\end{eqnarray}


\subsection{ Mean-First-Passage-Time $\tau(x,x_0)$ }

In the one-dimensional geometry, in order to go from $x_0$ to $x>x_0$,
one needs to visit sequentially all the sites between them,
i.e the first neighbor $(x_0+1)$, then the second neighbor $(x_0+2)$, etc up to the site $(x-1)$.
As a consequence, the Mean-First-Passage-Time $\tau(x,x_0)$ at $x$ when starting at  $x_0$
can be decomposed in the two regions $x>x_0$ and $x<x_0$ into the following sums 
over the links between them
\begin{eqnarray}
{\rm for } \ \ x>x_0 : \ \ \ \  \tau(x,x_0) && =\tau(x,x-1)+\tau(x-1,x-2)+ .. + \tau(x_0+1,x_0)
= \sum_{z=x_0 }^{x-1} \tau(z+1,z)
\nonumber \\
{\rm for } \ \ x<x_0 : \ \ \ \ \tau(x,x_0) && = \tau(x,x+1)+\tau(x+1,x+2)+ .. + \tau(x_0-1,x_0)
= \sum_{z=x}^{x_0-1} \tau(z,z+1)
 \label{tausum}
\end{eqnarray}
that involve the elementary Mean-First-Passage-Times $\tau(z + 1,z)$ 
and $\tau(z ,z+1) $ between two neighboring sites.

The backward equation of Eq. \ref{Ginvlefttau} for the Mean-First-Passage-Time $\tau(x,x_0)$ 
at $x$ when starting at $x_0$ can be then rewritten in these two regions as follows:

(i) for $x>x_0$, Eq. \ref{Ginvlefttau} reads
\begin{eqnarray}
{\rm for } \ \ x_0=1 : \ \ \ -1   
&& =    \left[   \tau(x,2)  - \tau(x,1) \right] w(2,1) 
 =    -   \tau(2,1) w(2,1) 
 \nonumber \\
{\rm for } \ \ x_0=2,..,x-1 : \ \ \ -1   
&& =    \left[   \tau(x,x_0+1)  - \tau(x,x_0) \right] w(x_0+1,x_0) 
+  \left[   \tau(x,x_0-1)  - \tau(x,x_0) \right] w(x_0-1,x_0) 
\nonumber \\
&& =    -   \tau(x_0+1,x_0) w(x_0+1,x_0) 
+     \tau(x_0,x_0-1) w(x_0-1,x_0) 
\nonumber \\
 \label{taubacksmaller}
\end{eqnarray}
The solution reads in terms of the potential $U(.)$ of Eq. \ref{steadyU}
and the diffusion coefficients $D(.)$ of Eq. \ref{DBdiffchain}
\begin{eqnarray}
  \tau(z+1,z) && = \frac{ 1 }{ D\left(z + \frac{1}{2} \right)  e^{- \frac{U(z)+U(z+1)}{2}} } \sum_{ y=1 }^{z}  e^{-U(y)}  
\label{taupUD}
\end{eqnarray}
or equivalently in terms of
the steady state of Eq. \ref{steady} and of the steady flow ${\cal F}_* \left(z + \frac{1}{2} \right) $
of Eq. \ref{DBflowchain}
\begin{eqnarray}
  \tau(z+1,z) && = \frac{ 1 }{{\cal F}_* \left(z + \frac{1}{2} \right)} \sum_{ y=1 }^{z}  P_*(y)  
    \equiv \frac{P_*([1,z]) } {{\cal F}_* \left(z + \frac{1}{2} \right)}
\label{taup}
\end{eqnarray}
where the numerator $P_*([1,z]) $ corresponds to the steady weight of the interval $[1,z]$, i.e. to the cumulative distribution associated to the steady state
\begin{eqnarray}
 P_*([1,z])  \equiv \sum_{ y=1 }^{z}  P_*(y)    
\label{cumul}
\end{eqnarray}

(ii) for $x<x_0$, Eq. \ref{Ginvlefttau} reads
\begin{eqnarray}
x_0=x+1,..,N-1 : \ \  -1    
&& =    \left[   \tau(x,x_0+1)  - \tau(x,x_0) \right] w(x_0+1,x_0) 
+  \left[   \tau(x,x_0-1)  - \tau(x,x_0) \right] w(x_0-1,x_0)
\nonumber \\
&& =      \tau(x_0,x_0+1)  w(x_0+1,x_0) 
  - \tau(x_0-1,x_0)  w(x_0-1,x_0) 
\nonumber \\
x_0=N : \ \  -1    
&& =     \left[   \tau(x,N-1)  - \tau(x,N) \right] w(N-1,N)
=       - \tau(N-1,N)  w(N-1,N) 
\label{taubackbigger}
\end{eqnarray}
The solution reads
in terms of the potential $U(.)$ of Eq. \ref{steadyU}
and the diffusion coefficients $D(.)$ of Eq. \ref{DBdiffchain}
as
\begin{eqnarray}
  \tau(z,z+1) && =   \frac{ 1 }{D\left(z + \frac{1}{2} \right)  e^{- \frac{U(z)+U(z+1)}{2}} }
 \sum_{y=z+1}^{N} e^{- U(y)}
\label{taumUD}
\end{eqnarray}
or equivalently
 in terms of the steady state of Eq. \ref{steady} and of the steady flow ${\cal F}_* \left(z + \frac{1}{2} \right) $
of Eq. \ref{DBflowchain}
\begin{eqnarray}
  \tau(z,z+1) && =   \frac{ 1 }{{\cal F}_* \left(z + \frac{1}{2} \right)}
 \sum_{y=z+1}^{N}  P_*(y)    = \frac{P_*([z+1,N]) } {{\cal F}_* \left(z + \frac{1}{2} \right)}
\label{taum}
\end{eqnarray}
where the numerator $ P_*([z+1,N])$ corresponds to the steady weight of the interval $[z+1,N]$, 
i.e. to the complementary cumulative distribution with respect to Eq. \ref{cumul}
\begin{eqnarray}
 P_*([z+1,N])  \equiv  \sum_{y=z+1}^{N}  P_*(y)= 1- \sum_{ y=1 }^{z}  P_*(y)    = 1-P_*([1,z]) 
\label{cumulcomplementary}
\end{eqnarray}

Plugging Eqs \ref{taup}
and \ref{taum}
into Eq. \ref{tausum}
yields the final result for the Mean-First-Passage-Times $\tau(x,x_0) $
\begin{eqnarray}
{\rm for } \ \ x_0<x : \ \ \ \  \tau(x,x_0) && = \sum_{z=x_0 }^{x-1} \tau(z+1,z)
=\sum_{z=x_0 }^{x-1} 
 \frac{ 1 }{{\cal F}_* \left(z + \frac{1}{2} \right)}
 \sum_{ y=1 }^{z}  P_*(y)  
\nonumber \\
{\rm for } \ \ x_0>x : \ \ \ \ \tau(x,x_0) && =\sum_{z=x}^{x_0-1} \tau(z,z+1)
= \sum_{z=x}^{x_0-1} 
\frac{ 1 }{{\cal F}_* \left(z + \frac{1}{2} \right)}
 \sum_{y=z+1}^{N}  P_*(y)
\label{tauDBM}
\end{eqnarray}


\subsection{ Real-space Kemeny time $\tau^{Space}_*$ }

\label{subsec_KemenyChainInterval}

The Real-space Kemeny time $\tau^{Space}_*$ of Eq. \ref{taukemeny}
does not depend on the initial configuration $x_0$
and can be thus evaluated for the special case $x_0=1$ using Eq. \ref{tauDBM}
 \begin{eqnarray}
\tau^{Space}_*  && =  \sum_{x=1}^N P_*(x) \tau(x,x_0=1) =  \sum_{x=2}^N P_*(x) \tau(x,x_0=1) 
=   \sum_{x=2}^N P_*(x) 
\sum_{z=1 }^{x-1}  \frac{ 1 }{{\cal F}_* \left(z + \frac{1}{2} \right)}
 \sum_{ y=1 }^{z}  P_*(y)
 \nonumber \\
 && = \sum_{z=1 }^{N-1}  \frac{ 1 }{{\cal F}_* \left(z + \frac{1}{2} \right)}
 \left[ \sum_{ y=1 }^{z}  P_*(y) \right] 
 \left[  \sum_{x=z+1}^N P_*(x) \right]
\equiv  \sum_{z=1 }^{N-1}  \Upsilon(z+1,z)
\label{taueqDBM}
\end{eqnarray}
In this sum over the $(N-1)$ links,
the elementary contribution $\Upsilon(z+1,z) $ of the link $(z,z+1)$ reads
 \begin{eqnarray}
\Upsilon(z+1,z)\equiv  \frac{ 1 }{{\cal F}_* \left(z + \frac{1}{2} \right)}
\left[ \sum_{ y=1 }^{z}  P_*(y) \right] 
 \left[  \sum_{x=z+1}^N P_*(x) \right]
= \frac{P_*([1,z])  P_*([z+1,N])}{{\cal F}_* \left(z + \frac{1}{2} \right)}
\label{upsilon}
\end{eqnarray}
The numerator contains the product of the cumulative distribution $P_*([1,z]) $ of Eq. \ref{cumul}
and of the complementary cumulative distribution $ P_*([z+1,N]) $ of Eq. \ref{cumulcomplementary},
while the denominator corresponds to the link steady flow ${\cal F}_* \left(z + \frac{1}{2} \right) $
of Eq. \ref{DBflowchain}.
The result of Eqs \ref{taueqDBM} and \ref{upsilon} is thus in direct correspondence with
the analog result for discrete-time Markov chains described in \cite{c_GW}.
If one replaces the steady state by Eq. \ref{steady}
and the steady flow by Eq. \ref{DBflowchain},
the elementary contribution $\Upsilon(z+1,z) $ of the link $(z,z+1)$
\begin{eqnarray}
\Upsilon(z+1,z)\equiv  \frac{ 1 }{ Z D\left(z + \frac{1}{2} \right)  e^{- \frac{U(z)+U(z+1)}{2}}}
\left[ \sum_{ y=1 }^{z}  e^{- U(y) }\right] 
 \left[  \sum_{x=z+1}^N e^{- U(x) } \right]
\label{upsilonUDZ}
\end{eqnarray}
involves the diffusion coefficient $D(.)$ of Eq. \ref{DBdiffchain}, the potential $U(.)$ of Eq. \ref{steadyU}
and its partition function $Z$ of Eq. \ref{steadynorma}.

In summary, for any Markov chain defined on the one-dimensional interval of $N$ sites $x=1,..,N$,
the explicit steady state $P_*(x)$ is given in Eq. \ref{steady} and
the Kemeny time can be computed via the real-space expression of Eq. \ref{taueqDBM}.
In particular, this gives an explicit result for the spectral Kemeny time of Eq. \ref{tauspectral}
even when the $(N-1)$ individual eigenvalues $\lambda_n$ are not explicit.


\subsection{ Example of the continuous-time symmetric random walk $w(x \pm 1,x)=1$  }

Let us consider the simplest example where all the rates have the same value unity
\begin{eqnarray}
w(x+1,x) =1 =w(x,x+1) \ \ \ {\rm for } \ x=1,2,..,N-1
\label{wuniform}
\end{eqnarray}
so the diffusion coefficient of Eq. \ref{DBdiffchain} is unity on all the links 
 \begin{eqnarray}
 D\left(x + \frac{1}{2} \right) \equiv \sqrt{w(x,x+1) w(x+1,x) } =1 \ \ \ {\rm for } \ x=1,2,..,N-1
\label{DBdiffchain1}
\end{eqnarray}
and the potential $U(x)$ of Eq. \ref{steadyU}  vanishes on all the sites
\begin{eqnarray}
 U(x) =  0  \ \ \ {\rm for } \ x=1,2,..,N
\label{steadyUzero}
\end{eqnarray}

\subsubsection{ Real-space Kemeny time  }

The steady state of Eqs \ref{steady} is uniform with $Z=N$
\begin{eqnarray}
P_*(x) && =   \frac{ 1  }{N} \ \ \ \ {\rm for } \ \ x=1,2,..,N
\label{steadymuzero}
\end{eqnarray}
so the real-space Kemeny time of Eq. \ref{taueqDBM} \ref{upsilonUDZ}
reduces to
 \begin{eqnarray}
\tau^{Space}_*   = \frac{1}{N} \sum_{z=1 }^{N-1}  
\left[ \sum_{ y=1 }^{z}  1\right] 
 \left[  \sum_{x=z+1}^N 1 \right]
  =\frac{1}{N} \sum_{z=1 }^{N-1}   z (N-z)
  = \frac{N^2-1}{6}
\label{spaceKemenymu0}
\end{eqnarray}
and displays the expected diffusive scaling $N^2$ for large size $N$.

\subsubsection{ Spectral Kemeny time  }

For the Markov chain of Eq. \ref{continuity} with the rates of Eq. \ref{wuniform}, 
the eigenvalue equation of Eq. \ref{defeigenlambdan}
reads for the right eigenvector $r(x)$ associated to the eigenvalue $\lambda$
\begin{eqnarray}
 x=1 :\ \ \  \ \ - \lambda r (1) && 
 =  r(2) -  r(1)  
\nonumber \\
x=2,..,N-1:   - \lambda r (x) && 
=  r(x-1) +  r(x+1)- 2 r(x)
\nonumber \\
x=N :\ \ \  \ \ - \lambda r (N) && 
=  r(N-1) -  r(N) 
\label{continuityr}
\end{eqnarray}
In the bulk $x=2,..,N-1$ where the recurrence is translation-invariant,
the general solution is a linear combination of the two Fourier solutions $e^{\pm i k x}$,
where the eigenvalue $\lambda$ is related to the wavevector $k$ via
\begin{eqnarray}
 \lambda(k) = 2- e^{-i k }- e^{i k } = 2 \left[ 1 -  \cos(k) \right]
\label{lambdakcos}
\end{eqnarray}
If one takes into account the satisfied bulk recurrence,
the two boundary conditions of Eq. \ref{continuity} for $x=1$ and $x=N$ 
are equivalent to $r(0)=r(1)$ and $r(N)=r(N+1)$.
The not-normalized linear combination $r_k(x)$ of the two Fourier solutions $e^{\pm i k x}$ 
satisfying $r(N)=r(N+1) $ can be written as 
\begin{eqnarray}
 r_k(x) =  \cos \left[ k \left(N+\frac{1}{2}-x\right)\right] \ \ { \rm for } \ \ x=1,..,N
\label{rkcos}
\end{eqnarray}
while the other boundary condition $r(0)=r(1)$ 
\begin{eqnarray}
0=r_k(0)-r_k(1)= 
 \cos \left[ k \left(N+\frac{1}{2}\right)\right] - \cos \left[ k \left(N-\frac{1}{2}\right)\right]
= - 2  \sin (kN) \sin \left(  \frac{k }{2}\right)
\label{rkcosboundary}
\end{eqnarray}
selects the $N$ possible wavevectors $k$ as a function of the size $N$ 
via the condition $\sin (kN)=0$ leading to
\begin{eqnarray}
k_n = n \frac{ \pi}{N} \ \ \ \ \ \ {\rm with } \ n=0,1,..,N-1
\label{rkcosselect}
\end{eqnarray}
with the corresponding eigenvalues of Eq. \ref{lambdakcos}
\begin{eqnarray}
 \lambda_n =   2 \left[ 1 -  \cos(k) \right] =  2 \left[ 1 - \cos\left( n  \frac{ \pi}{N}\right) \right]
 \ \ \ \ \ \ {\rm with } \ n=0,1,..,N-1
\label{lambdakcosn}
\end{eqnarray}
The spectral Kemeny time of Eq. \ref{tauspectral} that involves the $(N-1)$ non-vanishing eigenvalues $\lambda_n$ that can be computed using Eq. \ref{sumcos2}
 \begin{eqnarray}
\tau^{Spectral}_{*} 
\equiv  \sum_{n=1}^{N-1} \frac{1}{\lambda_n} 
= \frac{1}{2} \sum_{n=1}^{N-1} \frac{1}{1 - \cos\left( n  \frac{ \pi}{N}\right)} 
=\frac{N^2-1}{6}
\label{tauspectralcos}
\end{eqnarray}
is in agreement with the real-space Kemeny time of Eq. \ref{spaceKemenymu0} as it should.

\subsubsection{ Spectral Kemeny time in the presence of stochastic resetting at rate $\gamma$ }

In the presence of stochastic resetting at rate $\gamma$, the spectral Kemeny time of Eq. \ref{tauspectralcos}
is modified into Eq. \ref{KemenyReset} that can be computed using Eq. \ref{cos2sumhyperbolicres}
 \begin{eqnarray}
\tau^{[Reset]Spectral}_{*} 
&& =  \sum_{n=1}^{N-1} \frac{1}{\lambda_n+ \gamma}
=  \sum_{n=1}^{N-1} \frac{1}{(\gamma + 2) -2 \cos\left( n  \frac{ \pi}{N}\right)}
\nonumber \\
&& =  \frac{1}{  \sqrt{ \gamma ( 4 + \gamma ) } }
 \left[ N
   \frac{ \left( \sqrt{ \frac{\gamma}{4} }+ \sqrt{ \frac{\gamma}{4} +1}\right)^{4N} -1   }
   { \left( \sqrt{ \frac{\gamma}{4} }+ \sqrt{ \frac{\gamma}{4} +1}\right)^{4N} +1}
-    \frac{  \gamma +2 }{ \sqrt{ \gamma ( 4 + \gamma ) }} \right]
\label{KemenyResetcos}
\end{eqnarray}
In particular, the behavior for large $N$ of Eq. \ref{KemenyResetcos} is linear for any non-vanishing resetting rate $\gamma>0$
 \begin{eqnarray}
\tau^{[Reset]Spectral}_{*} 
\opsimeq_{N \to +\infty}   \frac{N}{\sqrt{ \gamma ( 4 + \gamma ) } } 
\label{KemenyResetcoslargeN}
\end{eqnarray}
in contrast to the diffusive scaling of Eq. \ref{tauspectralcos} when there is no resetting $\gamma=0$.
The crossover between these two scaling behaviors for large $N$ is analyzed in Eq. \ref{cos2sumhyperboliccrossovertranslation} of Appendix \ref{app_roots}.



\subsection{ Example of the continuous-time asymmetric random walk $w(x \pm 1,x)=e^{\pm \frac{\mu}{2} }$  }

Let us consider the biased Markov chain  
\begin{eqnarray}
w(x+1,x) && = e^{\frac{\mu}{2} } \ \ \ {\rm for } \ x=1,2,..,N-1
\nonumber \\
w(x,x+1) && = e^{-\frac{\mu}{2} }\ \ \ {\rm for } \ x=1,2,..,N-1
\label{wuniformmu}
\end{eqnarray}
so that the diffusion coefficient of Eq. \ref{DBdiffchain} is unity on all the links 
 \begin{eqnarray}
 D\left(x + \frac{1}{2} \right) \equiv \sqrt{w(x,x+1) w(x+1,x) } =1 \ \ \ {\rm for } \ x=1,2,..,N-1
\label{DBdiffchainmu}
\end{eqnarray}
while the potential $U(x)$ of Eq. \ref{steadyU} is linear 
\begin{eqnarray}
 U(x) =  - \mu (x-1) 
\label{steadyUlinear}
\end{eqnarray}
The steady state of Eqs \ref{steady} \ref{steadynorma}  is then exponential for $\mu \ne 0$
\begin{eqnarray}
P_*(x) && =  \frac{ e^{- U(x) }}{Z} = \frac{ e^{\mu (x-1)  }}{Z} 
\nonumber \\
Z && = \sum_{x=1}^N e^{\mu (x-1)  } = \frac{ e^{\mu N}-1}{ e^{\mu} -1}
\label{steadymu}
\end{eqnarray}
instead of the uniform steady state of Eq. \ref{steadymuzero} for $\mu=0$.


\subsubsection{ Real-space Kemeny time} 

The real-space Kemeny time of Eq. \ref{taueqDBM} \ref{upsilonUDZ}
reads
 \begin{eqnarray}
\tau^{Space}_*  && =  \sum_{z=1 }^{N-1}  \frac{ 1 }{ Z D\left(z + \frac{1}{2} \right)  e^{- \frac{U(z)+U(z+1)}{2}}}
\left[ \sum_{ y=1 }^{z}  e^{- U(y) }\right] 
 \left[  \sum_{x=z+1}^N e^{- U(x) } \right]
= \frac{1}{Z} \sum_{z=1 }^{N-1}  e^{ \frac{\mu}{2} - \mu z} 
\left[ \sum_{ y=1 }^{z}  e^{ \mu (y-1)  }\right] 
 \left[  \sum_{x=z+1}^N e^{\mu (x-1)  } \right]
  \nonumber \\ &&
 = \frac{e^{ \frac{\mu}{2}}}{Z (e^{\mu} -1 )^2 } \sum_{z=1 }^{N-1}  
  \left[   e^{\mu N} + 1 - e^{\mu z} - e^{\mu N} e^{ - \mu z}  \right]
= \frac{e^{ \frac{\mu}{2}}}{(e^{\mu N}-1) (e^{\mu} -1 ) }
  \left[   (e^{\mu N} + 1)(N-1)  - 2 \frac{ e^{\mu N} - e^{  \mu } }{ e^{  \mu } -1 }  \right]
 \nonumber \\
&& 
 =\frac{ 1}
{ 2 \sinh \left( \frac{\mu}{2} \right) }  
 \left[  \frac{ N }{\tanh \left( N \frac{\mu}{2} \right)}
  - \frac{ 1 }{ \tanh \left( \frac{\mu}{2} \right) }
  \right]
\label{spaceKemenymu}
\end{eqnarray}
The scaling for large $N$ is linear for any non-vanishing bias $\mu \ne 0$
 \begin{eqnarray}
\tau^{Space}_*  \opsimeq_{N \to + \infty} \frac{ N}
{ 2 \sinh \left( \frac{ \vert \mu \vert }{2} \right) }  
\label{spaceKemenymuNlarge}
\end{eqnarray}
in contrast to the diffusive scaling of Eq. \ref{spaceKemenymu0} for $\mu=0$.
The crossover between these two scaling behaviors for large $N$ is analyzed in Eq. \ref{cos2sumhyperboliccrossover} of Appendix \ref{app_roots} concerning spectral Kemeny times that we now describe.


\subsubsection{ Spectral Kemeny time } 

For the transition rates of Eq. \ref{wuniformmu},
the eigenvalue equation of Eq. \ref{defeigenlambdan}
reads for the right eigenvector $r(x)$ associated to the eigenvalue $\lambda$
\begin{eqnarray}
 x=1 :\ \ \  \ \ - \lambda r (1) && 
 =  e^{-\frac{\mu}{2} } r(2) - e^{\frac{\mu}{2} } r(1)  
\nonumber \\
x=2,..,N-1:   - \lambda r (x) && 
=  e^{\frac{\mu}{2} } r(x-1) + e^{-\frac{\mu}{2} } r(x+1)- \left[ e^{\frac{\mu}{2} }+e^{-\frac{\mu}{2} }\right]  r(x)
\nonumber \\
x=N :\ \ \  \ \ - \lambda r (N) && 
= e^{\frac{\mu}{2} } r(N-1) - e^{-\frac{\mu}{2} } r(N) 
\label{eigenrmu}
\end{eqnarray}
In the bulk $x=2,..,N-1$ where the recurrence is translation-invariant,
the general solution is a linear combination of the solutions $e^{ \frac{\mu}{2} x \pm i k x}$,
where the eigenvalue $\lambda$ is related to the bias and to the wavevector $k$ via
\begin{eqnarray}
 \lambda(k)=  2 \cosh \left(  \frac{\mu}{2} \right)  -2 \cos(k) 
\label{lambdakcosmu}
\end{eqnarray}
The not-normalized eigenstates $r_k(x)$ can be written as the linear combination
\begin{eqnarray}
 r_k(x) =  e^{ \frac{\mu}{2} x } \left[ A \cos(kx) + B \sin (kx)  \right] \ \ { \rm for } \ \ x=1,..,N
\label{rkcosmu}
\end{eqnarray}
When one takes into account the satisfied bulk recurrence,
the two boundary conditions of Eq. \ref{eigenrmu} for $x=1$ and $x=N$ 
are equivalent to $e^{\frac{\mu}{2} }r(0)=e^{-\frac{\mu}{2} }r(1)$ 
and $e^{\frac{\mu}{2} }r(N)=e^{-\frac{\mu}{2} }r(N+1)$, so that the coefficients $A$ and $B$ should satisfy
\begin{eqnarray}
0 && = r(0) - e^{- \mu }r(1) = A -  e^{ - \frac{\mu}{2}  } \left[ A \cos(k) + B \sin (k)  \right]
\nonumber \\
0 && =  r(N) - e^{- \mu }r(N+1) =
e^{ \frac{\mu}{2} N } \left[ A \cos(k N) + B \sin (k N)  \right]
- e^{ \frac{\mu}{2} N } e^{ - \frac{\mu}{2}  }\left[ A \cos(k (N+1)) + B \sin (k(N+1))  \right]
\label{rkcosboundarymu}
\end{eqnarray}
The possible wavevectors $k$ are selected via the condition of vanishing determinant for this system,
and one finally recovers exactly the same condition $\sin (kN)=0$ leading to the same values of Eq. \ref{rkcosselect}
of the case $\mu=0$, so that the corresponding eigenvalues of Eq. \ref{lambdakcosmu} read
\begin{eqnarray}
 \lambda_n =  2 \cosh \left(  \frac{\mu}{2} \right)  - 2 \cos\left( n  \frac{ \pi}{N}\right) 
 \ \ \ \ \ \ {\rm with } \ n=0,1,..,N-1
\label{lambdakcosnmu}
\end{eqnarray}
The spectral-Kemeny time of Eq. \ref{tauspectral}  that can be computed using Eq. \ref{cos2sumhyperbolic}
 \begin{eqnarray}
\tau^{Spectral}_{*} 
\equiv  \sum_{n=1}^{N-1} \frac{1}{\lambda_n} 
= \sum_{n=1}^{N-1} \frac{1}{2 \cosh \left(  \frac{\mu}{2} \right) - 2 \cos\left( n  \frac{ \pi}{N}\right)} 
=\frac{ 1}
{ 2 \sinh \left( \frac{\mu}{2} \right) }  
 \left[  \frac{ N }{\tanh \left( N \frac{\mu}{2} \right)}
  - \frac{ 1 }{ \tanh \left( \frac{\mu}{2} \right) }
  \right]
\label{tauspectralcosmu}
\end{eqnarray}
is in agreement with the real-space-Kemeny time of Eq. \ref{spaceKemenymu}.


\subsubsection{ Spectral Kemeny time in the presence of stochastic resetting at rate $\gamma$}

In the presence of stochastic resetting at rate $\gamma$, 
the excited eigenvalues of Eq. \ref{lambdakcosnmu}
are shifted via Eq. \ref{lambdanreset}
\begin{eqnarray}
 \lambda_n{[Reset]} =  \gamma + 2 \cosh \left(  \frac{\mu}{2} \right)  - 2 \cos\left( n  \frac{ \pi}{N}\right) 
 \ \ \ \ \ \ {\rm with } \ n=1,..,N-1
\label{lambdakcosnmuReset}
\end{eqnarray}
where the parameter $\mu$ and the resetting rate $\gamma$ only appear 
via the global variable $\left[  \gamma + 2 \cosh \left(  \frac{\mu}{2} \right)\right]$
and lead to the spectral Kemeny time
\begin{eqnarray}
\tau^{[Reset]Spectral}_* = \sum_{n=1}^{N-1} \frac{1}{\lambda_n^{[Reset]}}
 = \frac{\left[ N
   \frac{ \left(  \frac{\sqrt{\gamma +4 \sinh^2 \left(  \frac{\mu}{4} \right)}+\sqrt{\gamma +4 \cosh^2 \left(  \frac{\mu}{4} \right)} }{2} \right)^{4N} -1   }
   { \left(  \frac{\sqrt{\gamma +4 \sinh^2 \left(  \frac{\mu}{4} \right)}+\sqrt{\gamma +4 \cosh^2 \left(  \frac{\mu}{4} \right)} }{2} \right)^{4N} +1}
-    \frac{  \gamma + 2 \cosh \left(  \frac{\mu}{2} \right) }
{ \sqrt{ \left[ \gamma +4 \sinh^2 \left(  \frac{\mu}{4} \right)\right] 
\left[ \gamma +4 \cosh^2 \left(  \frac{\mu}{4} \right)\right] }} \right]}
 {  \sqrt{ \left[ \gamma +4 \sinh^2 \left(  \frac{\mu}{4} \right)\right] 
\left[ \gamma +4 \cosh^2 \left(  \frac{\mu}{4} \right)\right] } }
\label{lambdakcosnmuResetkemeny}
\end{eqnarray}


\subsection{ Example $w(x\pm 1,x)  = \frac{1}{ 2 \vartheta_x} $ of the symmetric trap model where each site $x$ has its own trapping time $\vartheta_x$ }

Trap models have been much studied in the context of anomalously slow glassy behaviors 
\cite{jpb_weak,dean,jp_ac,jp_pheno,bertinjp1,bertinjp2,trapsymmetric,trapnonlinear,trapreponse,trap_traj,c_ruelle,c_ring}.
The symmetric trap model on the one-dimensional interval $\{1,..,N\}$ is defined by the transition rates
\begin{eqnarray}
w(x+1,x) && = \frac{1}{ 2 \vartheta_x} \ \ {\rm for} \ \ x=1,2,..,N-1
\nonumber \\
w(x-1,x) && = \frac{1}{ 2 \vartheta_x} \ \ {\rm for} \ \ x=2,3,..,N
\label{wjumptrap}
\end{eqnarray}
So when the particle is on a bulk site $x=2,..N-1$ at time $t$,
 the escape-time $t \in [0,+\infty[$ from the site $x$ follows the exponential probability distribution
\begin{eqnarray}
p^{escape}_x(t) =   \frac{ 1 }{\vartheta_x}  e^{ - \frac{ t }{\vartheta_x} }
\ \ \ \ \ \ \text {with the normalization } \ \ \ \ \int_0^{+\infty} dt p^{escape}_x(t) =1
\label{escapejump}
\end{eqnarray}
whose averaged value is directly $\vartheta_x$ that will be called the trapping time of site $x$
\begin{eqnarray}
\int_0^{+\infty}dt  \ t \ p^{escape}_x(t) =   \vartheta_x
\label{escapejumpav}
\end{eqnarray}
and the particle jumps towards one of the two neighbors $(x \pm 1)$
with equal probabilities $(\frac{1}{2},\frac{1}{2})$.

The steady state $P_*(x)$ at site $x$ is simply proportional to its trapping time $\vartheta_x $
\begin{eqnarray}
P_*(x) && =  \frac{ \vartheta_x }{ Z}
\nonumber \\
Z && = \sum_{y=1}^L \vartheta_y
\label{steadySymTrap}
\end{eqnarray}
while the corresponding steady flows of Eq. \ref{DBflowchain} 
take the same value on all the links $x=1,..,N-1$
 \begin{eqnarray}
{\cal F}_*\left(x + \frac{1}{2} \right)  = w(x,x+1) P_*(x+1) =w(x+1,x) P_*(x) = \frac{1}{2 Z} 
\label{DBflowchainSymTrap}
\end{eqnarray}

The real-space Kemeny time of Eq. \ref{taueqDBM} 
reads
 \begin{eqnarray}
\tau^{Space}_*  && = \sum_{z=1 }^{N-1}  \frac{ 1 }{{\cal F}_* \left(z + \frac{1}{2} \right)}
 \left[ \sum_{ y=1 }^{z}  P_*(y) \right] 
 \left[  \sum_{x=z+1}^N P_*(x) \right]
\nonumber \\
&&  = \frac{2}{Z} \sum_{z=1 }^{N-1}  
 \left[ \sum_{ y=1 }^{z}   \vartheta_y \right]
 \left[  \sum_{x=z+1}^N   \vartheta_x \right]
 = \frac{2}{Z}  \sum_{ y=1 }^{N-1}   \vartheta_y
  \sum_{x=y+1}^N   \vartheta_x 
 \sum_{z=y }^{x-1}   1
 \nonumber \\
&& =  \frac{ \displaystyle \sum_{ y=1 }^{N-1}   \sum_{x=y+1}^N 2 (x-y)  \vartheta_y  \vartheta_x  }
{ \displaystyle \sum_{z=1}^N \vartheta_z }  
\label{taueqDBMsymTrap}
\end{eqnarray}
In the numerator, each pair $x<y$ of sites contributes with their trapping times $\vartheta_y $ and $\vartheta_x $, but also with their distance $(x-y)$, so that the positions of the trapping times within the sample is important, in contrast to the Directed version of the model that will be discussed later in Eqs \ref{kemenyDir} \ref{kemenyDirbis}.


\subsection{ Markov chains with explicit eigenvalues $\lambda_n$ associated to discrete orthogonal polynomials}

As a final remark, let us mention that the spectral Kemeny time $\tau^{Spectral}_{*}  $
can be computed for the Markov chains on intervals $x \in \{1,..,N\}$ 
with explicit eigenvalues $\lambda_n$ associated to discrete orthogonal polynomials
with the following possible dependencies with respect to $n$  (see \cite{SolvableBD} for more details):

(i) a first family of models is characterized by the quadratic and/or linear dependence in $n$
\begin{eqnarray}
 \lambda_n =  c_2 n^2+c_1 n
\label{family1}
\end{eqnarray}
The simplest examples involve rates $w(x \pm 1,x)$ that are linear or quadratic with respect to $x$.

(ii) a second family of models is characterized by the $(q^{-n}-1) $ and/or $(1-q^{n}) $ dependence in $n$
involving some parameter $q$
\begin{eqnarray}
 \lambda_n =  c_-(q^{-n}-1)+ c_+(1-q^{n})
\label{family2}
\end{eqnarray}
The simplest examples involve rates $w(x \pm 1,x)$ that are linear or quadratic with respect to $q^x$.


\section{ Diffusion processes on intervals $]x_L,x_R[$ (always reversible) }

\label{sec_DiffInterval}

In this section, we focus on diffusion processes on intervals $]x_L,x_R[$,
where the dynamics always satisfies the detailed-balance as a consequence of the vanishing
steady current at the two boundaries. The important technical difference with the previous section
 is that the space of configurations is continuous instead of discrete, 
 while the physical results are in direct correspondence as already discussed on a specific case in \cite{c_SkewDB}.
 Previous mathematical references on the Kemeny time for one-dimensional diffusion processes include 
 \cite{Pinsky_kemeny,MaoDiffusion2015,choi_diff,choi_stein,choi_diff_JacobiTaboo}
 (see Appendix \ref{app_SDE} for the translation of some notations and terminology).


\subsection{ Forward and backward equations for the propagator $p_t(x \vert x_0) $  }

The propagator $p_t(x \vert x_0) $ satisfies the 
Fokker-Planck equation with respect to the final position $x$
\begin{eqnarray}
 \partial_t p_t(x \vert x_0)  && =     \partial_{x}  \left[ - F(x)  + D (x)  \partial_{x} \right] p_t(x \vert x_0)
  \equiv {\cal L}_x p_t(x \vert x_0)
\label{fokkerplanckforward}
\end{eqnarray}
where the differential operator 
\begin{eqnarray}
 {\cal L}_x  =     \partial_{x}  \left[ - F(x)  + D (x)  \partial_{x} \right] 
\label{generator}
\end{eqnarray}
replaces the Markov matrix $w(.,.)$ of the previous sections.
The backward dynamics with respect to the initial position $x_0$
\begin{eqnarray}
 \partial_t p_t(x \vert x_0)  && =   {\cal L}^{\dagger}_{x_0} p_t(x \vert x_0)
\label{fokkerplanckbackward}
\end{eqnarray}
involves the adjoint operator of Eq. \ref{generator}
\begin{eqnarray}
 {\cal L}^{\dagger}_{x_0}  =  \left[  F(x_0)  +   \partial_{x_0} D (x_0) \right]    \partial_{x_0}  
\label{adjoint}
\end{eqnarray}

The Fokker-Planck Eq. \ref{fokkerplanckforward}
can be rewritten as a continuity equation analog to Eq. \ref{continuity}
\begin{eqnarray}
 \partial_t p_t(x\vert x_0)    =   -  \partial_{x}   j_t(x\vert x_0)
\label{fokkerplanck}
\end{eqnarray}
 where the current $j_t(x\vert x_0)$ associated to $p_t(x\vert x_0)$
involves the force $F(x) $ and the diffusion coefficient $D(x)$
\begin{eqnarray}
j_t(x\vert x_0)  = F(x)  p_t(x \vert x_0) - D (x)  \partial_{x} p_t(x\vert x_0)
\label{fokkerplanckj}
\end{eqnarray}
The conservation of the probability on the interval $]x_L,x_R[$
\begin{eqnarray}
 1 = \int_{x_L}^{x_R} dx p_t(x\vert x_0)   
\label{norma}
\end{eqnarray}
imposes the following boundary conditions.
The derivative of Eq. \ref{norma} with respect to time and the forward dynamics of 
Eqs \ref{fokkerplanckforward} \ref{fokkerplanck}
\begin{eqnarray}
 0 = \int_{x_L}^{x_R} dx \partial_t p_t(x\vert x_0)  
 = \int_{x_L}^{x_R} dx  {\cal L}_x  p_t(x\vert x_0)
  = - \int_{x_L}^{x_R} dx \partial_{x}   j_t(x\vert x_0)
 =  j_t(x_L\vert x_0) -  j_t(x_R\vert x_0)
\label{normaderitx}
\end{eqnarray}
yields that the current $j_t(x \vert x_0) $ should vanish at the two boundaries at $x=x_L$ and $x=x_R$
\begin{eqnarray}
 j_t(x_L\vert x_0) =0 =  j_t(x_R\vert x_0)
\label{currentxLxR}
\end{eqnarray}
as expected from the physical point of view.
At the technical level, this means that the second-order differential operator $ {\cal L}_x $ of Eq. \ref{generator}
acting on any function $r(x)$ on the interval $x \in ]x_L,x_R[$ 
has to be supplemented by the following vanishing-current boundary conditions at $x=x_L$ and at $x=x_R$
\begin{eqnarray}
{\rm B.C. \ \ for } \ \ {\cal L}_x : \ \ \ \ \ \    \left[  F(x) r(x) - D (x)  r'(x) \right]_{x=x_L} = 0 = \left[  F(x) r(x) - D (x)  r'(x) \right]_{x=x_R} 
\label{generatorCL}
\end{eqnarray}
in order to be fully defined.
Then the evaluation via two integrations by parts 
of the following integral involving another function $l(x)$
\begin{eqnarray}
  \int_{x_L}^{x_R} dx l(x)   {\cal L}_x r(x) 
  && =
  - \int_{x_L}^{x_R} dx l(x)  \frac{d}{dx}   \left[  F(x) r(x)  - D (x)   r'(x) \right]   
   = \int_{x_L}^{x_R} dx   \left[  F(x) r(x) - D (x)  r'(x) \right]        l'(x)
   \nonumber \\
 &&  = \int_{x_L}^{x_R} dx  r(x) \left[  F(x)    l'(x) + \partial_x \left( D (x)   l'(x) \right) \right]
- \left[  r(x)  D (x)   l'(x)\right]_{x=x_L}^{x=x_R}
 \nonumber \\
 && =  \int_{x_L}^{x_R} dx  r(x)  {\cal L}^{\dagger}_{x} l(x)
+  r(x_L)  D (x_L)   l'(x_L) -   r(x_R)  D (x_R)   l'(x_R)
\label{scalarprod}
\end{eqnarray}
yields that the proper definition of 
the action of the adjoint operator $ {\cal L}_x^{\dagger} $ on some function $l(x)$ 
should include, besides the explicit differential expression of Eq. \ref{generator} on the interval $x \in ]x_L,x_R[$,
the following boundary conditions at $x=x_L$ and $x=x_R$
\begin{eqnarray}
{\rm B.C. \ \ for } \ \ {\cal L}_x^{\dagger} : \ \ \ \ \ \    
\left[   D (x)  l'(x) \right]_{x=x_L} = 0 = \left[   D (x)  l'(x) \right]_{x=x_R} 
\label{adjointCL}
\end{eqnarray}


\subsection{ Steady-state $p_*(x)$ with its vanishing steady current $j_*=0$}

The steady-state $p_*(x)$ of the Fokker-Planck Eq. \ref{fokkerplanck} 
with the zero-current boundary conditions of Eq. \ref{currentxLxR}
can be found by the vanishing of the steady-current $j_*(x)=0$ on the interval $x \in ]x_L,x_R[$
\begin{eqnarray}
0= j_*(x) =  F(x)   p_*( x ) - D (x)   p_*'( x) 
\label{zFPJsteady}
\end{eqnarray}
i.e. as already mentioned for the Markov chain around Eq. \ref{zerosteadycurrent},
on a one-dimensional interval with boundaries one cannot avoid detailed-balance.
The steady state $p_*(x)$ can be rewritten as the Boltzmann distribution of Eq. \ref{eq}
\begin{eqnarray}
  p_*(x)  = \frac{ e^{ -U(x)} }{Z}
 \label{steadyeq}
\end{eqnarray}
where the potential $U(x)$ involves some arbitrary reference position $x_{ref} \in [x_L,x_R]$
that can be chosen as one wishes in specific examples
\begin{eqnarray}
U(x) && \equiv - \int_{x_{ref}}^x dy \frac{F(y)}{D(y)} 
\nonumber \\
U'(x) && = - \frac{F(x)}{D(x)}
 \label{UR}
\end{eqnarray}
while the normalization $Z$ corresponds to the partition function of the interval $ ]x_L,x_R[$ 
\begin{eqnarray}
Z=  \int_{x_L}^{x_R} dx e^{ -  U(x) } 
 \label{partitioneq}
\end{eqnarray}


\subsection{ Associated quantum supersymmetric Hamiltonian $H$}

\label{sub_susy}

The change of variables involving the steady state $p_*(.)$ of Eq. \ref{steadyeq} with its potential $U(.)$ of Eq. \ref{UR}
\begin{eqnarray}
p_t(x \vert x_0)  = \sqrt{   \frac{ p_*(x)}{p_*(x_0) } }  \psi_t(x \vert x_0) = 
 e^{  \frac{ U(x_0)-U(x)}{2} } \psi_t(x \vert x_0)
\label{ppsi}
\end{eqnarray}
transforms the forward dynamics of Eq. \ref{fokkerplanckforward} for the propagator $ p_t(x,x_0) $
 into an Euclidean Schr\"odinger equation for the quantum propagator $\psi_t(x \vert x_0)$
\begin{eqnarray}
-  \partial_t \psi_t(x \vert x_0)  = H \psi_t(x \vert x_0)
\label{schropsi}
\end{eqnarray}
The corresponding quantum Hermitian Hamiltonian 
\begin{eqnarray}
 H = H^{\dagger} =  - \frac{ \partial  }{\partial x} D(x) \frac{ \partial  }{\partial x} +V(x)
\label{hamiltonien}
\end{eqnarray}
involves the scalar potential
\begin{eqnarray}
V(x) && \equiv D(x)  \frac{ [U'(x)]^2 }{4 }  -D(x) \frac{U''(x)}{2} -D'(x) \frac{U'(x)}{2}
\nonumber \\
&& = \frac{ F^2(x) }{4 D(x) } + \frac{F'(x)}{2}
\label{vfromu}
\end{eqnarray}
that allows to factorize the Hamiltonian of Eq. \ref{hamiltonien}
into the well-known supersymmetric form (see the review \cite{review_susyquantum} and references therein)
\begin{eqnarray}
H \equiv    Q^{\dagger} Q
\label{hsusy}
\end{eqnarray}
involving the two first-order operators 
\begin{eqnarray}
Q  && \equiv    \sqrt{ D(x) }  \left( \frac{ d }{ d x}  +\frac{ U'(x)}{2 } \right)
\nonumber \\
Q^{\dagger}  &&\equiv  \left(   - \frac{ d }{ d x}  +\frac{ U'(x)}{2 } \right)\sqrt{ D(x) }
\label{qsusy}
\end{eqnarray}
The factorization of Eq. \ref{hsusy} can be considered as the continuous-space counterpart of Eq. \ref{Hproj}
concerning Markov chains with discrete configurations.

The consequences for the spectral properties have already been discussed
in Appendix \ref{app_DB} for the case of discrete configurations.
In particular, 
the positive groundstate wavefunction $ \phi_0(x) $ associated to the zero-eigenvalue $\lambda_0=0$
corresponds to the square-root of the steady state $p_*(x)$
as in Eq. \ref{psi0}
\begin{eqnarray}
  \phi_0 (x) = \sqrt{p_*(x) } = \frac{ e^{- \frac{U(x)}{2}  }}{\sqrt Z}
\label{psi0diff}
\end{eqnarray}
and is annihilated by the first-order differential operator $Q$ of Eq. \ref{qsusy} 
\begin{eqnarray}
Q     \phi_0(x)  =   \sqrt{ D(x) }  \left( \frac{ d }{ d x}  +\frac{ U'(x)}{2 } \right) \frac{ e^{ - \frac{ U(x)}{2} } }{\sqrt Z} =0
\label{qsusyeq}
\end{eqnarray}

The zero-current boundary conditions of Eqs \ref{currentxLxR} \ref{generatorCL}
mean that the current $j_n(x)$ associated to any right-eigenvector $r_n(x)$ of the Fokker-Planck generator
\begin{eqnarray}
 j_n(x)  \equiv F(x)  r_n(x) - D (x) r_n'(x)(x) = D(x) \left[ - U'(x) r_n(x) - r_n'(x)  \right] 
\label{jn}
\end{eqnarray}
should vanish at the two boundaries
\begin{eqnarray}
 j_n(x_L) =0 =  j_n(x_R)
\label{jnboundary}
\end{eqnarray}
Via the transformation of Eq. \ref{eigenLR}
\begin{eqnarray}
r_n(x)  && =\sqrt{ p_*(x)} \phi_n(x) =  \frac{ e^{- \frac{U(x)}{2}  }}{\sqrt Z} \phi_n(x)
\nonumber \\
 l_n (x) && = \frac{   \phi_n (x) }{ \sqrt{ p_*(x)}} =   \frac{ e^{ \frac{U(x)}{2}  }}{\sqrt Z} \phi_n(x)
\label{eigenRnpsin}
\end{eqnarray}
the current of Eq. \ref{jn} that should vanish at the two boundaries
reads in terms of the quantum eigenvector $\phi_n(x)$
\begin{eqnarray}
 j_n(x)  = F(x)  r_n(x) - D (x) r_n'(x) = 
 D(x) \frac{ e^{- \frac{U(x)}{2}  }}{\sqrt Z} \left[ -  \frac{U'(x)}{2} \phi_n(x) - \phi_n'(x)  \right]
\label{jnphi}
\end{eqnarray}
and in terms of the left eigenvector $l_n(x)$
\begin{eqnarray}
 j_n(x)  = F(x)  r_n(x) - D (x) r_n'(x) =  - D(x) l_n'(x)
\label{jnphiln}
\end{eqnarray}
so that one recovers the boundary conditions of Eq. \ref{adjointCL}
for the adjoint operator ${\cal L}_x^{\dagger} $ as it should for consistency.


\subsection{ Mean-First-Passage-Time $\tau(x,x_0)$ }

The backward equation of Eq. \ref{Ginvlefttau} for the Mean-First-Passage-Time $\tau(x,x_0)$ 
at $x$ when starting at $x_0$ now involves the adjoint operator $ {\cal L}^{\dagger}_{x_0} $ of Eq. \ref{adjoint}
that can be rewritten using $F(x_0)= -U'(x_0) D(x_0)  $ of Eq. \ref{UR} as
\begin{eqnarray}
-1 + \frac{ \delta(x_0-x)  }{ p_*(x)}   =  {\cal L}^{\dagger}_{x_0} \tau(x,x_0)
&& =  \left[  F(x_0)  +   \partial_{x_0} D (x_0) \right]    \partial_{x_0}  \tau(x,x_0)
\nonumber \\
&& =  \left[  -U'(x_0)    +   \partial_{x_0}  \right]  D (x_0)  \partial_{x_0}  \tau(x,x_0)
 \label{adjointtau}
\end{eqnarray}
while the boundary conditions of Eq. \ref{adjointCL}
for the adjoint operator ${\cal L}_{x_0}^{\dagger} $ reads
\begin{eqnarray}
   \left[   D (x_0) \partial_{x_0}  \tau(x,x_0) \right]_{x_0=x_L} = 0 
= \left[   D (x_0) \partial_{x_0}  \tau(x,x_0) \right]_{x_0=x_R} 
\label{adjointCLtau}
\end{eqnarray}

Since the final position $x$ plays the role of a parameter, 
it is convenient to introduce the auxiliary function
\begin{eqnarray}
 \chi^{[x]}(x_0) \equiv  D (x_0)  \partial_{x_0}  \tau(x,x_0)
   \label{nudifftau}
\end{eqnarray}
that should vanish at the two boundaries $x_0=x_L$ and $x=x_R$ as a consequence of Eq. \ref{adjointCLtau},
while Eq. \ref{adjointtau} becomes the first order differential equation for $\chi^{[x]}(x_0)$
\begin{eqnarray}
-1 + \frac{ \delta(x_0-x)  }{ p_*(x)}    =  \left[  -U'(x_0)    +   \partial_{x_0}  \right]   \chi^{[x]}(x_0)
 \label{adjointtaudiff}
\end{eqnarray}

In the two regions $x_0<x$ and $x_0>x$, the solutions that vanish at the boundaries $x_0=x_L$ and $x_0=x_R$ respectively can be rewritten in terms of 
the potential $U(.)$ or in terms of the steady state $p_*(.)$ of Eq. \ref{steadyeq} 
\begin{eqnarray}
 \chi^{[x]}(x_0) && = - e^{U(x_0)} \int_{x_L}^{x_0} dz e^{- U(z) } = - \frac{1}{p_*(x_0) }\int_{x_L}^{x_0} dz p_*(z) \ \ \ {\rm \ \ for} \ \  x_0 \in ]x_L,x[
\nonumber \\
   \chi^{[x]}(x_0) && =  e^{U(x_0)} \int_{x_0}^{x_R} dz e^{- U(z) } = \frac{1}{p_*(x_0) } \int_{x_0}^{x_R} dz p_*(z)\ \ \ {\rm \ \ for} \ \  x_0 \in ]x,x_R[
    \label{gfirstbigger}
\end{eqnarray}
The integration of Eq. \ref{adjointtaudiff} over $x_0$ between $(x-\epsilon)$ and $(x+\epsilon)$ for $\epsilon \to 0^+$ corresponds to the discontinuity
\begin{eqnarray}
 \frac{ 1  }{ p_*(x)}  &&  = \int_{x-\epsilon}^{x+\epsilon} dx_0 \left[  -U'(x_0)    +   \partial_{x_0}  \right]   \chi^{[x]}(x_0)
 =  \chi^{[x]}(x+\epsilon) - \chi^{[x]}(x-\epsilon)
\nonumber \\
&&  = \frac{1}{p_*(x) } \int_{x}^{x_R} dz p_*(z) + \frac{1}{p_*(x) }\int_{x_L}^{x} dz p_*(z)
=  \frac{1}{p_*(x) } \int_{x_L}^{x_R} dz p_*(z) 
 \label{adjointtaudiffdisc}
\end{eqnarray}
which is satisfied using the normalization of the steady state.

We may now use Eq. \ref{nudifftau}
\begin{eqnarray}
\partial_{x_0}  \tau(x,x_0) = \frac{ \chi^{[x]}(x_0)}{  D (x_0)  } 
   \label{nudifftauint}
\end{eqnarray}
with the solution of Eq. \ref{gfirstbigger} for $\chi^{[x]}(x_0) $
to obtain the Mean-First-Passage-Time $\tau(x,x_0)$ in the two regions $x_0<x$ and $x_0>x$
\begin{eqnarray}
{\rm \ \ for} \ \  x_0 \in ]x_L,x[ : \ 
 \tau(x,x_0) &&=\int_{x_0}^x dy \frac{1}{D(y) e^{-U(y)} }\int_{x_L}^{y} dz e^{-U(z)}
 = \int_{x_0}^x dy \frac{1}{D(y) p_*(y) }\int_{x_L}^{y} dz p_*(z) 
 \equiv \int_{x_0}^x dy \frac{ p_*([x_L,y]) }{D(y) p_*(y) }
\nonumber \\
{\rm \ \ for} \ \  x_0 \in ]x,x_R[: \  \tau(x,x_0) &&= \int_{x}^{x_0} dy \frac{1}{D(y)e^{-U(y)}  } \int_{y}^{x_R} dz e^{-U(z)}= \int_{x}^{x_0} dy \frac{1}{D(y) p_*(y) } \int_{y}^{x_R} dz p_*(z)  
\equiv \int_{x}^{x_0} dy \frac{p_*([y,x_R]) }{D(y) p_*(y) }  
\nonumber \\
\label{mftp}
\end{eqnarray}
where one recognizes the cumulative distribution $p_*([x_L,y]) $ 
and the complementary cumulative distribution $p_*([y,x_R]) $ analog to Eqs \ref{cumul} \ref{cumulcomplementary}
\begin{eqnarray}
p_*([x_L,y])  && \equiv \int_{x_L}^{y} dz p_*(z) 
\nonumber \\
p_*([y,x_R])  && \equiv \int_{y}^{x_R} dz p_*(z) = 1 - \int_{x_L}^{y} dz p_*(z) 
   \label{cumulcumul}
\end{eqnarray}
while the denominator $ D(y) p_*(y) $ is the continuous analog of the discrete-space flow of Eq. \ref{DBflowchain}.
The final results of Eqs \ref{mftp} are thus the direct continuous analogs of the discrete-space expressions 
of Eqs \ref{taupUD} \ref{taup} \ref{taumUD} \ref{taum}.


\subsection { Real-space Kemeny time }

The real-space Kemeny time of Eq. \ref{taukemeny} can be evaluated for the initial position 
at the left boundary $x_0=x_L$ using Eq. \ref{mftp}
 \begin{eqnarray}
\tau^{Space}_* && = \int_{x_L}^{x_R} dx p_*(x) \tau(x,x_0=x_L)    
= \int_{x_L}^{x_R} dx p_*(x) \int_{x_L}^x dy \frac{1}{D(y) p_*(y) }\int_{x_L}^{y} dz p_*(z) 
\nonumber \\
&& =\int_{x_L}^{x_R} dy \frac{1}{D(y) p_*(y) }
\left[ \int_{x_L}^{y} dz p_*(z) \right]
\left[ \int_{y}^{x_R} dx p_*(x) \right]
\nonumber \\
&& \equiv \int_{x_L}^{x_R} dy \frac{p_*([x_L,y]) p_*([y,x_R])}{D(y) p_*(y) }
\label{kemeny1d}
\end{eqnarray}
where the numerator involves the product of the cumulative distribution $p_*([x_L,y]) $ 
and the complementary cumulative distribution $p_*([y,x_R]) $ of Eq. \ref{cumulcumul},
while the denominator $ D(y) p_*(y) $ is the continuous analog of the discrete-space flow of Eq. \ref{DBflowchain}.
The final result of Eq. \ref{kemeny1d} is thus the direct continuous analog of the discrete-space expression of 
Eqs \ref{taueqDBM} \ref{upsilon}.

The conclusion is thus the same as at the end of subsection \ref{subsec_KemenyChainInterval}:
for any diffusion process defined on the interval $]x_L,x_R[$,
the steady state $p_*(x)$ is known via Eq. \ref{steadyeq} \ref{UR} and
the Kemeny time can be computed by the real-space expression of Eq. \ref{kemeny1d},
even when the individual eigenvalues $\lambda_n$ are not explicit.

From the physical point of view, one of the most important issue 
is the scaling of the Kemeny time $\tau_*$ with respect to the size $L=x_R-x_L$ of the interval.
Let us now describe various examples.


\subsection{ Example: diffusion coefficient $D(x)=1$ 
and no force $F(x)=0$ on the interval $[0,L]$ }

\label{subsec_DiffpureL}

\subsubsection{ Real-space Kemeny time  }

The steady state of Eq. \ref{steadyeq} is uniform
\begin{eqnarray}
  p_*(x)   = \frac{1}{L}  \ \ \ \ \ {\rm for } \ \ 0 \leq x \leq  L
 \label{steadyequni}
\end{eqnarray}
So the real-space Kemeny time Eq. \ref{kemeny1d} reduces to
 \begin{eqnarray}
\tau^{Space}_*    
 =\int_0^L dy \frac{1}{ p_*(y) }\int_{0}^{y} dz p_*(z) \int_{y}^{L} dx p_*(x) 
 = \frac{1}{L} \int_0^L dy y (L-y) = \frac{L^2}{6}
\label{kemeny1duni}
\end{eqnarray}
and displays the expected diffusive scaling in $L^2$ with respect to the size $L$.

\subsubsection{ Spectral Kemeny time  }

The Fourier modes $r(x)=\cos (kx)$ that have a vanishing derivative at $x=0$
should also have a vanishing derivative at $x=L$, so that the condition $\sin(kL)=0$ selects
 the series of possible wavevectors as a function of the size $L$
 \begin{eqnarray}
    k_n = \frac{ \pi n}{L} \ \ \ {\rm with } \ \  n=0,1,..,+\infty
\label{eigenexcitedunikn}
\end{eqnarray}
 with the corresponding eigenvalues
\begin{eqnarray}
\lambda_n =    k_n^2 = \frac{ \pi^2 n^2}{L^2} \ \ \ {\rm with } \ \  n=0,1,..,+\infty
\label{eigenexciteduni}
\end{eqnarray}
The spectral Kemeny time of Eq. \ref{tauspectral} involving the non-vanishing eigenvalues
 \begin{eqnarray}
\tau^{Spectral}_{*} =  \sum_{n=1}^{+\infty} \frac{1}{\lambda_n} = \frac{L^2}{ \pi^2 } \sum_{n=1}^{+\infty} \frac{1}{n^2}
= \frac{L^2}{ 6 } 
\label{tauspectraluni}
\end{eqnarray}
is in agreement with the real-space Kemeny time of Eq. \ref{kemeny1duni} as it should.


\subsubsection{ Spectral Kemeny time in the presence of stochastic resetting at rate $\gamma$ }

In the presence of stochastic resetting at rate $\gamma$, the spectral Kemeny time of Eq. \ref{tauspectraluni}
is modified into Eq. \ref{KemenyReset} 
 \begin{eqnarray}
\tau^{[Reset]Spectral}_{*} 
 =  \sum_{n=1}^{+\infty} \frac{1}{\lambda_n+ \gamma}
=  \sum_{n=1}^{+\infty} \frac{1}{ \gamma + \frac{ \pi^2 n^2}{L^2} }
    = \frac{1}{ 2 \gamma } \left[ \frac{ \sqrt{\gamma} L }{\tanh \left(\sqrt{\gamma} L \right)}  -1\right] 
\label{KemenyResetdiffuni}
\end{eqnarray}
In particular, the behavior for large $L$ is linear for any non-vanishing resetting rate $\gamma>0$
 \begin{eqnarray}
\tau^{[Reset]Spectral}_{*} 
\opsimeq_{L \to +\infty}   \frac{L}{2 \sqrt{ \gamma  } } 
\label{KemenyResetcoslargeNN}
\end{eqnarray}
in contrast to the diffusive scaling of Eq. \ref{kemeny1duni} when there is no resetting $\gamma=0$.

In order to describe the crossover between the two scaling behaviors as $L^2$ in Eq. \ref{kemeny1duni} 
and as $L$ in Eq. \ref{KemenyResetcoslargeNN} for large $L$, one can divide Eq. \ref{KemenyResetdiffuni}
by its value of Eq. \ref{kemeny1duni} without resetting $\gamma=0$
 \begin{eqnarray}
\frac{\tau^{[Reset]Spectral}_{*} }{\tau^{Spectral}_{*} }
    = \frac{3}{  (  \sqrt{\gamma} L)^2} \left[ \frac{ \sqrt{\gamma} L }{\tanh \left(\sqrt{\gamma} L \right)}  -1\right] 
    \equiv {\cal T} \left( {\hat L} = \sqrt{\gamma} L \right)
\label{KemenyResetdiffuniCrossover}
\end{eqnarray}
in order to obtain the following scaling function for the rescaled variable ${\hat L} \equiv  \sqrt{\gamma} L$
 \begin{eqnarray}
{\cal T} ({\hat L})    = \frac{3}{   {\hat L}^2} \left[ \frac{ {\hat L} }{\tanh {\hat L}}  -1\right] 
\label{KemenyResetdiffuniCrossoverCalT}
\end{eqnarray}


\subsection{ Example: diffusion coefficient $D(x)=1$ and force $F(x) = \pm \mu$ 
on two intervals $]0,\rho L[$ and $]\rho L,L[$ }

\label{subsec_sawtoothDB}

Let us now focus on the simple example on the interval $[0,L]$,
where the diffusion coefficient is constant $D(x)=1$
while the force $F(x)$ takes only two values $(\pm \mu)$ and changes at the position $(\rho L)$ parametrized by $\rho$
 \begin{eqnarray}
F(x) && = - \mu \ \ \ \ {\rm for } \ \ 0<x<\rho L
\nonumber \\
F(x) && =  \mu \ \ \ \ \ \ {\rm for } \ \  \rho L<x<L
\label{forcesaw}
\end{eqnarray}
The potential of Eq. \ref{UR} is then the saw-tooth potential
\begin{eqnarray}
U(x) && =  \mu x \ \ \ \ \ \ \ \ \ \ \ \ \ \ {\rm for } \ \ 0 \leq x \leq \rho L
\nonumber \\
U(x) && = 2 \mu \rho L  - \mu x \ \ \ {\rm for } \ \ \rho L \leq x \leq L
 \label{URsawtooth}
\end{eqnarray}
with the three important extremal values 
\begin{eqnarray}
U(x=0) && =0
\nonumber \\
U(x=\rho L) && =  \mu \rho L
\nonumber \\
U(x=L) && =  \mu (2 \rho -1 ) L
 \label{URsawtoothextreme}
\end{eqnarray}
that will allow to analyze different physical interesting 
cases depending on the two parameters $\mu$ and $\rho$.

The steady state of Eq. \ref{steadyeq} reads
\begin{eqnarray}
  p_*(x)  && = \frac{ e^{ - \mu x } }{Z}  \ \ \ \ \ {\rm for } \ \ 0 \leq x \leq \rho L
  \nonumber \\
p_*(x)  && = \frac{ e^{ \mu x - 2 \mu \rho L } }{Z} \ \ \ {\rm for } \ \ \rho L \leq x \leq L
 \label{steadyeqsaw}
\end{eqnarray}
with the partition function
\begin{eqnarray}
Z && =  \int_{0}^{L} dx e^{ -  U(x) } = \int_{0}^{\rho L} dx e^{ - \mu x } + e^{- 2 \mu \rho L} \int_{\rho L}^{L} dx e^{\mu x   }
 =   \frac{ 1+ e^{\mu (1- 2 \rho ) L  }-  2 e^{ -  \mu \rho L  }}{\mu}
 \label{partitioneqsaw}
\end{eqnarray}
The corresponding cumulative distributions of Eq. \ref{cumulcumul}
\begin{eqnarray}
{\rm for } \ 0 \leq y \leq \rho L : \ \ \ p_*([0,y])  && = \int_{0}^{y} dx p_*(x) 
= \int_{0}^{y} dx \frac{ e^{ - \mu x } }{Z} = \frac{ 1- e^{ - \mu y } }{ \mu Z}
\nonumber \\
{\rm for } \ \rho L \leq y \leq  L : \ \ \ p_*([y,L])  && =  \int_{y}^L dx p_*(x)
=  e^{- 2 \mu \rho L}  \   \frac{ e^{ \mu  L  } - e^{ \mu y  } }{ \mu Z}
   \label{cumulcumulsaw}
\end{eqnarray}
can be plugged into Eq. \ref{kemeny1d} to compute the real-space Kemeny time
 \begin{eqnarray}
\tau^{Space}_* && =\int_{0}^{\rho L} dy \frac{p_*([0,y]) \left[ 1- p_*([0,y])\right]}{ p_*(y) }
+ \int_{\rho L }^{L} dy \frac{p_*([y,L]) \left[ 1- p_*([y,L])\right]}{ p_*(y) }
\nonumber \\
&& = \int_{0}^{\rho L} dy \frac{ \frac{ 1- e^{ - \mu y } }{ \mu Z} \left[ 1-  \frac{ 1- e^{ - \mu y } }{ \mu Z}\right]}{ \frac{ e^{ - \mu y } }{Z} }
+ \int_{\rho L}^{L} dy 
\frac{  e^{- 2 \mu \rho L}    \frac{ e^{ \mu  L  } - e^{ \mu y  } }{ \mu Z} \left[ 1-  e^{- 2 \mu \rho L}    \frac{ e^{ \mu  L  } - e^{ \mu y  } }{ \mu Z}\right]}{ \frac{ e^{ \mu y - 2 \mu \rho L } }{Z} }
\label{kemeny1dsawcalcul}
\end{eqnarray}
These elementary integrals and the value of the partition function $Z$ of Eq. \ref{partitioneqsaw}
leads to the final result 
\begin{eqnarray}
\tau_* = \frac{  2 e^{\mu (1-  \rho ) L  }   
   +  e^{ \mu (1-2 \rho) L } [ \mu (1- 2\rho) L - 4 ]
 +   e^{ -  \mu \rho L  } [ 6  + 2   \mu  L ]
   + \mu ( 2\rho-1) L -4 }
   {\mu^2  \left[ 1+ e^{\mu (1- 2 \rho ) L  }-  2 e^{ -  \mu \rho L  } \right] } 
\label{kemeny1dsaw}
\end{eqnarray}
It is now interesting to discuss the scaling of the Kemeny time $\tau_*$
for large size $L$ as a function of the two parameters $(\rho,\mu)$ 
defining the saw-tooth potential of Eq. \ref{URsawtoothextreme}.


\subsubsection{ Case $\rho=0$: Diffusion $D(x)=1$ and constant force  $F(x)=\mu \ne 0 $ on the interval $[0,L]$ }

For $\rho=0$, the real-space Kemeny time of Eq. \ref{kemeny1dsaw} reduces to
\begin{eqnarray}
\tau^{Space}_* = \frac{    e^{ \mu  L } [ \mu  L - 2 ] +    [     \mu  L  +2]  }
   {\mu^2  \left[  e^{\mu  L  }-  1  \right] } 
   = \frac{2}{ \mu^2 } \left[ \frac{ \frac{\mu L}{2} }{\tanh \left(\frac{\mu L}{2} \right)}  -1\right] 
\label{kemeny1dsawrhozero}
\end{eqnarray}
so the leading behavior for large $L$ 
\begin{eqnarray}
\tau^{Space}_* \opsimeq_{L \to + \infty} 
    \frac{2}{ \mu^2 } \left[ \frac{ \frac{\mu L}{2} }{ {\rm sgn } (\mu ) }  \right] = \frac{L }{  \vert \mu \vert } 
\label{kemeny1dsawrhozeroscaling}
\end{eqnarray}
is linear for any non-vanishing force $\mu \ne 0$ instead of the diffusive scaling of Eq. \ref{kemeny1duni} for vanishing force $\mu=0$.
In order to describe the crossover between these two scaling behaviors for large $L$, one can divide Eq. \ref{kemeny1dsawrhozero}
by its value of Eq. \ref{kemeny1duni} without force $\mu=0$
 \begin{eqnarray}
\frac{\tau^{Space}_{*} }{ \frac{L^2}{6} }
    = \frac{3}{ \left(\frac{\mu L}{2} \right)^2} \left[ \frac{ \frac{\mu L}{2} }{\tanh \left(\frac{\mu L}{2} \right)}  -1\right] 
    \equiv {\cal T} \left( {\hat L} =\frac{\vert \mu \vert}{2}  L \right)
\label{kemeny1dsawrhozeroCrossover}
\end{eqnarray}
where one recognizes the scaling function ${\cal T} ({\hat L})  $ already encountered in Eq. \ref{KemenyResetdiffuniCrossoverCalT}
 but for the different rescaled variable ${\hat L} \equiv  \frac{\vert \mu \vert L}{2} L$,
 that involves the force $\mu$ instead of the resetting rate $\gamma$.
 In order to understand why the force $\mu$ and the resetting rate $\gamma$ 
 have similar effects on the Kemeny time,
 whereas they correspond to completely different processes
 with completely different steady states, 
 it is useful to consider the excited eigenvalues 
 in the presence of the force $\mu$ computed in Eq. \ref{eigenexcitedrhozero}
of Appendix \ref{app_EigenForceConstant}
\begin{eqnarray}
\lambda_n =  \frac{ \mu^2 }{4  } +  k_n^2 = \frac{ \mu^2 }{4  } + \frac{ \pi^2 n^2}{L^2} \ \ \ {\rm with } \ \  n=1,2,..,+\infty
\label{eigenexcitedrhozerotext}
\end{eqnarray}
that lead to the spectral Kemeny time 
\begin{eqnarray}
\tau^{Spectral}_*= \sum_{n=1}^{+\infty} \frac{1}{\lambda_n}
= \frac{L^2}{\pi^2} \sum_{n=1}^{+\infty} \frac{1}{  n^2+ \left( \frac{ \mu L}{2  \pi}\right)^2  }
  = \frac{2}{ \mu^2 } \left[ \frac{ \frac{\mu L}{2} }{\tanh \left(\frac{\mu L}{2} \right)}  -1\right] 
\label{eigenexcitedrhozerok}
\end{eqnarray}
 in agreement with the real-space Kemeny time of Eq. \ref{kemeny1dsawrhozero}.

In the presence of stochastic resetting at rate $\gamma$ in the presence of the force $\mu$, 
the excited eigenvalues of Eq. \ref{eigenexcitedrhozerotext}
are shifted via Eq. \ref{lambdanreset}
\begin{eqnarray}
\lambda_n^{[Reset]} = \gamma + \frac{ \mu^2 }{4  } + \frac{ \pi^2 n^2}{L^2} \ \ \ {\rm with } \ \  n=1,2,..,+\infty
\label{eigenexcitedrhozeroreset}
\end{eqnarray}
and lead to the spectral Kemeny time
\begin{eqnarray}
\tau^{[Reset]Spectral}_*= \sum_{n=1}^{+\infty} \frac{1}{\lambda_n^{[Reset]}}
=\frac{1}{ 2 \left(\gamma+  \frac{ \mu^2 }{4  } \right) } \left[ \frac{ \sqrt{\gamma+  \frac{ \mu^2 }{4  }} L }{\tanh \left(\sqrt{\gamma+  \frac{ \mu^2 }{4  }} L \right)}  -1\right]
\label{eigenexcitedrhozerokReset}
\end{eqnarray}
where the force $\mu$ and the resetting rate $\gamma$ only appear via the global variable $\sqrt{\gamma+  \frac{ \mu^2 }{4  }} $.


\subsubsection{ Case $\mu = -  \vert \mu \vert <0$: one valley with minimum $U(x=\rho L)  = -  \vert \mu \vert \rho L $ }

For $\mu = -  \vert \mu \vert <0$, the leading behavior of Eq. \ref{kemeny1dsaw}
for large $L$
\begin{eqnarray}
\tau_*^{[\mu<0]} \opsimeq_{L \to + \infty}  \frac{     e^{   \vert \mu \vert \rho L  } [   - 2     \vert \mu \vert  L ] }
   {\mu^2  \left[ -  2 e^{   \vert \mu \vert \rho L  } \right] } 
   = \frac{L }{  \vert \mu \vert } 
\label{kemeny1dsawmuneg}
\end{eqnarray}
is linear and does not depend on $\rho$, so that it coincides with Eq. \ref{kemeny1dsawrhozero}
corresponding to $\rho=0$.


\subsubsection{ Case $\mu>0$ and $0 <\rho< 1 $: maximum $U(x=\rho L)  =  \mu \rho L $
separating two valleys with minima at $x=0$ and $x=L$ }

\label{subsec_sawBarrier}

Here one expects that the result will depend on the smallest of the two barriers
\begin{eqnarray}
B_1(L) \equiv U(x=\rho L)  -U(x=0) && =  \mu \rho L
 \nonumber \\
B_2(L) \equiv  U(x=\rho L)  -U(x=L) && =  \mu (1-\rho)  L
\label{kemeny1dsaw2barr}
\end{eqnarray}
i.e. that it will depend on the sign of $\left(\rho-\frac{1}{2} \right)$ as follows:

(i) For $0 <\rho< \frac{1}{2} $, the leading behavior of Eq. \ref{kemeny1dsaw}
for large $L$
\begin{eqnarray}
\tau^{[\mu>0;0 <\rho< \frac{1}{2} ]}_* \opsimeq_{L \to + \infty}  \frac{  2 e^{\mu (1-  \rho ) L  }    }
   {\mu^2  \left[  e^{\mu (1- 2 \rho ) L  } \right] } = \frac{2}{\mu^2} e^{\mu \rho L} 
   = \frac{2}{\mu^2} e^{B_1(L)} 
\label{kemeny1dsawmuposrhosmall}
\end{eqnarray}
is exponentially large in $L$ and is dominated by the barrier $B_1(L)=\mu \rho L$ of Eq. \ref{kemeny1dsaw2barr}.

(ii) For $\frac{1}{2} <\rho< 1 $, the leading behavior of Eq. \ref{kemeny1dsaw}
for large $L$
\begin{eqnarray}
\tau_*^{[\mu>0;\frac{1}{2} <\rho< 1]} \opsimeq_{L \to + \infty}  \frac{  2 e^{\mu (1-  \rho ) L  }    }
   {\mu^2  \left[  1 \right] } = \frac{2}{\mu^2} e^{\mu (1-\rho) L} 
   = \frac{2}{\mu^2} e^{B_2(L)} 
\label{kemeny1dsawmuposrhobig}
\end{eqnarray}
is exponentially large in $L$ and is dominated by the barrier $B_2(L)=\mu (1-\rho) L$ of Eq. \ref{kemeny1dsaw2barr}.

(iii) For the symmetric case $\rho=\frac{1}{2}$ where the two barriers of Eq. \ref{kemeny1dsaw2barr}
coincide $B_1(L)=B_2(L)=\frac{\mu}{2} L $, Eq. \ref{kemeny1dsaw}
reduces to
\begin{eqnarray}
\tau_*^{[\mu>0;\rho=\frac{1}{2}]} = \frac{   e^{ \frac{\mu}{2} L  }       +   e^{ -   \frac{\mu}{2} L  } [ 3  + 1   \mu  L ]    -4 }
   {  \mu^2  \left[ 1-   e^{ -   \frac{\mu}{2} L  } \right] } 
   \opsimeq_{L \to + \infty} \frac{1}{\mu^2} e^{\frac{\mu}{2} L} 
\label{kemeny1dsawdemi}
\end{eqnarray}

In Appendix \ref{app_EigenSawTooth}, these three exponentially-large asymptotic behaviors of 
the real-space Kemeny times of Eqs \ref{kemeny1dsawmuposrhosmall} \ref{kemeny1dsawmuposrhobig}
and \ref{kemeny1dsawdemi} are explained by considering only the 
relaxation time $\tau_1 \equiv \frac{1}{\lambda_1 }$ 
associated to the lowest non-vanishing eigenvalue $\lambda_1$
(see Eqs \ref{group2hgapinverse} and \ref{tau1imaginary}).



\subsection{ Spectral Kemeny times for the diffusion processes with the explicit Jacobi spectra $\lambda_n= A n (n+B) $ }

The Pearson family of ergodic diffusions (see the mini review in section 2 of the mathematical paper \cite{pearson_class} 
and the recent physics preprint \cite{c_pearson} with references therein) contains the one-dimensional diffusions
 with a positive quadratic diffusion coefficient $D(x)$
and a linear force $F(x)$, i.e. equivalently a linear Ito force $ f^I(x) $ or a linear Stratonovich force $ f^S(x) $ as a consequence of Eq. \ref{fitostrato}. 
These technical simplifications produce six types of simple analytical steady states
and allow to compute the corresponding spectrum of eigenvalues \cite{pearson_class,c_pearson,pearson_wong}.
The only case with a finite Kemeny time 
corresponds to the Jacobi spectrum of eigenvalues
\begin{eqnarray}
\lambda_n= A n (n+B) \ \ \ {\rm for } \ \ \ n=0,1,...+\infty \ \ \ \ \ \ {\rm with \ parameters } \ \ A>0 \ \ B> -1
\label{jacobilambda}
\end{eqnarray}
and to the only Pearson diffusion defined on a finite interval.
Indeed, the other cases defined on $]-\infty,+\infty[$ or $[0,+\infty[$
are associated to the linear Ornstein-Uhlenbeck spectrum discussed in Eq. \ref{eigenOU}
or contain a continuum part in the spectrum besides a finite number of discrete states \cite{pearson_class,c_pearson,pearson_wong}, so that their Kemeny times diverge
as explained in Appendix \ref{app_infinite},
where the criteria for the convergence or the divergence for the
Kemeny times of diffusion processes defined on infinite intervals $(x_R-x_L) \to + \infty$ 
are discussed in detail in real-space and in the spectral domain.

The spectral Kemeny time associated to the Jacobi spectrum of Eq. \ref{jacobilambda}
\begin{equation}
      \tau_*^{Spectral[Jacobi]}  =   \sum_{n=1}^{+\infty} \frac{ 1 }{\lambda_n } = 
    \frac{1}{A}  \sum_{n=1}^{+\infty} \frac{ 1 }{n(n + B) }  
  =  \frac{H(B)}{A B} 
\label{kemenyjacobi}
\end{equation}
involves the Harmonic number function $H(B)$ that can be written as an integral 
or in terms of the $\Gamma$ function and of the Euler constant $\gamma_{Euler}$
\begin{equation}
 H(B) = \int_0^1 du \frac{1-u^B}{1-u}  = \gamma_{Euler}+ \frac{ \Gamma’(B+1) }{\Gamma(B+1) }
\label{harmonic}
\end{equation}
and that has a simple fractional value when the parameter $B$ is an integer
\begin{equation}
{\rm Integer } \ B =1,2,..,+\infty \ \ : \ \  H(B) = \int_0^1 du \left[ \sum_{k=0}^{B-1} u^k \right]   = \sum_{k=0}^{B-1}\frac{1}{k+1}
= \sum_{m=1}^B \frac{1}{m}
\label{harmonicBinteger}
\end{equation}

Let us now describe some examples characterized by the Jacobi spectrum of Eq. \ref{jacobilambda}
starting with the Jacobi process itself.

\subsubsection{ Jacobi process on $]0,1[$ converging towards the Beta-distribution of parameters $\alpha>0$ and $\beta>0$ with $D(x) =  x (1-x)  $ } 

For the Jacobi process \cite{pearson_class,c_pearson} defined on the finite interval $ x \in ]0,1[$, 
the diffusion coefficient is a positive quadratic polynomial
vanishing at the two boundaries $x=0$ and $x=1$
\begin{eqnarray}
D(x) = A x (1-x) \ \ \ {\rm with } \ A>0
\label{jacobiDx}
\end{eqnarray}
while the normalized steady state is the Beta-distribution of parameters $\alpha>0$ and $\beta>0$
\begin{eqnarray}
p_*(x) && = \frac{ x^{\alpha-1} (1-x)^{\beta-1} }{Z(\alpha,\beta)} \ \ \ {\rm for } \ x \in ]0,1[
\nonumber \\
Z(\alpha,\beta) && = \int_0^1 x^{\alpha-1} (1-x)^{\beta-1} = \frac{ \Gamma(\alpha) \Gamma(\beta) }{\Gamma(\alpha+\beta)}
\label{jacobiBeta}
\end{eqnarray}
Applications of the Jacobi process can be found
in mathematical finance~\cite{Gourieroux}, genetics~\cite{refEthier} as well as in neuroscience~\cite{Onofrio}.
The corresponding potential of Eq. \ref{steadyeq} involves a sum of logarithms
\begin{eqnarray}
U(x)  = - (\alpha-1) \ln (x) - (\beta-1) \ln (1-x)
\label{jacobiU}
\end{eqnarray}
while the force of Eq. \ref{UR} is linear 
\begin{eqnarray}
F(x) = - U'(x) D(x) =  (\alpha-1)A + (2-\alpha-\beta)A x
\label{jacobiF}
\end{eqnarray}
The eigenvalues correspond to the parameter $B=\alpha+\beta-1$ in the Jacobi spectrum of Eq. \ref{jacobilambda}
\begin{eqnarray}
\lambda_n= A n (n+\alpha+\beta-1 ) \ \ \ {\rm for } \ \ \ n=0,1,...+\infty 
\label{jacobilambdaBeta}
\end{eqnarray}
while the corresponding eigenvectors involve the Jacobi family of orthogonal polynomials.

The simplest example corresponds to the parameters $\alpha=1=\beta$ where the steady state $p_*(x)$ of Eq. \ref{jacobiBeta}
is uniform on $[0,1]$ 
\begin{eqnarray}
p_*(x) && = 1 \ \ {\rm for } \ \ 0 \leq x \leq 1
\label{jacobiBeta11}
\end{eqnarray}
and where the force of Eq. \ref{jacobiF} vanishes $F(x)=0$.
Then the spectral Kemeny time of Eq. \ref{kemenyjacobi} with $B=1$
and $H(1)= 1$ of Eq. \ref{harmonicBinteger}
\begin{equation}
      \tau_*^{Spectral[\alpha=1=\beta]}  = \frac{1}{A}  \sum_{n=1}^{+\infty} \frac{ 1 }{n (n+1) } 
      =  \frac{1}{A}    \sum_{n=1}^{+\infty} \frac{ 1 }{n(n + 1) }  
  =   \frac{H(1)}{A} = \frac{1}{A}
\label{kemenyjacobi11}
\end{equation}
coincides with the real-space Kemeny time of Eq. \ref{kemeny1d}
 \begin{eqnarray}
\tau_*^{Space[\alpha=1=\beta]}   =\int_0^1 dy \frac{1}{D(y)  } \left[ \int_{0}^{y} dz  \right] \left[ \int_{y}^{1} dx \right] =
\int_0^1 dy \frac{1}{ A y (1-y) }  y ( 1-y )
 = \frac{1}{A}
\label{kemeny1djacobi11}
\end{eqnarray}

As stressed in \cite{pearson_class,c_pearson}, the Jacobi eigenvalues of Eq. \ref{jacobilambdaBeta}
will be conserved via any change of variables $x \to y(x)$ that is twice-differentiable and invertible.
Among the many new processes $y(t)$ with the same spectrum that can be constructed 
via this procedure, one of the most natural is the process $y(t)$ that has a constant diffusion coefficient
instead of the quadratic polynomial of Eq. \ref{jacobiDx}, as described in the next subsection.


\subsubsection{ Change of variables $x \to y$ towards the process $y(t)$ on $[-\frac{L}{2},+\frac{L}{2}]$ with constant diffusion coefficient ${\cal D}(y)=1$ }

The change of variables from $x \in [0,1]$ to $ y \in [-\frac{L}{2},+\frac{L}{2}]$
 \begin{eqnarray}
x && = \frac{1+ \sin \left( \pi \frac{y}{L} \right)}{2}
\nonumber \\
dx && = dy \frac{ \pi   }{2 L}  \cos \left( \pi \frac{y}{L} \right)
= dy \frac{ \pi   }{2 L} \sqrt{ 1-  \sin^2 \left( \pi \frac{y}{L} \right) }
= dy \sqrt{ \frac{ \pi^2   }{ L^2}  x (1-x) }
\label{xtowardsy}
\end{eqnarray}
allows to transform the quadratic diffusion coefficient $D(x)=A x (1-x)$ of Eq. \ref{jacobiDx}
into the constant diffusion coefficient $ {\cal D}(y)=1$ for the new process $y(t)$ 
if one chooses the following value for the parameter $A$ 
 \begin{eqnarray}
A = \frac{ \pi^2   }{ L^2} 
\label{Achoice}
\end{eqnarray}
The normalized steady state ${\cal P}_*(y)$ $ y \in [-\frac{L}{2},+\frac{L}{2}]$ 
obtained from Eq. \ref{jacobiBeta} reads
\begin{eqnarray}
{\cal P}_*(y) && = p_*(x) \frac{dx}{dy} 
= \frac{\pi}{L Z(\alpha,\beta) } x^{\alpha-\frac{1}{2}} (1-x)^{\beta-\frac{1}{2}} 
\nonumber \\
 && = \frac{\pi \Gamma(\alpha+\beta)}{L \Gamma(\alpha) \Gamma(\beta) 2^{\alpha+\beta-1}} 
 \left[ 1+ \sin \left( \pi \frac{y}{L} \right)\right]^{\alpha-\frac{1}{2}} \left[ 1- \sin \left( \pi \frac{y}{L} \right)\right]^{\beta-\frac{1}{2}} 
\label{jacobiBetay}
\end{eqnarray}
The potential of Eq. \ref{steadyeq} associated to this steady state 
is 
\begin{eqnarray}
{\cal U}(y)  = - \left( \alpha-\frac{1}{2} \right) \ln  \left[ 1+ \sin \left( \pi \frac{y}{L} \right)\right]
-  \left( \beta-\frac{1}{2} \right) \ln  \left[ 1- \sin \left( \pi \frac{y}{L} \right)\right]
\label{jacobiUy}
\end{eqnarray}
The force of Eq. \ref{UR} 
\begin{eqnarray}
{\cal F}(y) && = -  {\cal U}'(y)  =  
\left( \alpha-\frac{1}{2} \right) \frac{\frac{\pi}{L} \cos \left( \pi \frac{y}{L} \right)} { 1+ \sin \left( \pi \frac{y}{L} \right) }
-  \left( \beta-\frac{1}{2} \right) \frac{\frac{\pi}{L} \cos \left( \pi \frac{y}{L}\right)} { 1- \sin \left( \pi \frac{y}{L} \right) }
\nonumber \\
&& = \frac{\pi \left[ (\alpha-\beta) - ( \alpha+\beta-1 )  \sin \left( \pi \frac{y}{L} \right)   \right]}{L \cos \left( \pi \frac{y}{L} \right)} 
\label{jacobiFy}
\end{eqnarray}
and its derivative
\begin{eqnarray}
{\cal F}'(y) &&  = \frac{\pi^2 \left[ (\alpha-\beta)\sin \left( \pi \frac{y}{L} \right) - ( \alpha+\beta-1 )    \right]}{L^2 \cos^2 \left( \pi \frac{y}{L} \right)} 
\label{jacobiFyderi}
\end{eqnarray}
allow to compute the quantum potential of Eq. \ref{vfromu} 
\begin{eqnarray}
{\cal V}(y)  && = \frac{ {\cal F}^2(y) }{4  } + \frac{{\cal F}'(y)}{2}
\nonumber \\
&& = \frac{\pi^2 }{2 L^2 } 
\left[ \frac{\left( \alpha-\frac{1}{2} \right)\left( \alpha-\frac{3}{2} \right)}{1+ \sin \left( \pi \frac{y}{L} \right)} 
+ \frac{\left( \beta-\frac{1}{2} \right)\left( \beta-\frac{3}{2} \right)}{1- \sin \left( \pi \frac{y}{L} \right)} 
- \frac{ (\alpha+\beta-1)^2}{2}  \right]
\label{vfromuJacobiy}
\end{eqnarray}

The Jacobi eigenvalues are given by Eq. \ref{jacobilambdaBeta} for the parameter $ A = \frac{ \pi^2   }{ L^2} $ of Eq. \ref{Achoice} 
\begin{eqnarray}
\lambda_n= \frac{ \pi^2   }{ L^2} n (n+\alpha+\beta-1 ) \ \ \ {\rm for } \ \ \ n=0,1,...+\infty 
\label{jacobilambdaBetay}
\end{eqnarray}
so that the spectral Kemeny time of Eq. \ref{kemenyjacobi} 
\begin{equation}
      \tau^{Spectral} _*  =   \sum_{n=1}^{+\infty} \frac{ 1 }{\lambda_n } =   L^2  \frac{H(\alpha+\beta-1)}{\pi^2 (\alpha+\beta-1)} 
\label{kemenyjacobiy}
\end{equation}
grows as $L^2$ with respect to the system size $L$, while the prefactor depends on the sum $(\alpha+\beta)$.

For the symmetric case $\beta=\alpha$ studied in detail in \cite{refAlainTaboo}, 
the steady state of Eq. \ref{jacobiBetay}
simplifies into a power of $\cos \left( \pi \frac{y}{L} \right) $
\begin{eqnarray}
{\cal P}_*^{[\beta=\alpha]}(y) 
 = \frac{\pi \Gamma(2\alpha)}{L \Gamma^2(\alpha)  2^{2\alpha-1}} 
 \left[ 1- \sin^2 \left( \pi \frac{y}{L} \right)\right]^{\alpha-\frac{1}{2}} 
  =   \frac{ \sqrt{\pi} \Gamma \left(\alpha+\frac{1}{2} \right)}{L \Gamma(\alpha)  } 
 \left[ \cos \left( \pi \frac{y}{L} \right)\right]^{2\alpha-1} 
 \label{jacobiBetaysim}
\end{eqnarray}
while the force of Eq. \ref{jacobiFy}
reduces to the tangent force 
\begin{eqnarray}
{\cal F}^{[\beta=\alpha]}(y)  = - ( 2\alpha -1 ) \frac{\pi }{L}  \tan \left( \pi \frac{y}{L} \right)
\label{jacobiFysym}
\end{eqnarray}
and the quantum potential of Eq. \ref{vfromuJacobiy} becomes
\begin{eqnarray}
{\cal V}^{[\beta=\alpha]}(y)   = \frac{\pi^2 }{2 L^2 } 
\left[ \frac{ 2 \left( \alpha-\frac{1}{2} \right)\left( \alpha-\frac{3}{2} \right)}{ \cos^2 \left( \pi \frac{y}{L} \right)}  
- \frac{ (2 \alpha-1)^2}{2}  \right]
\label{vfromuJacobiysym}
\end{eqnarray}

Let us now discuss the simplest values of the parameters $(\alpha,\beta)$ 
that produce a constant quantum potential ${\cal V}(y) $ in Eq. \ref{vfromuJacobiy}:

(i) the symmetric case $\beta=\alpha = \frac{1}{2} $ corresponds to the uniform steady state in Eq. \ref{jacobiBetaysim}
\begin{eqnarray}
{\cal P}_*^{[\beta=\alpha=\frac{1}{2}]}(y) 
 = \frac{ 1}{L} 
 \label{jacobiBetaysimuniform}
\end{eqnarray}
and one recovers via Eq. \ref{jacobilambdaBetay} the eigenvalues
\begin{eqnarray}
\lambda_n^{[\beta=\alpha=\frac{1}{2}]}= \frac{ \pi^2   }{ L^2} n^2 \ \ \ {\rm for } \ \ \ n=0,1,...+\infty 
\label{jacobilambdaBetaysymdemi}
\end{eqnarray}
discussed in Eq. \ref{eigenexciteduni} for the pure diffusion $D(x)=1$ with reflecting boundaries separated by the distance $L$.

(ii) the symmetric case $\beta=\alpha = \frac{3}{2} $ corresponds to the steady state in Eq. \ref{jacobiBetaysim}
\begin{eqnarray}
{\cal P}_*^{[\beta=\alpha=\frac{3}{2}]}(y) 
  =   \frac{ 2  \cos^2 \left( \pi \frac{y}{L} \right) }{L   }  =  \frac{ 1+  \cos \left( 2 \pi \frac{y}{L} \right) }{L   }
 \label{jacobiBetaytaboo}
\end{eqnarray}
where one recognizes the steady state of the Taboo process,
i.e. the pure diffusion $D(x)=1$ with absorbing boundaries at $x=-\frac{L}{2}$ and at $x=+\frac{L}{2}$
conditioned to survive forever (see \cite{refKnight,refPinskyTaboo,ref_GarbaczewskiTaboo,refAlainTaboo,ref_Adorisio} 
and the next subsection for more explanations),
with its eigenvalues of Eq. \ref{jacobilambdaBetay}
\begin{eqnarray}
\lambda_n^{[\beta=\alpha=\frac{3}{2}]}= \frac{ \pi^2   }{ L^2} n (n+2 ) \ \ \ {\rm for } \ \ \ n=0,1,...+\infty 
\label{jacobilambdaBetaytaboo}
\end{eqnarray}
The spectral Kemeny time of Eq. \ref{kemenyjacobiy}
\begin{equation}
      \tau_*^{Spectral[\beta=\alpha=\frac{3}{2}]}  =   \sum_{n=1}^{+\infty} \frac{ 1 }{\lambda_n } =   L^2  \frac{H(2)}{2 \pi^2 } 
      =   \frac{3}{4 \pi^2 } L^2
\label{kemenyjacobiytaboo}
\end{equation}
is in agreement with the real-space Kemeny time of Eq. \ref{kemeny1d}
based on the steady state of Eq. \ref{jacobiBetaytaboo}
 \begin{eqnarray}
\tau_*^{[\beta=\alpha=\frac{3}{2}]} &&  =\int_{-\frac{L}{2}}^{-\frac{L}{2}} dy \frac{1}{ {\cal P}_*(y) }
 \left[ \int_{-\frac{L}{2}}^{y} dz {\cal P}_*(z) \right] \left[\int_{y}^{\frac{L}{2}} dx {\cal P}_*(x) \right]
\nonumber \\
&& =\int_{-\frac{L}{2}}^{-\frac{L}{2}} dy \frac{L}{ 2  \cos^2 \left( \pi \frac{y}{L} \right) }
 \left[  \frac{ \frac{L}{2} +y + \frac{L}{2 \pi}  \sin \left( 2 \pi \frac{y}{L} \right) }{L   } \right] 
 \left[ \frac{ \frac{L}{2} -y - \frac{L}{2 \pi}  \sin \left( 2 \pi \frac{y}{L} \right) }{L   } \right] =   \frac{3}{4 \pi^2 } L^2
\label{kemeny1dtaboo}
\end{eqnarray}

(iii) the asymmetric case $[\alpha = \frac{3}{2};\beta= \frac{1}{2}] $ corresponds to the steady state in Eq. \ref{jacobiBetay}
\begin{eqnarray}
{\cal P}_*^{[\alpha = \frac{3}{2};\beta= \frac{1}{2}] }(y)  
= \frac{  1+ \sin \left( \pi \frac{y}{L} \right) }{L }  
\label{jacobiBetay3demi1demi}
\end{eqnarray}
and to the eigenvalues of Eq. \ref{jacobilambdaBetay}
\begin{eqnarray}
\lambda_n^{[\alpha = \frac{3}{2};\beta= \frac{1}{2}] }= \frac{ \pi^2   }{ L^2} n (n+1 ) \ \ \ {\rm for } \ \ \ n=0,1,...+\infty 
\label{jacobilambdaBetay3demi1demi}
\end{eqnarray}
The spectral Kemeny time of Eq. \ref{kemenyjacobiy}
\begin{equation}
      \tau_*^{Spectral[\alpha = \frac{3}{2};\beta= \frac{1}{2}] }  =   \sum_{n=1}^{+\infty} \frac{ 1 }{\lambda_n } =   L^2  \frac{H(1)}{\pi^2 } =   \frac{L^2}{\pi^2 }
\label{kemenyjacobiy3demi1demi}
\end{equation}
is in agreement with the real-space Kemeny time of Eq. \ref{kemeny1d}
based on the steady state of Eq. \ref{jacobiBetay3demi1demi}
 \begin{eqnarray}
\tau_*^{Space[\alpha = \frac{3}{2};\beta= \frac{1}{2}] } &&  
=\int_{-\frac{L}{2}}^{-\frac{L}{2}} dy \frac{1}{ {\cal P}_*(y) }
 \left[ \int_{-\frac{L}{2}}^{y} dz {\cal P}_*(z) \right] \left[\int_{y}^{\frac{L}{2}} dx {\cal P}_*(x) \right]
\nonumber \\
&& =\int_{-\frac{L}{2}}^{-\frac{L}{2}} dy \frac{L}{  1+ \sin \left( \pi \frac{y}{L} \right) }
 \left[ \frac{ \frac{L}{2} + y - \frac{L}{\pi}  \cos \left( \pi \frac{y}{L} \right) }{L }   \right] 
 \left[  \frac{ \frac{L}{2} - y + \frac{L}{\pi}  \cos \left( \pi \frac{y}{L} \right) }{L }     \right]=   \frac{L^2}{\pi^2 }
\label{kemeny1d3demi1demi}
\end{eqnarray}

The physical interpretation of the process $[\alpha = \frac{3}{2};\beta= \frac{1}{2}] $
can be guessed from the previous cases (i) and (ii): 
it should be the diffusion $D(x)=1$ with a reflecting boundary at $y=+\frac{L}{2}$
and an absorbing boundary at $y=-\frac{L}{2}$ conditioned to survive forever,
as will be checked in the next subsection.

(iv) the asymmetric case $[\alpha = \frac{1}{2};\beta= \frac{3}{2}] $
is similar to (iii) since it amounts to exchange the role of the two boundaries.


\subsection{ Kemeny time for diffusion processes with killing or absorption conditioned to survive forever}

For a diffusion process with killing or absorption characterized 
by the series of eigenvalues $\lambda_{n}^{[Abs]} >0 $
with $n=1,..,+\infty$,  
the process conditioned to survive forever will have the series of eigenvalues given by Eq. \ref{steadyCond}
\begin{eqnarray}
\lambda_m^{[Cond]}  = \lambda_{1+m}^{[Abs]} -\lambda^{[Abs]}_1 \ \ {\rm for } \ m=1,..,+\infty
\label{lambdaCondDiff}
\end{eqnarray}
with the corresponding spectral Kemeny time of Eq. \ref{KemenyCond}
 \begin{eqnarray}
\tau^{Spectral[Cond]}_{*} =     \sum_{m=1}^{+\infty} \frac{1}{\lambda^{[Cond]}_m} 
=  \sum_{m=1}^{+\infty} \frac{1}{\lambda_{1+m}^{[Abs]} -\lambda^{[Abs]}_1}
\label{KemenyCondDiff}
\end{eqnarray}
As simple examples, let us consider the Taboo processes already mentioned in (ii) and (iii) of the previous subsection.


\subsubsection{Diffusion $D(x)=1$ with absorbing boundaries at $x=-\frac{L}{2}$ and at $x=+\frac{L}{2}$
conditioned to survive forever }

When the two boundaries at $x= \pm \frac{L}{2}$ are absorbing, 
the Fourier mode $r(x)=\sin[ k (x+\frac{L}{2})] $
that already vanishes at $x=-\frac{L}{2}$ should also vanish at $x=+ \frac{L}{2}$
 \begin{eqnarray}
\sin[ k L] =0
\label{eqkTwoAbs}
\end{eqnarray}
The corresponding wavevectors and eigenvalues for this process with absorption 
 \begin{eqnarray}
 k_n^{[Abs]} && = n \frac{\pi}{L} \ \ \ {\rm for } \ \ \ n=1,2,..,+\infty
 \nonumber \\
 \lambda_n^{[Abs]} && = \left[ k_n^{[Abs]} \right]^2 = n^2 \frac{\pi^2}{L^2}
\label{kTwoAbs}
\end{eqnarray}
lead to the eigenvalues of Eq. \ref{lambdaCondDiff} for the process conditioned to survive forever
\begin{eqnarray}
{\rm for } \ m=1,..,+\infty \ \ \: \ \ \lambda_m^{[Cond]}  = \lambda_{1+m}^{[Abs]} -\lambda^{[Abs]}_1 
= \frac{\pi^2}{L^2} \left[ (m+1)^2-1 \right] = \frac{\pi^2}{L^2} ( m^2+2m )
\label{lambdaCondTwoAbs}
\end{eqnarray}
in agreement with the spectrum of Eq. \ref{jacobilambdaBetaytaboo}
with its Kemeny time of Eq. \ref{kemenyjacobiytaboo}.


\subsubsection{Diffusion $D(x)=1$ with reflecting boundary at $x=+\frac{L}{2}$
and absorbing boundary at $x=-\frac{L}{2}$ conditioned to survive forever}

When the boundary at $x= - \frac{L}{2}$ is absorbing, 
while the boundary at $y=+\frac{L}{2}$ is reflecting,
the Fourier mode $\sin[ k (x+\frac{L}{2})] $
that vanishes at $x=-\frac{L}{2}$ should have a vanishing derivative at $x=+ \frac{L}{2}$
 \begin{eqnarray}
\cos[ k L] =0
\label{eqkOneAbs}
\end{eqnarray}
The corresponding wavevectors and eigenvalues for this process with absorption 
 \begin{eqnarray}
 k_n^{[Abs]} && = \left( n - \frac{1}{2} \right)\frac{\pi}{L} \ \ \ {\rm for } \ \ \ n=1,2,..,+\infty
 \nonumber \\
 \lambda_n^{[Abs]} && = \left[ k_n^{[Abs]} \right]^2 = \left( n - \frac{1}{2} \right)^2 \frac{\pi^2}{L^2}
\label{kOneAbs}
\end{eqnarray}
lead to the eigenvalues of Eq. \ref{lambdaCondDiff} for the process conditioned to survive forever
\begin{eqnarray}
{\rm for } \ m=1,..,+\infty \ \ \: \ \ \lambda_m^{[Cond]}  = \lambda_{1+m}^{[Abs]} -\lambda^{[Abs]}_1 
= \frac{\pi^2}{L^2} \left[ \left( m+ \frac{1}{2} \right)^2-\left(  \frac{1}{2} \right)^2 \right] 
= \frac{\pi^2}{L^2} \left[ m^2+m \right]
\label{lambdaCondOneAbs}
\end{eqnarray}
in agreement with the spectrum of Eq. \ref{jacobilambdaBetay3demi1demi}
with its Kemeny time of Eq. \ref{kemenyjacobiy3demi1demi}.


\section{ Irreversible diffusion processes on the periodic Ring of length $L$}

\label{sec_DiffRing}

Instead of the interval $]x_L,x_R[$ with two boundaries considered in the previous section,
where one cannot avoid the detailed-balance,
we focus in the present section on diffusion processes on the periodic ring of length $L$ with $L+x \equiv x$
in order to have a non-equilibrium steady state with a non-vanishing steady-current along the ring
and to compute the corresponding real-space Kemeny time.


\subsection{ Steady-state $p_*(x)$ and steady-current $j_*$ along the ring }

On the periodic ring, the steady-state $p_*(x)$ of the Fokker-Planck Eq. \ref{fokkerplanck} 
corresponds to a uniform steady-current $j_*(x)=j_*$ along the ring
\begin{eqnarray}
 j_* =  F(x)   p_*( x ) - D (x)   p_*'( x) 
\label{FPJsteadyring}
\end{eqnarray}
It is still convenient to use the potential of Eq. \ref{UR} with reference $x_{ref}=0$ 
to have $U(0)=0$
\begin{eqnarray}
U(x) && \equiv - \int_{0}^x dy \frac{F(y)}{D(y)} \ \ \ {\rm for } \ \ x \in [0,L]
 \label{URring}
\end{eqnarray}
while the value $U(L)$ at $x=L$ depends
on the force $F(y)$ and on the diffusion coefficient $D(y)$ on the whole ring $0 \leq y \leq L$
\begin{eqnarray}
U(L)  && \equiv - \int_{0}^L dy \frac{F(y)}{D(y)} 
 \label{ULring}
\end{eqnarray}

Using $F(x)=-D(x) U'(x)$, Eq. \ref{FPJsteadyring} becomes
\begin{eqnarray}
 p_*'( x) +  U'(x)  p_*( x )   = - \frac{j_*}{D(x)}
\label{FPJsteadyringU}
\end{eqnarray}
with the solution
\begin{eqnarray}
p_*( x ) =  e^{ -U(x)}  \left[ K - j_*  \int_{0}^{x} d y\frac{e^{ U(y) } }{D(y)}   \right] 
\label{solReggene}
\end{eqnarray}
where the integration constant $K$ is fixed by the periodicity requirement 
for the steady state $p_*(x=0)=p_*(x=L)$
\begin{eqnarray}
 K =  e^{ -U(L)}  \left[ K - j_*  \int_{0}^L d y\frac{e^{ U(y) } }{D(y)}   \right] 
\label{solReggeneperio}
\end{eqnarray}
So one needs to distinguish whether the value $U(L)$ of Eq. \ref{ULring} vanishes or not.


\subsubsection{ Case of periodic potential $U(L)=U(0)=0$: equilibrium steady-state 
with vanishing steady current $j_{*}=0$  }

\label{sec_eqst}

When the potential $U(L)$ of Eq. \ref{ULring} vanishes $U(L)=0$,
Eq. \ref{solReggeneperio} yields that the steady state current vanishes $j_*=0$,
so the detailed-balance condition is satisfied.
Then the steady state 
reduces to the Boltzmann distribution in the potential $U(x)$ as in Eq. \ref{steadyeq}
\begin{eqnarray}
  p^{DB}_*(x) =  \frac{ e^{ -U(x)} }{Z}
 \label{steadyeqr}
\end{eqnarray}
where the normalization $Z$ corresponds to the partition function on the ring
\begin{eqnarray}
Z=  \int_{0}^L dx e^{ -  U(x) } 
 \label{partitioneqr}
\end{eqnarray}


\subsubsection{ Case of non-periodic potential $U(L) \ne U(0)=0$: non-equilibrium steady-state with a finite steady current  $j_* \ne 0$  }

\label{sec_noneqst}

When the potential $U(L)$ of Eq. \ref{ULring} does not vanish $U(L) \ne 0$,
Eq. \ref{solReggeneperio} yields the value of the constant $K$
\begin{eqnarray}
 K  =  j_*  \frac{   \int_{0}^L d y\frac{e^{ U(y) } }{D(y)}   }{ 1- e^{U(L)} }
\label{KsolReggeneperio}
\end{eqnarray}
so that the steady state of Eq. \ref{solReggene} reads
\begin{eqnarray}
p_*( x ) =  j_* \frac{ e^{ -U(x)} }{  e^{- U(L)} -1 } \left[ 
 \int_{0}^{x} d y\frac{e^{ U(y) } }{D(y)} 
+ e^{- U(L)}   \int_x^L d y\frac{e^{ U(y) } }{D(y)}  
   \right] 
\label{steadyring}
\end{eqnarray}
The steady-state current $j_*$ is determined by
the normalization of the steady state 
\begin{eqnarray}
1  = \int_{0}^{L}  dx p_*( x )
&& = 
 \frac{  j_* }{  e^{- U(L)} -1 } \left[ \int_{0}^{L}  dx e^{ -U(x)} \int_{0}^{x} d y\frac{e^{ U(y) } }{D(y)} 
+ e^{- U(L)} \int_{0}^{L}  dx e^{ -U(x)}  \int_x^L d y\frac{e^{ U(y) } }{D(y)}  
   \right] 
   \nonumber \\
&& = 
 \frac{  j_* }{  e^{- U(L)} -1 } 
 \int_{0}^L d y\frac{e^{ U(y) } }{D(y)}  \left[ \int_y^{L}  dx e^{ -U(x)} 
+ e^{- U(L)}    \int_{0}^y  dx e^{ -U(x)} 
   \right] 
\label{jsteadyringnorma}
\end{eqnarray}
leading to
\begin{eqnarray}
  j_*  = \frac{  e^{- U(L)} -1  }
{ \int_{0}^L d y\frac{e^{ U(y) } }{D(y)}  \left[ \int_y^{L}  dx e^{ -U(x)} + e^{- U(L)}    \int_{0}^y  dx e^{ -U(x)}    \right] }
\label{jsteadyring}
\end{eqnarray}


\subsection{ Interpretation of the non-equilibrium steady-state $p_*(x)$ via non-hermitian quantum mechanics  }

As explained in \cite{c_lyapunov}, it is interesting to analyze
the meaning of the non-equilibrium steady-state $p_*(x)$ associated to the 
non-vanishing steady current $j_* \ne 0$
in the quantum mechanical language of Eq. \ref{ppsi}.
Equation \ref{eigenLR} means that one should now distinguish
the left eigenvector $\phi_{0}^{[l]} (x_0) $ and the left eigenvector $\phi_{0}^{[r]} (x) $ for the quantum supersymmetric Hamiltonian 
$H$ of Eq. \ref{hamiltonien} \ref{vfromu} \ref{hsusy} \ref{qsusy}:

(i) the left eigenvector of Eq. \ref{eigenLR} (dropping out the constant factor involving $Z$)
\begin{eqnarray}
\phi_{0}^{[l]} (x) && = e^{ -  \frac{ U(x)}{2} } l_0(x) =  e^{ -  \frac{ U(x)}{2} }
\label{eigenLring}
\end{eqnarray}
is still annihilated by the operator $Q$ of Eq. \ref{qsusy}
\begin{eqnarray}
Q \phi_{0}^{[l]} (x) =    \sqrt{ D(x) }  \left( \frac{ d }{ d x}  +\frac{ U'(x)}{2 } \right)e^{ -  \frac{ U(x)}{2} } =0
\label{qsusyleft}
\end{eqnarray}

(ii) the right eigenvector of Eq. \ref{eigenLR} (dropping out the constant factor involving $Z$)
\begin{eqnarray}
\phi_{0}^{[r]} (x)  && =     e^{   \frac{ U(x)}{2} } r_0(x) =     e^{   \frac{ U(x)}{2} } p_*(x)
\nonumber \\
&& = j_* \frac{e^{  - \frac{ U(x)}{2} }  }{  e^{- U(L)} -1 } \left[ 
 \int_{0}^{x} d y\frac{e^{ U(y) } }{D(y)} 
+ e^{- U(L)}   \int_x^L d y\frac{e^{ U(y) } }{D(y)}  
   \right] 
\label{eigenRring}
\end{eqnarray}
is not annihilated by the operator $Q$ of Eq. \ref{qsusy} anymore
\begin{eqnarray}
Q  \phi_{0}^{[r]} (x)  && = \sqrt{ D(x) }  \left( \frac{ d }{ d x}  +\frac{ U'(x)}{2 } \right)\phi_{0}^{[r]} (x) 
=  \frac{(- j_*)}{ \sqrt{ D(x) }} e^{  \frac{ U(x) }{ 2} } 
   \label{qsusynoneq}
\end{eqnarray}
but this state is annihilated by the operator $Q^{\dagger}$ of Eq. \ref{qsusy} 
\begin{eqnarray}
Q^{\dagger} \left( Q  \phi_{0}^{[r]} (x)  \right)  =
\left(   - \frac{ d }{ d x}  +\frac{ U'(x)}{2 } \right)\sqrt{ D(x) } \left( \frac{(- j_*)}{ \sqrt{ D(x) }} e^{  \frac{ U(x) }{ 2} }\right) =0
   \label{qsusynoneqdagger}
\end{eqnarray}
as it should to produce the ground-state of energy zero of the Hamiltonian of Eq. \ref{hsusy}.
So even if the quantum Hamiltonian of Eq. \ref{hsusy} is Hermitian, 
the periodic boundary conditions on the Fokker-Planck right and left eigenvectors 
$l_0(L)=l_0(0)$ and $r_0(L)=r_0(0)$
induces a difference between the associated left and right quantum eigenvectors
of Eqs \ref{eigenLring}
and \ref{eigenRring}
that are submitted to the following different boundary conditions 
whenever the potential $U(.)$ is not periodic $U(L) \ne U(0)=0$
\begin{eqnarray}
\frac{ \phi_{0}^{[l]} (x=L) }{ \phi_{0}^{[l]} (x=0)} &&  =  e^{ -  \frac{ U(L)}{2} } 
\nonumber \\
\frac{ \phi_{0}^{[r]} (x=L) }{ \phi_{0}^{[r]} (x=0)} &&  =  e^{ +  \frac{ U(L) }{2} } 
\label{eigenLringbc}
\end{eqnarray}
These boundary conditions make the quantum problem non-hermitian
and can be interpreted as the Aharonov-Bohm effect for an imaginary magnetic field
that produces an imaginary flux through the ring  (see \cite{c_lyapunov} for more details).
In higher dimensions $d \geq 2$, the link between non-equilibrium diffusion processes
and non-hermitian quantum mechanics associated to imaginary magnetic fields
is described in detail in \cite{us_gyrator}.


\subsection{ Mean-First-Passage-Times $\tau(x,x_0)$ }

The backward equation of Eq. \ref{adjointtau} for the Mean-First-Passage-Time $\tau(x,x_0)$ 
at $x$ when starting at $x_0$ can be again rewritten for the auxiliary function $\chi^{[x]}(x_0)  $ of Eq. \ref{nudifftau} as Eq. \ref{adjointtaudiff}
\begin{eqnarray}
-1 + \frac{ \delta(x_0-x)  }{ p_*(x)}    =  \left[  -U'(x_0)    +   \partial_{x_0}  \right]   \chi^{[x]}(x_0)
 \label{adjointtaudiffring}
\end{eqnarray}
that should now be solved with the periodic boundary condition $\chi^{[x]}(x_0=0)=\chi^{[x]}(x_0=L)$
so that one obtains 
\begin{eqnarray}
{\rm \ \ for} \ \  x_0 \in [0,x[ : \ \ \ 
&& \chi^{[x]}(x_0)  =   \frac{  e^{U(x_0)} }{  e^{- U(L)} -1 }
 \left[ \frac{e^{-U(x) }}{p_*(x)} 
  - e^{- U(L)}  \int_0^{x_0} dz e^{- U(z) }
 - \int_{x_0}^L dz e^{- U(z) }  \right] 
\nonumber \\
{\rm \ \ for} \ \  x_0 \in ]x,L] : \ \ \
&&   \chi^{[x]}(x_0)  =  \frac{  e^{U(x_0)} }{  e^{- U(L)} -1 } \left[ 
 e^{- U(L)}  \frac{e^{-U(x) }}{p_*(x)}  
  - e^{- U(L)}  \int_0^{x_0} dz e^{- U(z) }
 - \int_{x_0}^L dz e^{- U(z) }
    \right] 
    \label{nuring}
\end{eqnarray}

Plugging this solution into Eq. \ref{nudifftau}
\begin{eqnarray}
\partial_{x_0}  \tau(x,x_0) = \frac{ \chi^{[x]}(x_0)}{  D (x_0)  } 
   \label{nudifftauintring}
\end{eqnarray}
yields the Mean-First-Passage-Time $\tau(x,x_0)$ vanishing at $x_0=x$
\begin{eqnarray}
{\rm \ \ for} \ \  x_0 \in [0,x] : \ \ \ 
&& \tau(x,x_0) = - \int_{x_0}^x dy \frac{\chi^{[x]}(y)}{D(y)  } 
\label{mftpring1}
 \\
&&  =  \frac{  1 }{  e^{- U(L)} -1 } 
 \left[ - \frac{e^{-U(x) }}{p_*(x)} \int_{x_0}^x dy    \frac{  e^{U(y)} }{ D(y)  }
  +  \int_{x_0}^x dy    \frac{  e^{U(y)} }{ D(y)  }  \left( e^{- U(L)} \int_0^{y} dz e^{- U(z) }
 + \int_{y}^L dz e^{- U(z) } \right) \right]  
\nonumber
\end{eqnarray}
and
\begin{eqnarray}
{\rm \ \ for} \ \ 
&&  x_0 \in [x,L]: \ \ \  \tau(x,x_0) 
= \int_{x}^{x_0} dy \frac{\chi^{[x]}(y)}{D(y)  } 
\label{mftpring2} \\
&& =  \frac{  1 }{  e^{- U(L)} -1 }  \left[ 
 e^{- U(L)}  \frac{e^{-U(x) }}{p_*(x)}  \int_{x}^{x_0} dy \frac{  e^{U(y)} }{ D(y)  }
  -  \int_{x}^{x_0} dy \frac{  e^{U(y)} }{ D(y)  } \left( e^{- U(L)} \int_0^y dz e^{- U(z) }
 + \int_y^L dz e^{- U(z) } \right) \right]
\nonumber
\end{eqnarray}


\subsection { Real-space Kemeny time }

The real-space Kemeny time of Eq. \ref{taukemeny} can be evaluated for the initial position 
at $x_0=0$ using Eq. \ref{mftpring1} 
 \begin{eqnarray}
\tau^{Space}_* && = \int_{0}^{L} dx p_*(x) \tau(x,x_0=0)    
\label{kemenyringcalcul} \\
&&  = -  \frac{  1 }{  e^{- U(L)} -1 } \int_{0}^{L} dx 
  e^{-U(x) } \int_0^x dy    \frac{  e^{U(y)} }{ D(y)  }
 + \frac{ 1  }{  e^{- U(L)} -1 } 
\int_{0}^{L} dx p_*(x)   \int_0^x dy    \frac{  e^{U(y)} }{ D(y)  } 
\left[ e^{- U(L)} \int_0^{y} dz e^{- U(z) } + \int_{y}^L dz e^{- U(z) }  \right]
 \nonumber \\
&&  = -  \frac{  1 }{  e^{- U(L)} -1 } 
\int_0^L dy    \frac{  e^{U(y)} }{ D(y)  }\int_y^{L} dx   e^{-U(x) } 
 + \frac{  1 }{  e^{- U(L)} -1 } 
 \int_0^L dy    \frac{  e^{U(y)} }{ D(y)  }   
 \left[ e^{- U(L)} \int_0^{y} dz e^{- U(z) } + \int_{y}^L dz e^{- U(z) }  \right] \int_y^{L} dx p_*(x) 
 \nonumber
\end{eqnarray}
In the last contribution, it is convenient to replace $\int_y^L dx p_*(x) =1- \int_{0}^y dx p_*(x)$ 
in order to obtain the final result
 \begin{eqnarray}
\tau^{Space}_*   && =  \frac{ 1   }{  e^{- U(L)} -1 } 
 \int_0^L dy    \frac{  e^{U(y)} }{ D(y)  }  
 \left[ e^{- U(L)}  \left( \int_0^{y} dz e^{- U(z) } \right) \left( \int_y^L dx p_*(x) \right)
-    \left( \int_{y}^L dz e^{- U(z) }  \right) \left( \int_{0}^y dx p_*(x) \right) \right]
\label{kemenyring}
\end{eqnarray}
One may also change the order of the integrals to obtain the alternative form
 \begin{eqnarray}
\tau^{Space}_*   && =  \frac{ 1   }{  e^{- U(L)} -1 }  \int_0^L dx p_*(x) 
 \left[ e^{- U(L)}   \int_0^x dy    \frac{  e^{U(y)} }{ D(y)  } \int_0^{y} dz e^{- U(z) }   
-    \int_x^L dy    \frac{  e^{U(y)} }{ D(y)  }  \int_{y}^L dz e^{- U(z) }     \right]
\label{kemenyringalter}
\end{eqnarray}
So once again, 
the real-space expressions Eqs \ref{kemenyring} \ref{kemenyringalter} give
 explicit results for the spectral Kemeny time 
even if the individual eigenvalues $\lambda_n$ are not known.


\subsection{ Example with diffusion coefficient $D(x)=1$ and constant force $F(x)=\mu$ on the ring of length $L$}

\label{subsec_Ringmu}

The simplest example corresponds to the diffusion coefficient $D(x)=1$ 
and to the constant force $F(x)=\mu$ on the ring.
Then the potential of Eq. \ref{URring} is linear
\begin{eqnarray}
U(x) && \equiv - \mu x \ \ \ {\rm for } \ \ x \in [0,L]
 \label{URringlin}
\end{eqnarray}
The steady current of Eq. \ref{jsteadyring} reduces to
\begin{eqnarray}
  j_*  = \frac{ \mu}{L} 
\label{jsteadyringuni}
\end{eqnarray}
and the steady state of Eq. \ref{steadyring} is uniform
\begin{eqnarray}
p_*( x ) =  \frac{ 1}{L} 
\label{steadyringuni}
\end{eqnarray}

\subsubsection{ Real-space Kemeny time}

The real-space Kemeny time of Eq. \ref{kemenyring} reduces to
 \begin{eqnarray}
\tau^{Space}_*   && = \frac{ 1   }{ \mu L  ( 1- e^{-\mu L} )  } 
 \int_0^L dy    \left[   L  -   L  e^{- \mu y}     - ( 1- e^{-\mu L} ) y  \right]
 =  \frac{  \left( \frac{\mu L}{2} -1\right)  + e^{-\mu L}  \left( \frac{\mu L}{2} +1\right)    }{ \mu^2   ( 1- e^{-\mu L} )  }     
\nonumber \\
&& =
  \frac{1}{\mu^2 } \left[ \frac{ \frac{\mu L}{2 } }{ \tanh \left( \frac{\mu L}{2 } \right)} -1 \right]        
\label{kemenyringmu}
\end{eqnarray}

For any non-vanishing $\mu \ne 0$, the Kemeny time $\tau_*$ displays the linear scaling for large $L$
 \begin{eqnarray}
\tau_*^{[\mu \ne 0]} && \opsimeq_{L \to + \infty} \frac{ L  }{2 \vert \mu \vert}    
\label{kemenyringmuspectrallargeL}
\end{eqnarray}
while for $\mu \to 0$ corresponding to detailed-balance, one recovers the diffusive scaling
 \begin{eqnarray}
 \ \ \tau^{[\mu \to 0]}_* = \frac{L^2}{12}
\label{kemenyringmuspectralDB}
\end{eqnarray}
The present prefactor $1/12$ corresponding to periodic boundary conditions on the ring
is different from the constant $1/6 $ found in Eq. \ref{kemeny1duni} corresponding to reflecting boundary conditions.

\subsubsection{ Spectral Kemeny time}

The generator ${\cal L}_x $ of Eq. \ref{generator}
has for right eigenvectors the periodic Fourier modes 
 \begin{eqnarray}
r_n(x)=e^{i n 2 \pi \frac{x}{L}}\ \ \ {\rm with } \ \ n \in \mathbb{Z}
\label{rnxperio}
\end{eqnarray}
The eigenvalues $\lambda_n$ computed from
 \begin{eqnarray}
- \lambda_n e^{i n 2 \pi \frac{x}{L}} 
=  {\cal L}_x e^{i n 2 \pi \frac{x}{L}} 
=  \partial_{x}  \left[ - \mu  +   \partial_{x} \right] e^{i n 2 \pi \frac{x}{L}}
= i n  \frac{2 \pi}{L} \left[ - \mu  +   i n  \frac{2 \pi}{L} \right] e^{i n 2 \pi \frac{x}{L}}
\label{eigenringmueq}
\end{eqnarray}
read
 \begin{eqnarray}
\lambda_n  =  n  \frac{2 \pi}{L}  \left[ +  i\mu  +    n  \frac{2 \pi}{L} \right] 
=   \frac{4 \pi^2}{L^2}  \left[     n^2 +  i n \frac{\mu L}{2 \pi}  \right]
\label{eigenringmu}
\end{eqnarray}

Besides the vanishing eigenvalue $\lambda_{n=0}=0$, 
all the other eigenvalues appear in complex-conjugate pairs
 \begin{eqnarray}
 \lambda_{(-n)}=  \frac{4 \pi^2}{L^2}  \left[     n^2 -  i n \frac{\mu L}{2 \pi}  \right] = \lambda_n^*
\label{eigenringmucc}
\end{eqnarray}
The spectral Kemeny time of Eq. \ref{tauspectral} 
 \begin{eqnarray}
\tau^{Spectral}_* && = \sum_{n=1}^{+\infty} \left( \frac{1}{\lambda_n} +\frac{1}{\lambda_n^*} \right)  
= \sum_{n=1}^{+\infty}\frac{\lambda_n+\lambda_n^*}{\lambda_n \lambda_n^*} 
=  \frac{L^2}{2 \pi^2} \sum_{n=1}^{+\infty}  \frac{      1 }
{      n^2 +  \left( \frac{\mu L}{2 \pi} \right)^2 }  
\nonumber \\
&& =     \frac{1}{\mu^2 } \left[ \frac{ \frac{\mu L}{2 } }{ \tanh \left( \frac{\mu L}{2 } \right)} -1 \right]
\label{kemenyringmuspectral}
\end{eqnarray}
is in agreement with the real-space Kemeny time of Eq. \ref{kemenyringmu}.


\subsubsection{ Spectral Kemeny time in the presence of stochastic resetting at rate $\gamma$}

The eigenvalues $\lambda_n$ without resetting computed in Eq. \ref{eigenringmu}
leads to the following non-vanishing eigenvalues of Eq. \ref{lambdanreset}
for the process with resetting
\begin{eqnarray}
\lambda_n^{[Reset]} = \lambda_n+ \gamma = 
 \frac{4 \pi^2}{L^2}      n^2 +  i  \mu \frac{2 \pi} { L} n  +\gamma 
 \ \ \ {\rm with } \ \  n=\pm 1,\pm 2, \pm 3, ...
\label{lambdanresetDiff}
\end{eqnarray}
that appear in complex-conjugate pairs as before in Eq. \ref{eigenringmucc}
\begin{eqnarray}
 \lambda_{(-n)}^{[Reset]}=   \frac{4 \pi^2}{L^2}      n^2 -  i  \mu \frac{2 \pi} { L} n  +\gamma  = (\lambda_n^{[Reset]} )^*
\label{eigenringmuccreset}
\end{eqnarray}
The spectral Kemeny time of Eq. \ref{KemenyReset} 
 \begin{eqnarray}
\tau^{[Reset]Spectral}_{*} && =   
  \sum_{n=1}^{+\infty} \left( \frac{1}{\lambda^{[Reset]}_n} + \frac{1}{\lambda^{[Reset]}_{-n}}\right)
=  \sum_{n=1}^{+\infty} 
 \left( \frac{1}{ \frac{4 \pi^2}{L^2}      n^2 +\gamma+  i  \mu \frac{2 \pi} { L} n  } 
 + \frac{1}{\frac{4 \pi^2}{L^2}      n^2 +\gamma-  i  \mu \frac{2 \pi} { L} n}\right)
\nonumber \\
&& =  - \frac{1}{\gamma} + \frac{L}{2 \sqrt{\mu^2+4 \gamma} } 
\left[ \coth \left( L \frac{\sqrt{\mu^2+4 \gamma}+ \mu}{4}\right) 
+ \coth \left( L \frac{\sqrt{\mu^2+4 \gamma}- \mu}{4}\right)\right]
\nonumber \\
&& =  - \frac{1}{\gamma} 
+ \frac{L \sinh \left( \frac{L }{2} \sqrt{\mu^2+4 \gamma}\right)}{2 \sqrt{\mu^2+4 \gamma}  
\sinh \left( L \frac{\sqrt{\mu^2+4 \gamma}+ \mu}{4}\right) 
\sinh \left( L \frac{\sqrt{\mu^2+4 \gamma}- \mu}{4}\right)}
\label{KemenyResetDiff}
\end{eqnarray}
is in agreement with the real-space Kemeny time computed in Eq. \ref{taukemenyresetspace} 
of Appendix \ref{app_JumpDiff}.

The asymptotic linear behavior for large $L$ reads
\begin{eqnarray}
\tau^{[Reset]Spectral}_{*} \opsimeq_{L \to + \infty} 
  = \frac{L }{   \sqrt{\mu^2+4 \gamma} } 
\label{KemenyResetDifflarge}
\end{eqnarray}
So for any non-vanishing drift $\mu \ne 0$, the presence of resetting $\gamma>0$
only changes the coefficient of asymptotic linear behavior for large $L$ of Eq. \ref{kemenyringmuspectrallargeL}.

For vanishing force $\mu = 0$, the Kemeny time of Eq. \ref{KemenyResetDiff} reduces to
 \begin{eqnarray}
\mu=0 : \ \ \ \tau^{[Reset]Spectral}_{*} 
&& =  - \frac{1}{\gamma} 
+ \frac{L \sinh \left( L  \sqrt{ \gamma}\right)}{4 \sqrt{ \gamma}  
\sinh^2 \left( \frac{L}{2}  \sqrt{ \gamma}\right)}
=  - \frac{1}{\gamma} 
+ \frac{L  \cosh \left( \frac{L}{2}  \sqrt{ \gamma}\right)}
{2 \sqrt{ \gamma}  \sinh \left( \frac{L}{2}  \sqrt{ \gamma}\right)}
\nonumber \\
&& =  \frac{1}{ \gamma } \left[ 
\frac{  \frac{L}{2}  \sqrt{\gamma} }{\tanh \left(  \frac{L}{2}  \sqrt{\gamma} \right)}  -1\right] 
\label{KemenyResetDiffmu0}
\end{eqnarray}
So the presence of resetting $\gamma>0$
produces the asymptotic linear behavior 
\begin{eqnarray}
\mu=0 : \ \ \ \tau^{[Reset]Spectral}_{*} \opsimeq_{L \to + \infty} 
  = \frac{L }{  2 \sqrt{ \gamma} } 
\label{KemenyResetDifflargemu0}
\end{eqnarray}
instead of the diffusive scaling of Eq. \ref{kemenyringmuspectralDB} for $\gamma=0$.

As already stressed after Eq. \ref{KemenyReset}, the Kemeny time does not depend on
the resetting probability $\Pi(y)$. In the present example,
a natural choice is the uniform resetting probability 
\begin{eqnarray}
\Pi(y) = \frac{1}{L} \ \ {\rm for } \ \ 0 \leq y \leq L
\label{Piuniform}
\end{eqnarray}
so that the steady state $p^{[Reset]}_*(x)$ with resetting of Eqs \ref{steadyreset} \ref{proparesetlimit}
is also uniform and coincides with the steady state $p_*(x) $ without resetting of Eq. \ref{steadyringuni}
\begin{eqnarray}
p^{[Reset]}_*(x) = p_*(x) = \frac{1}{L} \ \ {\rm for } \ \ 0 \leq x \leq L
\label{steadyresetuni}
\end{eqnarray}
Then the conclusion is that the convergence towards the uniform distribution is always better
in the presence of resetting.


\section{ Directed Trap model on the periodic Ring of $N$ sites  }

\label{sec_DirectedTrap}

For Markov chains on the periodic ring of $N$ sites, one can write general formula
as in the previous section concerning irreversible diffusions on the ring.
In order to avoid repetitions, we focus in the present section
on the specific example of the Directed Trap model
in order to compare with the reflecting boundary conditions considered in section 
\ref{sec_ChainInterval}.


\subsection{ Directed Trap model with arbitrary site-dependent trapping times $\vartheta_x $ on the periodic Ring of $N$ sites}

The model is defined on a ring of $L$ sites with periodic boundary conditions $x+L \equiv x$
with the master equation
\begin{eqnarray}
\partial_t P_t(x) =   \frac{ P_t(x-1) }{\vartheta_{x-1}}   - \frac{ P_t(x) }{\vartheta_x}
\label{wjumptrapDIR}
\end{eqnarray}
So $\vartheta_x $ represents the trapping time already discussed in Eqs \ref{escapejump}
\ref{escapejumpav}, but the difference with respect to the symmetric trap model discussed
in Eq. \ref{wjumptrap} is that now, when the particle escapes from site $x$, it jumps towards the
 the right neighbor $(x+1)$.

The steady state $P_*(x)$ at site $x$ is simply proportional to its trapping time $\vartheta_x $
\begin{eqnarray}
P_*(x) =  \frac{ \vartheta_x }{ \displaystyle \sum_{y=1}^N \vartheta_y}
\label{steadyDIR}
\end{eqnarray}
as in Eq. \ref{steadySymTrap} for the symmetric trap model.

The Mean-First-Passage-Time $\tau(x+1,x)$ at site $(x+1)$ when starting at $x$
 is directly the trapping time $\vartheta_x $
\begin{eqnarray}
\tau(x+1,x) =\vartheta_x 
\label{mftpDirected}
\end{eqnarray}
As a consequence, the Mean-First-Passage-Time $\tau(x,x_0)$ at $x$ when starting at  $x_0$
can be decomposed in the two regions $x>x_0$ and $x<x_0$ into the following sums 
that take into account the directed motion along the ring
\begin{eqnarray}
{\rm for } \ \ x_0 \in \{1,..,x-1\} : \ \ \ \  \tau(x,x_0) && =\tau(x,x-1)+\tau(x-1,x-2)+ .. + \tau(x_0+1,x_0)
= \sum_{z=x_0 }^{x-1} \tau(z+1,z) = \sum_{z=x_0 }^{x-1} \vartheta_z
\nonumber \\
{\rm for } \ \ x_0 \in \{x+1,..,N\} : \ \ \ \ \tau(x,x_0) && = \tau(x,x-1)+.. \tau(2,1) + \tau(1,N) + \tau(N,N-1) + .. \tau(x_0+1,x_0)
\nonumber \\
&& = \sum_{z=1 }^{x-1} \tau(z+1,z) +  \sum_{z=x_0 }^{N} \tau(z+1,z)
= \sum_{z=1 }^{x-1} \vartheta_z +  \sum_{z=x_0 }^{N} \vartheta_z
 \label{tausumDir}
\end{eqnarray}

The real-space Kemeny time $\tau^{Space}_*$ of Eq. \ref{taukemeny}
can be thus evaluated for the special case $x_0=1$ using Eq. \ref{tausumDir}
 \begin{eqnarray}
\tau^{Space}_*   && =  \sum_{x=1}^N p_*(x) \tau(x,x_0=1) 
=  \sum_{x=2}^N p_*(x) \tau(x,x_0=1) 
\nonumber \\
&& =   \frac{ \displaystyle \sum_{x=2}^N \vartheta_x  \sum_{y=1 }^{x-1} \vartheta_y}
{ \displaystyle \sum_{z=1}^N \vartheta_z}
= \frac{ \displaystyle \sum_{y=1 }^{N-1}  \sum_{x=y+1}^N \vartheta_y \vartheta_x  }
{ \displaystyle \sum_{z=1}^N \vartheta_z}
\label{kemenyDir}
\end{eqnarray}
that is always smaller than the Kemeny time of Eq. \ref{taueqDBMsymTrap}
for the symmetric version of the trap model with reflecting boundary conditions
involving the same trapping times.
In addition, Eq. \ref{kemenyDir} can be rewritten only in terms of the sum of all the trapping times
and of the sum of their squares
 \begin{eqnarray}
\tau^{Space}_*   && 
=  \frac{ \displaystyle \left[ \sum_{y=1 }^{N} \vartheta_y \right]^2- \left[ \sum_{1}^N \vartheta^2_y \right]  }
{2 \displaystyle \sum_{z=1}^N \vartheta_z}
\label{kemenyDirbis}
\end{eqnarray}
so that it does not depend on the positions of the trapping times along the ring as expected.


\subsection{ Special case of the pure Directed Trap model $\vartheta_x = \frac{1}{2} $ for $x=1,..,N$ }

When the trapping times take the same value along the ring 
\begin{eqnarray}
\vartheta_x = \frac{1}{2}  \ \ \ {\rm for } \ \ x=1,2,..,N
\label{trapDIRuni}
\end{eqnarray}
 one obtains the Directed version of the symmetric
continuous-time random walk on the interval with reflecting boundary conditions
of Eq. \ref{wuniform}.

\subsubsection{ Real-space Kemeny time}

The steady state of Eq. \ref{steadyDIR} is uniform
\begin{eqnarray}
p_*(x) =  \frac{ 1 }{ N} 
\label{steadyDIRuni}
\end{eqnarray}
and coincides with the steady state of Eq. \ref{steadymuzero} for the symmetric model.
The real-space Kemeny time of Eq. \ref{kemenyDir}
 \begin{eqnarray}
\tau^{Space}_*  = \frac{1}{2} \frac{ \displaystyle \sum_{x=2}^N   \sum_{z=1 }^{x-1} 1}
{ \displaystyle \sum_{y=1}^N 1} = \frac{ N-1}{4}
\label{kemenyDiruni}
\end{eqnarray}
is linear with respect to the size $N$, in contrast to the diffusive scaling of Eq. \ref{spaceKemenymu0}
for the symmetric model.

\subsubsection{ Spectral Kemeny time}

For the master equation of Eq. \ref{wjumptrapDIR}, the eigenvalue equation reads
 \begin{eqnarray}
-\lambda r(x) =  2 r(x-1) -  2  r(x)
\label{eigenvecDir}
\end{eqnarray}
On this periodic ring, the eigenvectors are simply the Fourier modes
 \begin{eqnarray}
r_n (x) = e^{i 2 \pi n \frac{x}{N}} \ \ \ {\rm with } \ \ n=0,1,..,N-1
\label{fourierDir}
\end{eqnarray}
with the complex eigenvalues
 \begin{eqnarray}
\lambda_n  =   2 \left( 1- e^{- i 2 \pi  \frac{n}{N} } \right) \ \ \ {\rm with } \ \ n=0,1,..,N-1
\label{eigenDir}
\end{eqnarray}
so that the spectral Kemeny time of Eq. \ref{tauspectral} 
that can be computed using Eq. \ref{polynomialderilog1}
 \begin{eqnarray}
\tau^{Spectral}_{*} =  \sum_{n=1}^{N-1} \frac{1}{\lambda_n} 
= \frac{1}{2} \sum_{n=1}^{N-1} \frac{1}{ 1 -  e^{- i 2 \pi  \frac{n}{N} }}  = \frac{N-1}{4}
\label{tauspectralDir}
\end{eqnarray}
is in agreement with the real-space Kemeny time of Eq. \ref{kemenyDiruni}.

\subsubsection{ Spectral Kemeny time in the presence of stochastic resetting at rate $\gamma$ }

In the presence of stochastic resetting at rate $\gamma$, the spectral Kemeny time of Eq. \ref{tauspectralDir}
is modified into Eq. \ref{KemenyReset} that can be computed using Eq. \ref{polynomialderilogz}
 \begin{eqnarray}
\tau^{[Reset]Spectral}_{*} 
=  \sum_{n=1}^{N-1} \frac{1}{\gamma+\lambda_n}
=\frac{1}{2} \sum_{n=1}^{N-1} \frac{1}{ \left( \frac{\gamma}{2}+1\right)  -  e^{- i 2 \pi  \frac{n}{N} }}
=  \frac{ N   }{ \left( \gamma +2\right)
 \left[ 1-\left( \frac{\gamma}{2}+1\right)^{-N} \right] } - \frac{ 1 }{  \gamma } 
\label{KemenyResetDirTrap}
\end{eqnarray}
so that the behavior for large $N$ is linear with a modified prefactor with respect to Eq. \ref{tauspectralDir}
 \begin{eqnarray}
\tau^{[Reset]Spectral}_{*} 
\opsimeq_{N \to + \infty}  \frac{ N   }{  \gamma +2}
\label{KemenyResetDirTrapNlarge}
\end{eqnarray}


\section{ Generalizations of the Kemeny time for absorbing Markov processes }

\label{sec_Absorbing}

In subsection \ref{subsec_forever}, we have considered the Kemeny times of
 absorbing Markov process conditioned to survive forever.
 In the present section, we discuss the analogs of the Kemeny time for the initial unconditioned processes.

\subsection{ Analogs of the Kemeny times for absorbing Markov processes }

For the Markov process that gets absorbed in the dead configuration $0$,
it is interesting to analyze the analogs of the spectral Kemeny time and of the real-space Kemeny times
  \cite{Mao2006Transient,Mao2010Irreversible}:

(i) on one hand, one can introduce the analog of the spectral Kemeny time of Eq. \ref{tauspectral}
 \begin{eqnarray}
\tau^{Spectral[Abs]}_{*} 
\equiv  \sum_{n=1}^{N} \frac{1}{\lambda^{[Abs]}_n} 
\label{tauspectralAbs}
\end{eqnarray}
that involves the sum of the inverse of the $N$ absorbing-eigenvalues $\lambda^{[Abs]}_n $ 
associated to the propagator of Eq. \ref{propagatorAbs}.

(ii) on the other hand, one can introduce the analog of the real-space Kemeny time Eq. \ref{taukemeny}:
however it reduces to the single term $x=0$ associated
to the absorbing steady state $P^{[Abs]}_*(x)=\delta_{x,0}$
 \begin{eqnarray}
  \sum_x P^{[Abs]}_{*}(x) \tau(x,x_0) = \tau^{[Abs]}(0,x_0) \ \ \ \ { \rm dependent \ \ of } \ \ x_0
\label{taukemenyAbs}
\end{eqnarray}
so that it will depend
 on the initial point $x_0$ in contrast to the ergodic case of Eq. \ref{taukemeny}.
Note that this Mean-First-Passage-Time $\tau^{[Abs]}(0,x_0) $ at the absorbing site $x=0$ 
when starting at $x_0$,
represents the Mean-Absorption-Time where the dynamics stops.

To analyze the properties of the two times of Eqs \ref{tauspectralAbs} and \ref{taukemenyAbs}
it is useful to consider the properties of the Green function when the Markov process
is absorbing, and to mention the changes with respect to the properties of the Green function
of ergodic cases described in section \ref{sec_green}.


\subsection{ Properties of the Green function $G^{[Abs]}(x,x_0)  $ for the absorbing Markov chain} 

From the absorbing propagator of Eq. \ref{propagatorAbs} that converge towards the absorbing state at $x=0$
\begin{eqnarray}
P^{[Abs]}_*(x)=\delta_{x,0}
 \label{steadysabs}
\end{eqnarray}
 one can define the absorbing Green function
via the analog of Eq. \ref{defgreen} 
\begin{eqnarray}
G^{[Abs]}(x,x_0) && \equiv \int_0^{+\infty} dt \left[ P^{[Abs]}_t( x \vert x_0 )  - \delta_{x,0}\right]
= \int_0^{+\infty} dt \left[  \sum_{n=1}^{N} e^{- t \lambda_n^{[Abs]} } \langle x \vert r_n \rangle \langle l_n  \vert x_0 \rangle \right]
\nonumber \\
&& =\sum_{n=1}^{N}  \frac{ \langle x \vert r_n \rangle \langle l_n  \vert x_0 \rangle }
{\lambda_n^{[Abs]}} 
\label{defgreenAbs}
\end{eqnarray}
The forward and backward equations of Eq. \ref{eqG} become 
\begin{eqnarray}
\sum_y w^{[Abs]}(x,y)  G^{[Abs]}(y,x_0) && = \delta_{x,0} - \delta_{ x,x_0}  
 \nonumber \\
\sum_{y_0}  G^{[Abs]}(x,y_0)  w^{[Abs]}(y_0,x_0) && =\delta_{x,0} - \delta_{ x,x_0}  
 \label{eqGAbs}
\end{eqnarray}
while the orthogonality relations of Eq. \ref{Gorthog} read
 \begin{eqnarray}
&& {\rm for \ any \ x_0 } : \ \ \ \ 
 0=  \langle l_0 \vert  G^{[Abs]} \vert x_0 \rangle  =  \sum_{x=0}^N  G^{[Abs]}(x,x_0)  
 \nonumber \\
&& {\rm for \ any \ x } : \ \ \ \ 
 0=  \langle x \vert   G^{[Abs]} \vert r_0 \rangle  
 =  \sum_{x_0=0}^N   G^{[Abs]}(x,x_0) \delta_{x_0,0} = G^{[Abs]}(x,0)
\label{GorthogAbs}
\end{eqnarray}
The second relation is obvious from the definition of 
the first expression Eq. \ref{defgreenAbs} when the starting point is the absorbing point itself $x_0=0$
from which one cannot move.

Using $G^{[Abs]}(x,0)=0 $ and the orthonormalization of Eq. \ref{fermeturej},
the trace of the Green function of Eq. \ref{defgreenAbs}
that can be computed as a sum of the diagonal elements $G(x,x)$ over living configurations $x \ne 0$
\begin{eqnarray}
{\rm Tr} ( G^{[Abs]}) = \sum_{x=1}^N G^{[Abs]}(x,x)  =\sum_{n=1}^{N}  
\frac{ \sum_x \langle l_n  \vert x \rangle \langle x \vert r_n \rangle}
{\lambda_n^{[Abs]}} 
=\sum_{n=1}^{N}  \frac{  \langle l_n   \vert r_n \rangle}
{\lambda_n^{[Abs]}} 
= \sum_{n=1}^{N}  
\frac{  1}{\lambda_n^{[Abs]}} =\tau^{Spectral[Abs]}_{*} 
\label{tracegreenAbs}
\end{eqnarray}
corresponds to the sum of the inverses of the eigenvalues introduced in Eq. \ref{tauspectralAbs}.


 \subsection{ Link with the averaged time $t^{[Abs]}_{[0,+\infty[}(x,x_0)$ spent in the living configuration $x$ 
when starting at $x_0$  }

The averaged time $t^{[Abs]}_{[0,T]}(x,x_0)$ spent in the living configuration $x \ne 0$ 
when starting at $x_0$ remains finite for $T \to +\infty$  
(in contrast to the extensive behavior in $T$ of Eq. \ref{avlocaltimeExtensive} for the ergodic dynamics)
\begin{eqnarray}
t^{[Abs]}_{[0,+\infty[}(x,x_0) \equiv \int_0^{+\infty} dt  P^{[Abs]}_t( x \vert x_0 ) = G^{[Abs]}(x,x_0) \ \ {\rm for } \ x \ne 0
\label{avlocaltimeAbs}
\end{eqnarray}
and directly corresponds to the Green function $G^{[Abs]}(x,x_0) $.
It is only for the absorbing configuration $x=0$ that one still has the extensive behavior in $T$ of Eq. \ref{avlocaltimeExtensivefinite} 
\begin{eqnarray}
t^{[Abs]}_{[0,T]}(0,x_0) \equiv \int_0^{T} dt  P^{[Abs]}_t( x=0 \vert x_0 )
\opsimeq_{T \to +\infty}  T  +G(0,x_0) + o(T^0)
\label{avlocaltimeExtensivefiniteAbs}
\end{eqnarray}


 \subsection{ Link with the Mean-First-Passage-time $\tau^{[Abs]}(0,x_0)$ at the dead configuration $x=0$
when starting at $x_0$  }

The averaged time $t^{[Abs]}_{[0,T]}(0,x_0) $ spent in the dead configuration $x=0$
is directly given by the difference between the total time $T$
and the Mean-First-Passage-time $\tau^{[Abs]}(0,x_0)$
\begin{eqnarray}
t^{[Abs]}_{[0,T]}(0,x_0) =  T  -\tau^{[Abs]}(0,x_0) 
\label{avlocaltimeDiffAbs}
\end{eqnarray}
so the comparison with Eq. \ref{avlocaltimeExtensivefiniteAbs} yields the simple relation
\begin{eqnarray}
\tau^{[Abs]}(0,x_0)  = - G^{[Abs]}(0,x_0)    
\label{Gtau0Abs}
\end{eqnarray}
Plugging Eq. \ref{Gtau0Abs} into the backward Eq. \ref{eqGAbs} for $G^{[Abs]}(x=0,y_0)  $ yields
the standard backward equation of Eq. \ref{Ginvlefttau}
for the Mean-First-Passage-time $\tau^{[Abs]}(0,x_0) $
\begin{eqnarray}
\sum_{y_0}  \tau^{[Abs]}(0,y_0)  w^{[Abs]}(y_0,x_0)  =  -1
 \label{eqtauAbs}
\end{eqnarray}
while plugging Eq. \ref{Gtau0Abs} into the first orthogonality relation of Eq. \ref{GorthogAbs}
yields that the Mean-First-Passage-time $\tau^{[Abs]}(0,x_0) $ 
\begin{eqnarray}
\tau^{[Abs]}(0,x_0)  = - G^{[Abs]}(0,x_0)   
=  \sum_{x \ne 0} G^{[Abs]}(x,x_0) =\sum_{x \ne 0} t^{[Abs]}_{[0,+\infty[}(x,x_0)
\label{tau0AbsumG}
\end{eqnarray}
can be rewritten as the sum of the averaged times $t^{[Abs]}_{[0,+\infty[}(x,x_0) $ of Eq. \ref{avlocaltimeAbs}
spent in all the living configurations $x \ne 0$ before absorption, as expected.


 \subsection{ Link with the probability $q(x \vert x_0) $ to visit $x$ before absorption at $0$ when starting at $x_0$ }

For the living configuration $x \ne 0$, it is useful to introduce the probability distribution $Q_{t_1}(x \vert x_0)$ of the first-passage time $t_1$ at $x$ when starting at $x_0 \ne x $,
whose normalization over $t_1 \in ]0,+\infty[$
\begin{eqnarray}
\int_0^{+\infty} dt_1 Q_{t_1}(x \vert x_0) = q(x \vert x_0) \in [0,1]
\label{Qt1norma}
\end{eqnarray}
represents the probability $ q(x \vert x_0)$ to visit $x$ before absorption at $0$ when starting at $x_0$.

\subsubsection{ Time-decomposition of the propagator $P^{[Abs]}_t( x \vert x_0 ) $ 
between two living distinct configurations $0 \ne x \ne x_0 \ne 0 $}

The propagator $P^{[Abs]}_t( x \vert x_0 ) $ can be decomposed 
for two living distinct configurations $0 \ne x \ne x_0 \ne 0 $
with respect to the first-passage time $t_1$ at $x$ via the convolution
\begin{eqnarray}
P^{[Abs]}_t( x \vert x_0 ) = \int_0^{t} dt_1 Q_{t_1}(x \vert x_0) P^{[Abs]}_{t-t_1}( x \vert x ) 
\ \ \ {\rm for } \  0 \ne x \ne x_0 \ne 0
\label{PAbsQ}
\end{eqnarray}
The integration over time $t \in [0,+\infty[$ yields that
 for two different living configurations $ 0 \ne x \ne x_0 \ne 0$,
the Green function $G^{[Abs]}(x,x_0) $ of Eq. \ref{defgreenAbs} 
\begin{eqnarray}
G^{[Abs]}(x,x_0) = \int_0^{+\infty} dt  P^{[Abs]}_t( x \vert x_0 )  
= \int_0^{+\infty} dt_1 Q_{t_1}(x \vert x_0)  \int_0^{+\infty} ds    P^{[Abs]}_{s}( x \vert x )
= q(x \vert x_0) G^{[Abs]}(x,x)
\label{GGqAbs}
\end{eqnarray}
reduces to the product of the probability $q(x \vert x_0) $ of Eq. \ref{Qt1norma}
and of the Green function $G^{[Abs]}(x,x) $ at coinciding points,
that will be analyzed in the next subsection.

Plugging Eq. \ref{GGqAbs} into the backward 
Eq. \ref{eqGAbs} for $G^{[Abs]}(x,y_0)  $ with $ 0 \ne x \ne x_0 \ne 0$
yields the backward equation for the probability 
$q(x \vert y_0)$ to visit $x$ before $0$ when starting at $y_0$
\begin{eqnarray}
 \sum_{y_0}  q(x \vert y_0)  w^{[Abs]}(y_0,x_0)  =0
 \label{backwardq}
\end{eqnarray}
which is indeed a standard method to compute $q(x \vert y_0)$ 
with the boundary conditions $q(x \vert x_0=x)=1 $ and $q(x \vert x_0=0)=0 $
(see for instance the textbooks \cite{gardiner,vankampen,risken}).


\subsubsection{ Time-decomposition of the propagator $P^{[Abs]}_t( x \vert x ) $ 
for a living configuration $x \ne 0  $}

Let us now write the analog of the decomposition of Eq. \ref{PAbsQ} for $x_0=x \ne 0$
\begin{eqnarray}
P^{[Abs]}_t( x \vert x ) = e^{ t w^{[Abs]}(x,x) } 
+ \int_0^{t} dt_1  e^{ t_1 w^{[Abs]}(x,x) } \sum_{y \ne x} w^{[Abs]}(y,x) 
\int_0^{t-t_1} dt_2 Q_{t_2}(x \vert y) P^{[Abs]}_{t-t_1-t_2}( x \vert x ) 
\label{PAbsQxx}
\end{eqnarray}
with the following physical meaning:

(1) the first term $ e^{ t w^{[Abs]}(x,x) }$ represents the probability to have remained in the configuration $x$
during the time $t$, where the opposite of the diagonal element of Eq. \ref{wdiag}
\begin{eqnarray}
- w^{[Abs]}(x,x) \equiv   \sum_{y \ne x} w^{[Abs]}(y,x) >0
\label{wdiagAbs}
\end{eqnarray}
represents the total rate out of the configuration $x$.

(2) the second term of Eq. \ref{PAbsQxx} is the time-convolution of three functions:

(2a) the first function $\left[ e^{ t_1 w^{[Abs]}(x,x) } w^{[Abs]}(y,x) \right] $ represents the probability 
that the first escape from configuration $x$ occurs at time $t_1$ towards the configuration $y$.

(2b) the second function $ Q_{t_2}(x \vert y)$ represents the probability distribution of the 
first-passage-time $t_2$ at configuration $x$ when starting at configuration $y$, 
as introduced in Eq. \ref{Qt1norma}

(2c) the third function $ P^{[Abs]}_{t-t_1-t_2}( x \vert x )$ represents the propagator from $x$ to $x$
 in the remaining available time $(t-t_1-t_2)$.

The integration over time $t \in [0,+\infty[$ of Eq. \ref{PAbsQxx}
yields that
the Green function $G^{[Abs]}(x,x) $ at coinciding points of Eq. \ref{defgreenAbs} satisfies for $x \ne 0$
the consistency equation
\begin{eqnarray}
G^{[Abs]}( x \vert x ) = \frac{1}{ [- w^{[Abs]}(x,x) ] } \left[ 1 + \sum_{y \ne x} w^{[Abs]}(y,x)  q(x \vert y) G^{[Abs]}( x \vert x ) \right]
\label{GAbsQxx}
\end{eqnarray}
So $G^{[Abs]}( x \vert x ) $ 
can be rewritten in terms of the probabilities $[1-q(x \vert y) ]$ of never returning to $x$
when starting at configurations $y \ne x$
that are accessible from $x$ via the strictly positive transition rates $w^{[Abs]}(y,x)  >0$
\begin{eqnarray}
G^{[Abs]}( x \vert x ) = \frac{1}{ \displaystyle \sum_{y \ne x} w^{[Abs]}(y,x) \left[ 1- q(x \vert y)\right]  } 
\label{GAbsQxxsol}
\end{eqnarray}


 \subsection{ Conclusion for the spectral Kemeny time $\tau^{Spectral[Abs]}_{*} $ 
 and the Mean-First-Passage-time $\tau^{[Abs]}(0,x_0) $ }

In summary,
 the spectral Kemeny time $ \tau^{Spectral[Abs]}_{*}$ of Eq. \ref{tracegreenAbs} 
 corresponds to the sum of $G^{[Abs]}( x \vert x )  $ over $x \ne 0$,
while the absorbing time $\tau^{[Abs]}(0,x_0) $ of Eq. \ref{tau0AbsumG}  when starting from $x_0$
can be rewritten as the sum of Eq. \ref{GGqAbs}
over the living configurations $x \ne 0$ as
\begin{eqnarray}
\tau^{[Abs]}(0,x_0)  =  \sum_{x \ne 0} G^{[Abs]}(x,x_0) 
=\sum_{x \ne 0} G^{[Abs]}(x,x) q(x \vert x_0) 
\label{tau0AbsumGGG}
\end{eqnarray}

Since the probability $q(x \vert x_0) $ of Eq. \ref{Qt1norma} is smaller than unity $q(x \vert x_0) \leq 1$,
the general conclusion is that the spectral Kemeny time $\tau^{Spectral[Abs]}_{*}  $ 
is an upper bound for the absorbing time $\tau^{[Abs]}(0,x_0) $ for any starting configuration $x_0$ \cite{Mao2006Transient,Mao2010Irreversible}
\begin{eqnarray}
{\rm for \ any } \ x_0 : \tau^{[Abs]}(0,x_0)  
\leq  \sum_{x \ne 0} G^{[Abs]}(x,x) =  \sum_{n=1}^{N}  \frac{  1}{\lambda_n^{[Abs]}}
\equiv \tau^{Spectral[Abs]}_{*} 
\label{absSmallerSpectralineq}
\end{eqnarray}

This inequality becomes an equality only if the starting point $x_0$ satisfies 
$q(x \vert x_0) = 1$ for any other living configuration $x \ne 0$
\begin{eqnarray}
 \tau^{[Abs]}(0,x_0)  =
 \tau^{Spectral[Abs]}_{*} \equiv \sum_{n=1}^{N}  \frac{  1}{\lambda_n^{[Abs]}}
\ \ {\rm \ \ only \ \  if \ for \ any } \ x \ne 0  : \ \ q(x \vert x_0) = 1
\label{absSmallerSpectral}
\end{eqnarray}
In the following subsections, we focus on such examples in the one-dimensional geometry.


 \subsection{Application to Markov chains on the interval $\{0,1,..,N\}$ with absorption at $x=0$ }

Let us consider the Markov chains on the interval $\{0,1,..,N\}$ with absorption at $x=0$
\begin{eqnarray}
\partial_t P_t(0) && = w(0,1) P_t(1)
\nonumber \\
\partial_t P_t(1) && = w(1,2) P_t(2) - \left[ w(0,1) +w(2,1) \right] P_t(1)
\nonumber \\
\partial_t P_t(x) && = w(x,x-1) P_t(x-1) +w(x,x+1) P_t(x+1) - \left[ w(x-1,x) +w(x+1,x) \right] P_t(x) 
 \ \ \ {\rm for } \ x=2,..,N-1
\nonumber \\
\partial_t P_t(N) && =w(N,N-1) P_t(N-1)  -  w(N-1,N) P_t(N)
\label{Absorbibg1d}
\end{eqnarray}


\subsubsection{ Mean-First-Passage-time $ \tau^{[Abs]}(0,x_0=N) $ at the dead configuration $x=0$ when starting at the other boundary $x_0=N$}

When the starting point is $x_0=N$, 
the particle cannot reach the absorbing configuration $0$ 
without visiting all the other living configurations $x=N-1,..,1$, so
the probability $q(x \vert N) $ of Eq. \ref{Qt1norma}
is unity
 \begin{eqnarray}
   q(x \vert N)    = 1 \ \ \ {\rm for } \ \ x=1,2,..,N-1,N
 \label{qendunity}
\end{eqnarray}
Eq. \ref{absSmallerSpectral} then
yields that the Mean-First-Passage-time $ \tau^{[Abs]}(0,x_0=N) $ when starting at $x_0=N$
coincides with the spectral Kemeny time 
\begin{eqnarray}
 \tau^{[Abs]}(0,x_0=N)  =
 \tau^{Spectral[Abs]}_{*} \equiv \sum_{n=1}^{N}  \frac{  1}{\lambda_n^{[Abs]}}
\label{tauAbsEndSpectral}
\end{eqnarray}

Besides this result concerning the Mean-First-Passage-time $ \tau^{[Abs]}(0,x_0=N) $
on which we focus here as an example of the generalization of the Kemeny
 time for absorbing Markov chains,
let us mention that the whole probability distribution $Q_{t_1}(x=0 \vert x_0=N)$ of the first-passage time $t_1$ at $x=0$ when starting at $x_0 =N $, 
whose normalization over $t_1 \in ]0,+\infty[$ of Eq. \ref{Qt1norma} is unity,
has a very simple Laplace transform in terms of the $N$ eigenvalues $\lambda_n^{[Abs]}$
(see \cite{Mao2012hitting} and references therein)
\begin{eqnarray}
\int_0^{+\infty} dt_1 e^{-s t_1} Q_{t_1}(x=0 \vert x_0=N) 
= \sum_{n=1}^N \frac{\lambda_n^{[Abs]}}{s+\lambda_n^{[Abs]}}
\label{Qt1laplace}
\end{eqnarray}
This means that $Q_{t_1}(x=0 \vert x_0=N)$ is the convolution of the $N$ exponential distributions
${\cal E}_n(t)= \lambda_n^{[Abs]} e^{- t  \lambda_n^{[Abs]} }$.

Let us now return to Eq. \ref{tauAbsEndSpectral}.
The Mean-First-Passage-Time $\tau^{[Abs]}(0,y_0)$
can be computed as a function of the starting point $y_0$
via the backward Eq. \ref{eqtauAbs}
with the boundary condition $\tau^{[Abs]}(0, 0) =0$
\begin{eqnarray}
-1 && =   \left[ \tau^{[Abs]}(0, 2)-\tau^{[Abs]}(0, 1) \right]  w(2,1) 
+\left[ 0  - \tau^{[Abs]}(0, 1)\right] w(0,1)
   \nonumber \\
-1 && =   \left[ \tau^{[Abs]}(0, x_0+1)-\tau^{[Abs]}(0, x_0) \right]  w(x_0+1,x_0) 
+\left[ \tau^{[Abs]}(0, x_0-1)  - \tau^{[Abs]}(0, x_0)\right] w(x_0-1,x_0)
   \ \ \ {\rm for } \ x_0=2,..,N-1
   \nonumber \\
-1 && = \left[ \tau^{[Abs]}(0, N-1)  - \tau^{[Abs]}(0, N)\right] w(N-1,N) 
 \label{eqtauAbschain}
\end{eqnarray}
The solution
\begin{eqnarray}
\tau^{[Abs]}(0, x_0) = \sum_{x=1}^{x_0} \frac{1}{w(x-1,x)} \left[ 1+ \sum_{z=x}^{N-1}  \prod_{y=x}^z \frac{w(y+1,y)  }{w(y,y+1)}\right]
 \label{eqtauAbschainSol}
\end{eqnarray}
reads for $x_0=N$ 
\begin{eqnarray}
\tau^{[Abs]}(0, N) = \sum_{x=1}^{N} \frac{1}{w(x-1,x)} \left[ 1+ \sum_{z=x}^{N-1}  \prod_{y=x}^z \frac{w(y+1,y)  }{w(y,y+1)}\right]
 \label{AbsN}
\end{eqnarray}
So even if the individual eigenvalues $ \lambda_n^{[Abs]}$ are not known,
Eq. \ref{eqtauAbschainSol} gives
an explicit expression for arbitrary transitions rates $w(.,.)$ of the spectral Kemeny time $ \tau^{Spectral[Abs]}_{*} $ of Eq. \ref{tauAbsEndSpectral}.


\subsubsection{ Simplest example of the continuous-time random walk with rates $w(x\pm 1,x)  =1 $
with absorption at $0$}

Let us consider the simplest example where all the rates have the same value unity
\begin{eqnarray}
w(x+1,x) && =1  \ \ \ {\rm for } \ x=1,2,..,N-1
\nonumber \\
w(x,x+1) && =1  \ \ \ {\rm for } \ x=0,1,2,..,N-1
\label{wuniformabs}
\end{eqnarray}
Then the Mean-First-Passage-Time of Eq. \ref{AbsN}
reduces to
\begin{eqnarray}
\tau^{[Abs]}(0, N) = \sum_{x=1}^{N}  \left[ 1+ \sum_{z=x}^{N-1} 1 \right]
= \sum_{x=1}^{N}  \left[ 1+ N-x \right] = \frac{N^2+N}{2}
 \label{AbsNuni}
\end{eqnarray}
Let us now compare with the spectral analysis.
For the absorbing Markov chain of Eq. \ref{continuity} with the rates of Eq. \ref{wuniformabs}, 
the eigenvalue equation 
reads for the right eigenvector $r(x)$ associated to the eigenvalue $\lambda $
\begin{eqnarray}
x=0 :\ \ \  \ \ - \lambda r (0) && =  r (1)
\nonumber \\
 x=1 :\ \ \  \ \ - \lambda r (1) && = r(2) - 2 r (1)
\nonumber \\
x=2,..,N-1:   - \lambda r (x) && 
=  r(x-1) +  r(x+1)- 2 r(x)
\nonumber \\
x=N :\ \ \  \ \ - \lambda r (N) && 
=  r(N-1) -  r(N) 
\label{eigenrabs}
\end{eqnarray}
Since $\lambda=0$ is associated to the absorbing steady state $r_0(x)=\delta_{x,0}$,
let us now focus on the non-vanishing eigenvalues $\lambda \ne 0$,
where the first equation simply allows to compute $r(0)$, so we can now focus on the living configurations
$x=1,..,N$.
The bulk recurrence for $x=2,..,N-1$ and the boundary condition at $x=N$
are exactly the same as in Eq. \ref{continuity}, 
so the not-normalized eigenvector $r_k(x)$ associated to the eigenvalue $ \lambda(k) $ of Eq. \ref{lambdakcos}
is given by Eq. \ref{rkcos}, while the boundary equation for $x=1$ in Eq. \ref{eigenrabs}
\begin{eqnarray}
0= [2- \lambda(k) ] r_k (1) -  r_k(2) = 2 \cos(k) \cos \left[ k \left(N-\frac{1}{2}\right)\right] 
 - \cos \left[ k \left(N-\frac{3}{2}\right)\right]
 = \cos \left[ k \left(N+\frac{1}{2}\right)\right] 
\label{eigenrabsb}
\end{eqnarray}
selects the $N$ possible wavevectors $k$ as a function of the size $N$ 
\begin{eqnarray}
k_n = \frac{ - \frac{\pi}{2} + n \pi}{N+\frac{1}{2}} = \pi \frac{  2 n -1}{2N+1} 
\ \ \ \ \ \ {\rm with } \ n=1,..,N
\label{rkcosselectabs}
\end{eqnarray}
with the corresponding eigenvalues of Eq. \ref{lambdakcos}
\begin{eqnarray}
 \lambda^{[Abs]}_n =   2 \left[ 1 -  \cos(k) \right] =  2 \left[ 1 - \cos\left(  \pi \frac{  2 n -1}{2N+1}  \right) \right]
 \ \ \ \ \ \ {\rm with } \ n=1,..,N
\label{lambdakcosnabs}
\end{eqnarray}
The spectral Kemeny time of Eq. \ref{tauAbsEndSpectral} that involves 
these $N$ non-vanishing eigenvalues
$\lambda^{[Abs]}_n$ 
 \begin{eqnarray}
  \tau^{Spectral[Abs]}_{*} \equiv \sum_{n=1}^{N}  \frac{  1}{\lambda_n^{[Abs]}}
  = \frac{1}{2} \sum_{n=1}^{N} \frac{1}{1 - \cos\left(  \pi \frac{  2 n -1}{2N+1} \right)} = \frac{N^2+N}{2}
\label{tauAbsEndSpectraluni}
\end{eqnarray}
is in agreement with the 
Mean-First-Passage-Time $\tau^{[Abs]}(0, N) $ of Eq. \ref{AbsNuni} as it should.


 \subsection{Diffusion processes on the interval $]0,L]$ with absorption at $x=0$ and reflecting boundary at $x=L$}


\subsubsection{ Mean-First-Passage-time $ \tau^{[Abs]}(0,x_0=L) $ at the dead configuration $x=0$ when starting at the other boundary $x_0=L$}
 
 Let us now consider the analog of Eq. \ref{tauAbsEndSpectral}
 for diffusion processes on the interval $]0,L]$ with absorption at $x=0$ and reflecting boundary at $x=L$
\begin{eqnarray}
 \tau^{[Abs]}(0,x_0=L)  =
 \tau^{Spectral[Abs]}_{*} \equiv \sum_{n=1}^{+\infty}  \frac{  1}{\lambda_n^{[Abs]}}
\label{tauAbsEndSpectralDiff}
\end{eqnarray}
 where the Mean-First-Passage-Time $ \tau^{[Abs]}(0,x_0=L)  $ at $x=0$ when starting at $x_0=L$
 was written in Eq. \ref{mftp}
 \begin{eqnarray}
  \tau^{[Abs]}(0,L) &&= \int_{0}^{L} dy \frac{1}{D(y)e^{-U(y)}  } \int_{y}^{L} dz e^{-U(z)}
 \label{mftpAbs}
\end{eqnarray}
in terms of the diffusion coefficient $D(y)$ and in terms of the potential
$U(x)$ of Eq. \ref{UR} with $x_{ref}=L$ 
 \begin{eqnarray}
U(x) && \equiv  \int_{x}^L dy \frac{F(y)}{D(y)} 
\nonumber \\
U'(x) && = - \frac{F(x)}{D(x)}
 \label{URAbs}
\end{eqnarray}


\subsubsection{ Simplest example with diffusion coefficient $D(x)=1$ 
and no force $F(x)=0$}

The simplest example corresponds to the diffusion coefficient $D(x)=1$ 
and to the vanishing force $F(x)=0$ leading to the vanishing potential $U(x)=0$.
Then the Mean-First-Passage-Time of Eq. \ref{mftpAbs}
reduces to
\begin{eqnarray}
\tau^{[Abs]}(0, L) =  \int_{0}^{L} dy  \int_{y}^{L} dz 
= \int_{0}^{L} dy (L-y) =  \frac{L^2}{2}
 \label{mftpAbsuni}
\end{eqnarray}
For this diffusion with reflecting boundary condition at $L$ and absorbing boundary condition at $0$,
the eigenvalues have already been written in Eq. \ref{kOneAbs}
 \begin{eqnarray}
 \lambda_n^{[Abs]}  = \left( n - \frac{1}{2} \right)^2 \frac{\pi^2}{L^2} \ \ \ {\rm for } \ \ \ n=1,2,..,+\infty
\label{kOneAbsL}
\end{eqnarray}
The corresponding spectral Kemeny time
\begin{eqnarray}
 \tau^{Spectral[Abs]}_{*} \equiv \sum_{n=1}^{+\infty}  \frac{  1}{\lambda_n^{[Abs]}}
 = \frac{L^2} {\pi^2}\sum_{n=1}^{+\infty}  \frac{  1}{ \left( n - \frac{1}{2} \right)^2}=  \frac{L^2}{2}
\label{tauAbsEndSpectralDiffuni}
\end{eqnarray}
is in agreement with the MFTP of Eq. \ref{mftpAbsuni}.


\section{ Conclusions }

\label{sec_conclusions}

In this paper, we have first described the general properties of the Kemeny time $\tau_*$ needed to converge towards the steady state $P_*(x)$ for continuous-time Markov processes. We have then given many illustrative examples where the Kemeny time can be explicitly computed as a function of the system-size, either via its real-space definition where one needs to 
average the Mean-First-Passage-Time $\tau(x,x_0) $ over the final configuration $x$ drawn with the steady state $P_*(x)$ (which turns out to be independent of the initial configuration $x_0$), 
or via its spectral definition where one needs to know all the eigenvalues $\lambda_n \ne 0 $ of the opposite generator. We have considered both reversible processes satisfying detailed-balance, where the eigenvalues are real, and irreversible processes characterized by non-vanishing steady currents, where the eigenvalues can be complex. We have also emphasized the specific properties of the Kemeny times for Markov processes with stochastic resetting at constant rate.
Finally, for Markov processes with killing or absorption that converge towards a dead configuration, 
we have analyzed the Kemeny time of the process conditioned to survive forever,
while we have also discussed what happens to the analogs of the real-space Kemeny time
and of the spectral Kemeny time for the initial unconditioned process.

In the illustrative examples of these various aspects, 
we have chosen to focus only on the simplest continuous-time Markov processes 
involving jumps and/or diffusion
in the one-dimensional geometry, either on intervals with reflecting boundary conditions,
or on rings with periodic boundary conditions. But it is of course very interesting to consider
other geometries for the space of configurations:  

(i) In the field of Monte-Carlo sampling, in order to accelerate the convergence towards the steady state with respect to the detailed-balance dynamics,  the idea of 'lifting' consists in duplicating the configuration space into two copies with stochastic switching between the two copies and in imposing directed flows in each copy:
the simplest geometry is thus a ladder, 
and the real-space Kemeny time has been explicitly computed, 
either for continuous-time Markov processes \cite{c_SkewDB} 
or for discrete-time Markov processes \cite{c_GW}. Note that this type of intermittent Markov processes
have been also much studied recently with many other motivations under the name "run-and-tumble" processes.

(ii) In dimensions $d>1$, whenever the eigenvalues in dimension $d$ can be computed from some underlying one-dimensional eigenvalues, they can be used to evaluate the spectral Kemeny time  \cite{c_GW}.

As a final remark, let us mention that some results of the present paper
can be used to obtain the Kemeny times for
boundary-driven one-dimensional non-equilibrium Markov dynamics \cite{c_susyboundarydriven}.


\section*{ Acknowledgements }

C. Monthus thanks W. Krauth and G. Robichon for the joint recent work \cite{c_GW}, 
where many discussions and examples concerning the Kemeny time of discrete-time Markov chains can be found.


\appendix


\section{ Relations with the spectral zeta-function and the spectral partition function }

\label{app_zeta}

In this Appendix, we describe how the spectral Kemeny time $\tau^{Spectral}_*$ 
of Eq. \ref{tauspectral}
is related to the spectral zeta-function and to the spectral partition function.


\subsection{ Spectral zeta-function $\zeta^{Spectral}(s)$ 
and spectral partition function ${\cal Z}^{Spectral}(t) $ }

The spectral Kemeny time $\tau^{Spectral}_*$ of Eq. \ref{tauspectral} 
corresponds to the special case $s=1$ 
 \begin{eqnarray}
\tau^{Spectral}_* \equiv \sum_{n>0} \frac{1}{\lambda_n}  = \zeta^{Spectral}(s=1)  
\label{spectralzeta1}
\end{eqnarray}
of the spectral zeta-function  
 \begin{eqnarray}
\zeta^{Spectral}(s) \equiv   \sum_{n > 0} \frac{1}{\lambda_n^s} 
\label{spectralzeta}
\end{eqnarray}
defined for any complex $s$ that makes sense,
 as in the Riemann zeta-function corresponding to the sum of the integers 
 $\lambda_n=n$ over $n=1,2,..,+\infty$.
Such spectral zeta-functions have been much studied, in particular for Laplacians on compact manifolds and for Schr\"odinger operators with an infinite discrete spectrum $N=+\infty$
(see the reviews \cite{Voros1992,Zeta2017,Zeta2019} and references therein)
together with the spectral partition function
 \begin{eqnarray}
{\cal Z}^{Spectral}(t) \equiv   \sum_{n >0} e^{- t \lambda_n} 
\label{TraceceHeat}
\end{eqnarray}
that is directly related to the spectral zeta-function of Eq. \ref{spectralzeta}
via the Mellin transform
 \begin{eqnarray}
\zeta^{Spectral}(s)  = \frac{1}{\Gamma(s) } \int_0^{+\infty} dt t^{s-1} {\cal Z}^{Spectral}(t) 
\label{spectralzetaZ}
\end{eqnarray}
In particular, the spectral Kemeny time of Eq. \ref{spectralzeta1} corresponds to the 
integral of the spectral partition function over time
 \begin{eqnarray}
\tau^{Spectral}_* \equiv \sum_{n>0} \frac{1}{\lambda_n}  = \int_0^{+\infty} dt  {\cal Z}^{Spectral}(t) 
\label{tauspectralpartition}
\end{eqnarray}


\subsection{ Spectral zeta-function $\zeta^{Spectral}(s)$ for integer $s$ in terms of Mean-First-Passage-Times}

It is interesting to consider the spectral zeta function $\zeta^{Spectral}(s)$ of Eq. \ref{spectralzeta}
when $s$ is an integer $s=2,3,...$ 
higher than $s=1$ corresponding to the Kemeny time $\tau_*=\zeta^{Spectral}(s=1)$
to better understand their physical meanings in terms of Mean-First-Passage-Times.
For $s=2$, Eq. \ref{spectralzeta2} corresponds to the trace of the square $G^2$ 
of the Green matrix $G$
and can be evaluated using Eq. \ref{greenandfirst} 
and the Kemeny time $\tau_*={\rm Trace} (G)$ 
\begin{eqnarray}
\zeta^{Spectral}(s=2) && =   \sum_{n>0} \frac{1}{\lambda_n^2} =   {\rm Trace}(G^2) 
= \sum_{x=1}^N \sum_{y=1}^N G(x,y) G(y,x) 
= \sum_{x=1}^N \sum_{y=1}^N \left[G(x,x)  - P_{*}(x) \tau(x,y)   \right] \left[G(y,y)  - P_{*}(y) \tau(y,x)   \right] 
\nonumber \\
&& = \sum_{x=1}^N \sum_{y=1}^N \left[G(x,x) G(y,y) - P_{*}(x) \tau(x,y)   G(y,y)  
- P_{*}(y) \tau(y,x) G(x,x) +  P_{*}(x) \tau(x,y)  P_{*}(y) \tau(y,x) \right] 
\nonumber \\
&& = \sum_{x=1}^N \sum_{y=1}^N  P_{*}(x) P_{*}(y) \tau(x,y)   \tau(y,x) - \tau_*^2
\label{spectralzeta2}
\end{eqnarray}
So $\zeta^{Spectral}(s=2) $ contains the information on 
the average of the product of the two MFPT $ \tau(x,y) $ and $  \tau(y,x) $
when both configurations $x$ and $y$ are drawn with the steady state distribution $P_*(.)$.

This computation can be generalized for any higher integer $s=3,4,..$ 
\begin{eqnarray}
{\rm Integer \ s} : \ \ \zeta^{Spectral}(s) && =   \sum_{n>0} \frac{1}{\lambda_n^s} =   {\rm Trace}(G^s) 
= \sum_{x_1} \sum_{x_2} ... \sum_{x_s} G(x_s,x_{s-1}) ...  G(x_2,x_1)G(x_1,x_s)
\nonumber \\
&& = (-1)^s \left[ \sum_{x_1} \sum_{x_2} ... \sum_{x_s}  \left[ \prod_{k=1}^s P_{*}(x_k) \right]  \tau(x_s,x_{s-1}) ... 
 \tau(x_2,x_1)  \tau(x_1,x_s)- \tau_*^s \right]
\label{spectralzetaInteger}
\end{eqnarray}
So $\zeta^{Spectral}(s) $ for integer $s$
contains the information on the average of the product of the $s$ MFPT 
along a cycle of $s$ configurations $(x_1,..,x_s)$
when these $s$ configurations $(x_1,..,x_s)$ are all drawn with the steady state distribution $P_*(.)$.



\subsection{ Spectral partition function ${\cal Z}(t) $ for the convergence of the propagator towards the steady state}

It is also interesting to mention the physical interpretation of the spectral partition function of Eq. \ref{TraceceHeat} via the computation of the following observable
measuring the convergence of the propagator $P_.(. \vert .)$ towards the steady state $P_*(.)$ for two configurations $(x,y)$
 \begin{eqnarray}
\sum_{x}   \sum_{y} P_*(x) P_*(y) \bigg[ \frac{P_{\frac{t}{2}}(x \vert y) } {P_*(x) } -1 \bigg] 
\bigg[\frac{P_{\frac{t}{2}}(y \vert x)  }{ P_*(y) }-1\bigg] 
&& =
\sum_{x}   \sum_{y}  \bigg[ P_{\frac{t}{2}}(x \vert y) - P_*(x) \bigg] 
\bigg[P_{\frac{t}{2}}(y \vert x)  - P_*(y) \bigg]
\nonumber \\
&& = \sum_{x}   \sum_{y} \bigg[ P_{\frac{t}{2}}(x \vert y)P_{\frac{t}{2}}(y \vert x) 
- P_*(x) P_{\frac{t}{2}}(y \vert x)
-  P_{\frac{t}{2}}(x \vert y) P_*(y)
+ P_*(x) P_*(y) \bigg]
\nonumber \\
&& = \sum_{x}    P_{t}(x \vert x) - 1
= {\rm Trace} (e^{w  t} ) -1
=  \sum_{n > 0} e^{-t \lambda_n} = {\cal Z}(t)
 \label{spectralZt}
\end{eqnarray}
As a consequence, the Kemeny time $\tau_*=\zeta(s=1)$ of Eq. \ref{spectralzetaZ}
also corresponds to the integration over time $t \in [0,+\infty[$ of Eq. \ref{spectralZt}
 \begin{eqnarray}
\tau_*  =  \int_0^{+\infty} dt  {\cal Z}(t) 
&& = \int_0^{+\infty} dt \sum_{x}   \sum_{y} P_*(x) P_*(y) \bigg[ \frac{P_{\frac{t}{2}}(x \vert y) } {P_*(x) } -1 \bigg] 
\bigg[\frac{P_{\frac{t}{2}}(y \vert x)  }{ P_*(y) }-1\bigg]
\nonumber \\
&& = \int_0^{+\infty} dt \sum_{x}   \sum_{y}  \bigg[ P_{\frac{t}{2}}(x \vert y) - P_*(x) \bigg] 
\bigg[P_{\frac{t}{2}}(y \vert x)  - P_*(y) \bigg]
\label{kemenyspectralzetaZ}
\end{eqnarray}


\section{ Specific spectral properties of reversible Markov chains satisfying detailed-balance }

\label{app_DB}

In this Appendix, we focus on reversible Markov chains satisfying detailed-balance
in order to recall their very specific spectral properties.

\subsection{ Markov matrix $w$ satisfying detailed-balance in terms of the potential $U(x)$ and  $D(x,y)=\sqrt{w(x,y) w(y,x) }$}

The detailed-balance condition means that for any pair of configurations $(x,y)$,
the steady flow ${\cal F}(x,y) \equiv w(x,y) P_*(y) $ from $y$ to $x$
is equal to the steady flow ${\cal F}(y,x) \equiv w(y,x) P_*(x) $ from $x$ to $y$
 \begin{eqnarray}
{\cal F}(x,y) \equiv w(x,y) P_*(y) = w(y,x) P_*(x) \equiv {\cal F}(y,x)
\label{DBdef}
\end{eqnarray}
so that the steady current corresponding to the difference of the two flows vanishes on all links.
It is convenient to write the steady state as the Boltzmann equilibrium in the potential $U(x)$
\begin{eqnarray}
P_*(x) = \frac{ e^{- U(x)} }{ Z}
\label{eq}
\end{eqnarray}
with the normalization given by the partition function
\begin{eqnarray}
Z =  \sum_x e^{- U(x)}
\label{zeq}
\end{eqnarray}
On a given link $(x,y)$, since the ratio of the two transition rates is fixed by the
detailed-balance condition of Eq. \ref{DBdef}
 \begin{eqnarray}
\frac{w(x,y)}{ w(y,x)}  = \frac{P_*(x)}{P_*(y) } = e^{U(y)- U(x)}
\label{DBratio}
\end{eqnarray}
it is useful to introduce the symmetric positive function 
 \begin{eqnarray}
 D(x,y) =\sqrt{w(x,y) w(y,x) } = D(y,x)
\label{DBdiff}
\end{eqnarray}
that can be considered as the diffusion coefficient associated to the link $(x,y)$,
in order to rewrite the transition rates as
 \begin{eqnarray}
w(x,y)  =  D(x,y) \sqrt{\frac{P_*(x)}{P_*(y) } } = D(x,y) e^{- \frac{U(x)-U(y)}{2}} 
\label{DBpara}
\end{eqnarray}
The two equal flows of Eq. \ref{DBdef} on the link $(x,y)$ correspond to the symmetric function
 \begin{eqnarray}
{\cal F}(x,y) = {\cal F}(y,x) = w(x,y) P_*(y) = D(x,y) \sqrt{P_*(x) P_*(y)  } = D(x,y) \frac{ e^{- \frac{U(x)+U(y)}{2}} }{Z}
\label{DBflow}
\end{eqnarray}
The diagonal element of Eq. \ref{wdiag} reads
\begin{eqnarray}
w(x,x) \equiv  - \sum_{y \ne x} w(y,x) = - \sum_{y \ne x} D(x,y) e^{- \frac{U(y)-U(x)}{2}} 
\label{wdiagDB}
\end{eqnarray}

The parametrization of the Markov matrix elements of 
Eqs \ref{DBpara} and \ref{wdiagDB} allows to rewrite the Markov matrix 
using bra and ket notations as a sum over links
\begin{eqnarray}
w   = - \sum_{y } \sum_{x > y }  D(x,y) e^{ - \frac{ U(x)  + U(y)}{2} } 
\left[    \vert x \rangle -     \vert y \rangle   \right] 
 \left[     e^{  U(x) } \langle x \vert -    e^{ U(y)  }  \langle y \vert  \right] 
\label{wproj}
\end{eqnarray}
where it is now obvious on each link contribution that the left and right eigenvectors of Eq. \ref{mastereigen0}
are given by Eq. \ref{markovleft} and by Eq. \ref{markovright} with Eq. \ref{eq}.

The form of Eq. \ref{wproj} that involves the the potential $U(.)$ and the function $D(.,.)$ 
computed from the initial Markov matrix elements via Eqs \ref{DBratio} \ref{DBdiff},
can be also rewritten in terms of the steady state of Eq. \ref{eq}
and the link steady flows of Eq. \ref{DBflow} as
\begin{eqnarray}
w   = - \sum_{y } \sum_{x > y }  {\cal F}(x,y)
\left[    \vert x \rangle -     \vert y \rangle   \right] 
 \left[     \frac{ \langle x \vert }{P_*(x) } -    \frac{  \langle y \vert }{P_*(y)} \right] 
\label{wprojsteady}
\end{eqnarray}


\subsection{ Similarity transformation towards a quantum Hermitian Hamiltonian $H$ 
}
 
The change of variables 
involving the steady state $P_*(.)$ of Eq. \ref{eq} 
or the potential $U(.)$ 
\begin{eqnarray}
P_t(x \vert x_0)  = \sqrt{   \frac{ P_*(x)}{P_*(x_0) } }  \psi_t(x \vert x_0) = 
 e^{  - \frac{ U(x)}{2} } \psi_t(x \vert x_0) e^{   \frac{ U(x_0)}{2} }
\label{DBppsi}
\end{eqnarray}
transforms the forward dynamics of Eq. \ref{forward} for the propagator $ P_t(x,x_0) $
 into the following Euclidean Schr\"odinger equation for the quantum propagator $\psi_t(x \vert x_0)$
\begin{eqnarray}
-  \partial_t \psi_t(x \vert x_0)  = \sum_y H(x,y)  \psi_t(y \vert x_0)
\label{schropsiDB}
\end{eqnarray}
where the matrix elements of the quantum Hermitian Hamiltonian $H$ 
are obtained from the Markov matrix $w$ via the similarity transformation
\begin{eqnarray}
H(x,y) =   -  e^{   \frac{ U(x)}{2} } w(x,y)  e^{ -  \frac{ U(y)}{2} }
\label{Hwsimilaritysimi}
\end{eqnarray}
i.e. at the matrix level using bra and ket notations, 
Eq. \ref{wproj} translates 
\begin{eqnarray}
H  && =  \sum_{y } \sum_{x > y }  D(x,y) e^{ - \frac{ U(x)  + U(y)}{2} } 
\left[  e^{   \frac{ U(x)}{2} }   \vert x \rangle -   e^{   \frac{ U(y)}{2} }   \vert y \rangle   \right] 
 \left[     e^{   \frac{ U(x)}{2} }  \langle x \vert -    e^{   \frac{ U(y)}{2} }   \langle y \vert  \right] 
 \nonumber \\
 && \equiv \sum_{y } \sum_{x > y } \vert Q_{x,y} \rangle \langle Q_{x,y} \vert
\label{Hproj}
\end{eqnarray}
into a sum over links of the not-normalized projectors $ \left( \vert Q_{x,y} \rangle \langle Q_{x,y} \vert \right)$,
where the elementary ket $\vert Q_{x,y} \rangle $ localized on the two sites $(x,y)$ 
can be also rewritten in terms of the transition rates of Eq. \ref{DBpara}
or in terms of steady state of Eq. \ref{eq}
and the link steady flow of Eq. \ref{DBflow}
\begin{eqnarray}
\vert Q_{x,y} \rangle  && =  \sqrt{ D(x,y) e^{ - \frac{ U(x)  + U(y)}{2} } }
\left[  e^{   \frac{ U(x)}{2} }   \vert x \rangle -   e^{   \frac{ U(y)}{2} }   \vert y \rangle   \right] 
 \nonumber \\
 &&     =  \sqrt{ w(y,x) }  \vert x \rangle  - \sqrt{ w(x,y)  }     \vert y \rangle  
  \nonumber \\
 && =  \sqrt{ {\cal F}(x,y) }
\left[  \frac{  \vert x \rangle }{\sqrt{P_*(x)}} -  \frac{   \vert y \rangle }{\sqrt{P_*(y)}}  \right] 
\label{ketQxy}
\end{eqnarray}
with the corresponding different expressions for its norm 
\begin{eqnarray}
\langle Q_{x,y} \vert Q_{x,y} \rangle   
&&= D(x,y) \left[ e^{  \frac{ U(x)  - U(y)}{2} }   + e^{  \frac{ U(y)  - U(x)}{2} } \right]
\nonumber \\
&& =  w(y,x) + w(x,y)  
  \nonumber \\
 && =   {\cal F}(x,y) 
\left[  \frac{  1 }{P_*(x)} +  \frac{   1 }{P_*(y)}  \right]   
\label{ketQxynorm}
\end{eqnarray}


\subsection{ Consequences for the spectral properties of the Markov matrix satisfying detailed-balance}

The similarity transformation of Eq. \ref{Hwsimilaritysimi} yields that the $N$ eigenvalues $\lambda_n$ of $[-w]$
are the eigenvalues of the hermitian Hamiltonian $H$.
So the $(N-1)$ non-vanishing eigenvalues $\lambda_n$ for $n \in \{1,..,N-1\}$
are  real and positive
\begin{eqnarray}
\lambda_n \in ]0,+\infty[ \ \ \ { \rm for } \ \ n \in \{1,..,N-1\}
\label{spectreHreal}
\end{eqnarray}
The corresponding orthonormalized basis of eigenvectors $ \vert \phi_n \rangle $ 
\begin{eqnarray}
H  \vert  \phi_n \rangle && = \lambda_n \vert \phi_n>
 \\ \nonumber 
\langle \phi_n \vert \phi_m \rangle && \equiv \delta_{n,m}
\label{spectreH}
\end{eqnarray}
appears in the spectral decomposition of the quantum propagator 
\begin{eqnarray}
 \psi_t(x \vert x_0) \equiv \langle x \vert e^{-t H} \vert x_0 \rangle 
 = \sum_{n=0}^{N-1} e^{-t \lambda_n } \langle x \vert  \phi_n \rangle \langle  \phi_n \vert x_0 \rangle
\label{psipropagator}
\end{eqnarray}
that translated via Eq. \ref{DBppsi} into the following spectral decomposition for the propagator
\begin{eqnarray}
P_t(x \vert x_0)   = 
\sqrt{   \frac{ P_*(x)}{P_*(x_0) } } \psi_t(x \vert x_0) 
 =  \sum_{n=0}^{N-1} e^{-t \lambda_n } \left[ \sqrt{ P_*(x)}\langle x \vert  \phi_n \rangle \right]
 \left[ \frac{ \langle  \phi_n \vert x_0 \rangle }{ \sqrt{ P_*(x_0)}}  \right]
\label{DBppsispectral}
\end{eqnarray} 
The comparison with Eq. \ref{propagatorspectral}
allows to compute the right and left eigenvectors of $[-w]$ in terms of the eigenvectors $\phi_n$ of the quantum Hamiltonian $H$
\begin{eqnarray}
\langle x \vert r_n \rangle && =\sqrt{ P_*(x)} \langle x \vert  \phi_n \rangle=   \frac{e^{  - \frac{ U(x)}{2} }}{\sqrt{Z} }\langle x \vert  \phi_n \rangle 
     \nonumber \\
\langle l_n  \vert x_0 \rangle && = \frac{ \langle  \phi_n \vert x_0 \rangle }{ \sqrt{ P_*(x_0)}} =  \langle  \phi_n \vert x_0 \rangle \frac{ e^{   \frac{ U(x_0)}{2} }}{\sqrt{Z} }
\label{eigenLR}
\end{eqnarray}
In particular for $n=0$, the quantum ground-state associated to the zero eigenvalue $\lambda=0$ 
is the square-root of the steady state $P_*(x)$
\begin{eqnarray}
  \phi_0 (x) = \sqrt{P_*(x) } = \frac{ e^{- \frac{U(x)}{2}  }}{\sqrt Z}
\label{psi0}
\end{eqnarray}


\subsection{ Consequence of detailed-balance for the spectral partition function ${\cal Z}(t)  $ and for the Kemeny time $\tau_*$}

At the level of finite-time propagators, the detailed-balance condition of Eq. \ref{DBdef}
translates into the property
 \begin{eqnarray}
P_t(x \vert x_0) P_*(x_0) = P_t(x_0 \vert x) P_*(x)
\label{DBpropagator}
\end{eqnarray}
As a consequence, the observable of Eq. \ref{spectralZt} corresponding to the 
spectral partition function can be rewritten as
 \begin{eqnarray}
{\cal Z}(t) && = \sum_{x}   \sum_{x_0} P_*(x) P_*(x_0) \bigg[ \frac{P_{\frac{t}{2}}(x \vert x_0) } {P_*(x) } -1 \bigg] 
\bigg[\frac{P_{\frac{t}{2}}(x_0 \vert x)  }{ P_*(x_0) }-1\bigg] 
\nonumber \\
&&  =\sum_{x}   \sum_{x_0} P_*(x) P_*(x_0) \bigg[ \frac{P_{\frac{t}{2}}(x \vert x_0) } {P_*(x) } -1 \bigg]^2
\nonumber \\
&&  =   \sum_{x_0}  P_*(x_0) \frac{ \bigg[ P_{\frac{t}{2}}(x \vert x_0) -P_*(x)   \bigg]^2 }{P_*(x) }
\equiv \sum_{x_0}  P_*(x_0) \chi^2_{x_0} (t)
 \label{spectralZtDB}
\end{eqnarray}
i.e. it corresponds to the average over the initial position $x_0$ drawn with the steady state $P_*(x_0)$
of the chi-squared function
 \begin{eqnarray}
\chi^2_{x_0}(t) \equiv \frac{ \bigg[ P_{\frac{t}{2}}(x \vert x_0) -P_*(x)   \bigg]^2 }{P_*(x) }
\label{chi2}
\end{eqnarray}
that measures the convergence of the propagator $P_{\frac{t}{2}}(x \vert x_0) $ 
towards the steady state $P_*(x) $ as a function of the time $\frac{t}{2}$ for a given initial configuration $x_0$.
Then the Kemeny time $\tau_*=\zeta(s=1)$ of Eq. \ref{spectralzetaZ}
also corresponds to the integration over time $t \in [0,+\infty[$ of Eq. \ref{spectralZtDB}
 \begin{eqnarray}
\tau_*  =  \int_0^{+\infty} dt  {\cal Z}(t) 
&& = \int_0^{+\infty} dt \sum_{x}   \sum_{x_0} P_*(x) P_*(x_0) \bigg[ \frac{P_{\frac{t}{2}}(x \vert x_0) } {P_*(x) } -1 \bigg]^2
\nonumber \\
&& = \int_0^{+\infty} dt \sum_{x_0}  P_*(x_0) \frac{ \bigg[ P_{\frac{t}{2}}(x \vert x_0) -P_*(x)   \bigg]^2 }{P_*(x) }
\label{kemenyspectralZtDB}
\end{eqnarray}


  \section{ Statistical properties of time-additives observables during $[0,T]$ in terms of the Green function}
  
  \label{app_additive}
  
  In this Appendix, we recall how the Green function described in section \ref{sec_green}
  allows to analyze the statistical properties of 
  all the time-additives observables of the Markov process.
  
  \subsection{ Time-additive observables of trajectories that can be reconstructed from the empirical density} 
  
For each Markov trajectory $x(0 \leq t \leq T) $, the empirical density $\rho_{[0,T]}(x,x_0)$ 
represents the fraction of the time spent at configuration $x$
during the time-window $[0,T]$ when starting at configuration $x_0$
\begin{eqnarray}
\rho_{[0,T]}(x,x_0) \equiv \frac{1}{T} \int_0^{T} dt  \delta_{x(t),x}  
\label{defrho}
\end{eqnarray} 
It plays an essential role since it
 allows to reconstruct any time-additive observable ${\cal O}_{[0,T]}(x_0) $ 
 parametrized by some function $O(x)$ 
\begin{eqnarray}
{\cal O}_{[0,T]}(x_0) \equiv \frac{1}{T} \int_0^{T} dt  O \left(  x(t)  \right) = \sum_{x=1}^N O(x) \rho_{[0,T]}(x,x_0)
\label{defadditive}
\end{eqnarray} 

 \subsection{ Role of the Green function for the convergence of the averaged values of time-additive observables } 

The averaged value of the empirical density of Eq. \ref{defrho}
over the trajectories $x(0 \leq t \leq T)  $ 
corresponds to the averaged time $t_{[0,T]}(x,x_0) $ of Eq. \ref{avlocaltimeExtensivefiniteAbs}
divided by $T$
\begin{eqnarray}
\langle \rho_{[0,T]}(x,x_0) \rangle_{Traj} \equiv \frac{t_{[0,T]}(x,x_0)}{T} 
 \opsimeq_{T \to +\infty}   P_*(x) +\frac{G(x,x_0)}{T} + o\left( \frac{1}{T}\right)
\label{rhoav}
\end{eqnarray} 
This allows to obtain the averaged values of the time-additive observables of Eq. \ref{defadditive}
\begin{eqnarray}
\langle{\cal O}_{[0,T]}(x_0) \rangle_{Traj}  = \sum_{x=1}^N O(x) \langle \rho_{[0,T]}(x,x_0) \rangle_{Traj}
 \opsimeq_{T \to +\infty} \sum_{x=1}^N O(x) P_*(x) +\frac{1}{T} \left[ \sum_{x=1}^N O(x) G(x,x_0)\right] + o\left( \frac{1}{T}\right)
\label{avadditive}
\end{eqnarray} 
So the Green function $G(x,x_0)$ governs the convergence of order $1/T$
towards the value $O_*$
computed from the steady state $P_*(x)$
\begin{eqnarray}
O_* \equiv \sum_{x=1}^N O(x) P_*(x) 
\label{steadyadditive}
\end{eqnarray}

 \subsection{ Role of the Green function for the large-time variances of time-additive observables }

The variance of the time-additive observable ${\cal O}_{[0,T]}(x_0)$ of Eq. \ref{defadditive} can be also computed 
at the leading order $1/T$ in terms of the Green function $G$ (see \cite{c_SkewDB} for more detailed calculations):
  \begin{eqnarray}
 \langle  {\cal O}^2_{[0,T]}(x_0) \rangle_{Traj} -  \langle{\cal O}_{[0,T]}(x_0) \rangle_{Traj}^2 
\opsimeq_{T \to +\infty} \frac{2}{T}
 \sum_x  \sum_y O(x) G(x,y) O(y) P_*(y)
\label{varcv}
\end{eqnarray}
For the special case of the function $O(x)=\delta_{x,a}$ that corresponds in Eq. \ref{defadditive} to 
the empirical density $\rho_{[0,T]}(a,x_0) $ at configuration $a$, 
Eq. \ref{varcv} yields 
 that the variance of the empirical density $\rho_{[0,T]}(a,x_0) $ at configuration $a$
 involves the Green function $G(a,a)$ at coinciding points
  \begin{eqnarray}
   \langle  \rho_{[0,T]}^2(a,x_0)  \rangle_{Traj} 
   -  \langle \rho_{[0,T]}(a,x_0) \rangle_{Traj}^2 
 \opsimeq_{T \to +\infty} \frac{2  G(a,a)  P_*(a)}{T}  
\label{rhovarcv}
\end{eqnarray}
As a consequence, the Kemeny time $\tau_*$
corresponding to the trace of the Green function of Eq. \ref{tautraceG}
 can be recovered if one computes
  \begin{eqnarray}
 \frac{T}{2 } \sum_{x=1}^N \frac{ \langle  \rho_{[0,T]}^2(x,x_0)  \rangle_{Traj} 
   -  \langle \rho_{[0,T]}(x,x_0) \rangle_{Traj}^2 }{P_*(x) } 
 \opsimeq_{T \to +\infty} \sum_{x=1}^N G(x,x)  = \tau_*
\label{Kemenyrhoempirical}
\end{eqnarray}

\subsection{Discussion }

The Green function also allows to analyze the averaged value and the variance of
more general time-additive observables than Eq. \ref{defadditive}
that involve not only the time spent in the various configurations
but also all the jumps between two different configurations that occur during the time-window $[0,T]$,
as described in \cite{c_ruelle}.
Beyond the averaged values of Eq. \ref{steadyadditive} and the variance of Eq. \ref{varcv},
the Green function allows to analyze also the higher cumulants, for instance of order three or four.
However if one wishes to study the whole series of higher cumulants,
it is better to use the theory of large deviations 
 (see the reviews \cite{oono,ellis,review_touchette} and references therein)
that has become the unifying language in the field of non-equilibrium processes
(see the reviews with different scopes \cite{derrida-lecture,harris_Schu,searles,harris,mft,sollich_review,lazarescu_companion,lazarescu_generic,jack_review}, 
the PhD Theses \cite{fortelle_thesis,vivien_thesis,chetrite_thesis,wynants_thesis,chabane_thesis,duBuisson_thesis} 
 and the Habilitation Thesis \cite{chetrite_HDR}).
 In particular, many works have been devoted recently
to the large deviations properties of many interesting time-additive observables for various Markov processes 
 \cite{peliti,derrida-lecture,sollich_review,lazarescu_companion,lazarescu_generic,jack_review,vivien_thesis,lecomte_chaotic,lecomte_thermo,lecomte_formalism,lecomte_glass,kristina1,kristina2,jack_ensemble,simon1,simon2,tailleur,simon3,Gunter1,Gunter2,Gunter3,Gunter4,chetrite_canonical,chetrite_conditioned,chetrite_optimal,chetrite_HDR,touchette_circle,touchette_langevin,touchette_occ,touchette_occupation,garrahan_lecture,Vivo,c_ring,c_detailed,chemical,derrida-conditioned,derrida-ring,bertin-conditioned,touchette-reflected,touchette-reflectedbis,c_lyapunov,previousquantum2.5doob,quantum2.5doob,quantum2.5dooblong,c_ruelle,lapolla,c_east,chabane,us_gyrator,duBuissonGyrator}.


\section{ Parametrization of one-dimensional diffusion processes with two functions  }

\label{app_SDE}

The Fokker-Planck Eq. \ref{fokkerplanck} for one-dimensional diffusion processes
involves two functions: the force $F(x) $ and the diffusion coefficient $D(x)$.
In this Appendix, we describe various other choices for the two functions
that can be chosen to parametrize diffusion processes.

\subsection{Replacing the Fokker-Planck force $F(x)$ 
by the Stratonovich force $f^S(x) $ or the Ito force $f^I(x) $ }

When the diffusion coefficient $D(x)$ depends explicitly on the position $x$,
the corresponding Langevin Stochastic Differential Equation involving a Brownian motion $B(t)$
will depend on the Stratonovich or the Ito interpretations
\begin{eqnarray}
dx(t) && =  f^S( x (t) ) \ dt + \sqrt{ 2 D ( x (t) ) } \ dB(t)
\ \ \ \ \ \ \ \ \ \ \ \ \ \ [{\rm Stratonovich \ Interpretation}]
\nonumber \\
dx(t) &&=  f^I( x (t) ) \ dt + \sqrt{ 2 D ( x (t) ) }  \ dB(t)
\ \ \ \ \ \ \ \ \ \ \ \ \ \ [{\rm Ito \ Interpretation}]
 \label{langevin}
\end{eqnarray}
where the Stratonovich force $f^S(x) $ and the Ito force $f^I(x) $ 
have be to computed from the Fokker-Planck force $F(x)$ and the diffusion coefficient $D(x)$ via
\begin{eqnarray}
   f^S(x) && = F(x) +  \frac{D' (x)}{2}  
\nonumber \\
 f^I(x) && = F(x) + D' (x)
\label{fitostrato}
\end{eqnarray}
The mathematical literature is always written in terms of the Ito force $  f^I(x)$,
so that the adjoint operator of Eq. \ref{adjointtranslation} reads
\begin{eqnarray}
 {\cal L}^{\dagger}_{x_0}  =  \left[  F(x_0)  +   \partial_{x_0} D (x_0) \right]    \partial_{x_0}  
 =  f^I(x_0)  \partial_{x_0}  + D (x_0) \partial^2_{x_0} 
\label{adjointtranslation}
\end{eqnarray}

Note that when the Fokker-Planck force vanishes $F(x)=0$, 
the operator ${\cal L} $ and its adjoint ${\cal L}^{\dagger} $ coincide
\begin{eqnarray}
{\rm when } \ F(x) =0  \ \ : {\cal L}_{x}  =     \partial_{x} D (x)  \partial_{x}  = {\cal L}^{\dagger}_{x}
\label{zeroforcegene}
\end{eqnarray}
and the corresponding steady state is uniform $p_*(x)= {\rm cst} $.
This explains why the Fokker-Planck force $F(x)$ is often more convenient 
from a physical point of view.


\subsection{ Replacing the Fokker-Planck force $F(x)$ by the potential $U(x)$ with derivative $U'(x)  = - \frac{F(x)}{D(x)} $}

As explained in the main text, 
it is convenient to replace the Fokker-Planck force $F(x)$ by the potential $U(x)$ of Eq. \ref{UR} 
both in the reversible and in the irreversible cases:

(i) In the reversible diffusion processes satisfying detailed-balance discussed in section \ref{sec_DiffInterval}, 
it is convenient to further replace the potential $U(x)$
by the normalized steady state $p_*(x)  = \frac{ e^{ -U(x)} }{Z}$ of Eq. \ref{steadyeq}.
Then the diffusion process is characterized by its steady state $p_*(x) $ and its diffusion coefficient $D(x)$,
that are the two functions that we have chosen to write the real-space Kemeny time of Eq. \ref{kemeny1d}.
Note that in the mathematical literature (see \cite{pearson_class} for instance), 
the non-normalized steady state is called "the speed density $m(x)$",
while the other function defining the process is called "the scale density $s(x) \propto \frac{1}{  D(x) m(x)} $".

(ii) In the irreversible diffusion processes on the ring discussed in section \ref{sec_DiffRing},
then the steady state $p_*(x) $ is given by the non-equilibrium expression of Eq. \ref{steadyring},
and it is thus more convenient to keep the potential $U(x)$ and the diffusion coefficient $D(x)$
as the two functions defining the non-equilibrium diffusion process.


\section{ Eigenvalues for the Diffusion $D(x)=1$ with the constant force  $F(x)=\mu $ on the interval $[0,L]$ }

\label{app_EigenForceConstant}

In this Appendix, we describe the eigenvalues for the Diffusion $D(x)=1$ with the constant force  $F(x)=\mu $ on the interval $[0,L]$ for various boundary conditions.


\subsection{ Similarity transformation towards the free quantum Hamiltonian }

For the diffusion coefficient $D(x)=1$
and the constant force $ F(x)=\mu$, the potential $U(x)$ of Eq. \ref{UR} is linear
\begin{eqnarray}
U(x)  = - \mu x
\label{URlin}
\end{eqnarray}
while the quantum potential of Eq. \ref{vfromu} reduces to the constant
\begin{eqnarray}
V(x)  = \frac{ F^2(x) }{4  } + \frac{F'(x)}{2} = \frac{ \mu^2 }{4  } 
\label{vfromulin}
\end{eqnarray}
The eigenvalue Eq. \ref{spectreH} for the quantum Hamiltonian of Eq. \ref{hamiltonien} 
with the potential of Eq. \ref{vfromusaw} reduces to
\begin{eqnarray}
\lambda  \phi(x) = H \phi(x)  =  - \phi'' (x) + \frac{ \mu^2 }{4  }  \phi(x)
\label{hamiltonienlin}
\end{eqnarray}
while the zero-current boundary condition of Eqs \ref{jnboundary} \ref{jnphi} at $x=L$ reads
\begin{eqnarray}
0 =   \frac{\mu}{2} \phi(L) - \phi'(L) 
\label{jnphilin}
\end{eqnarray}
As a consequence, the solution $\phi(x)$ can be written as the linear combination
\begin{eqnarray}
\phi(x)  = \phi(L) \bigg( \cos[k (L-x) ] - \frac{\mu}{2k}  \sin[k (L-x) ]  \bigg)  
\label{phinlin}
\end{eqnarray}
in terms of the wavevector 
\begin{eqnarray}
k \equiv \sqrt{ \lambda - \frac{ \mu^2 }{4  } }
\label{wavevector}
\end{eqnarray}
associated to the eigenvalue
\begin{eqnarray}
\lambda= \frac{\mu^2 }{4  } + k^2
\label{wavevectoreigen}
\end{eqnarray}
Now one needs to impose the boundary condition at $x=0$ to determine the possible wavevectors $k$
and the corresponding eigenvalues $\lambda$.
In the two following subsections, we consider two different boundary conditions.


\subsection{ Eigenvalues for the reflecting boundary condition at $x=0$ }

Let us impose the zero-current boundary condition of Eqs \ref{jnboundary} \ref{jnphi} at $x=0$ 
to the solution of Eq. \ref{phinlin}
\begin{eqnarray}
0  && =     \frac{\mu}{2} \phi(0) - \phi'(0)
= \frac{\mu}{2}  \phi(L) \bigg( \cos(kL) - \frac{\mu}{2k}  \sin(kL)  \bigg)  
-  \phi(L) \bigg( k \sin(kL) + \frac{\mu}{2}  \cos(kL)  \bigg)  
\nonumber \\
&& = - \frac{\phi(L)}{k} \left( \frac{\mu^2}{ 4} + k^2 \right)  \sin(kL)  
\label{j0philin}
\end{eqnarray}
So apart from the vanishing eigenvalue $\lambda=0$ in Eq. \ref{wavevectoreigen} corresponding
to $k=i \frac{\mu}{2}$, the other wavevectors should satisfy
\begin{eqnarray}
 \sin(kL) =0
\label{kexcitedrhozero}
\end{eqnarray}
and thus read 
\begin{eqnarray}
 k_n = n \frac{ \pi }{L} \ \ \ {\rm with } \ \  n=1,2,..,+\infty
\label{knexcitedrhozero}
\end{eqnarray}
with the corresponding excited eigenvalues
\begin{eqnarray}
\lambda_n =  \frac{ \mu^2 }{4  } +  k_n^2 = \frac{ \mu^2 }{4  } + n^2 \frac{ \pi^2 }{L^2} \ \ \ {\rm with } \ \  n=1,2,..,+\infty
\label{eigenexcitedrhozero}
\end{eqnarray}
that are used to compute the spectral Kemeny time given in Eq. \ref{eigenexcitedrhozerok}
of the main text.


\subsection{ Eigenvalues for the absorbing boundary condition at $x=0$ }

Let us impose the absorbing boundary condition at $x=0$ 
to the solution of Eq. \ref{phinlin}
\begin{eqnarray}
0   =    \phi(0) 
=   \phi(L) \bigg( \cos(kL) - \frac{\mu}{2k}  \sin(kL)  \bigg)  
\label{abslin}
\end{eqnarray}
So the wavevectors $k^{[abs]}_n $ are the solutions of
\begin{eqnarray}
 \tan(k^{[Abs]}_n L) = \frac{2 }{\mu} k^{[Abs]}_n
\label{abslinear}
\end{eqnarray}
and will give the absorbing eigenvalues
\begin{eqnarray}
\lambda^{[Abs]}_n =  \frac{ \mu^2 }{4  } +  \left[ k^{[Abs]}_n \right]^2 
\label{abslineareigen}
\end{eqnarray}


\section{ Eigenvalues for the diffusion in the saw-tooth potential on the interval $[0,L]$ 
of subsection \ref{subsec_sawtoothDB}}

\label{app_EigenSawTooth}

In this Appendix, we analyze the eigenvalues for the diffusion in 
the saw-tooth potential $U(x)$ on the interval $[0,L]$ considered in section \ref{subsec_sawtoothDB}
of the main text.

\subsection{ Spectral analysis of the quantum Hamiltonian associated to the saw-tooth potential}

The quantum potential of Eq. \ref{vfromu} associated to the diffusion coefficient $D(x)=1$
and to the force of Eq. \ref{forcesaw}
\begin{eqnarray}
V(x)  = \frac{ F^2(x) }{4  } + \frac{F'(x)}{2} = \frac{ \mu^2 }{4  } + \mu \delta(x-\rho L)
\label{vfromusaw}
\end{eqnarray}
reduces to the constant $\frac{ \mu^2 }{4  } $ and to a delta impurity of strength $\mu$ located at $x=\rho L$.
The eigenvalue Eq. \ref{spectreH} for the quantum Hamiltonian of Eq. \ref{hamiltonien} 
with the potential of Eq. \ref{vfromusaw} reads
\begin{eqnarray}
\lambda  \phi(x) = H \phi(x)  =  - \phi'' (x) +\left[ \frac{ \mu^2 }{4  } + \mu \delta(x-\rho L) \right] \phi(x)
\label{hamiltoniensaw}
\end{eqnarray}
with the zero-current boundary conditions of Eqs \ref{jnboundary} \ref{jnphi} at $x=0$ and $x=L$ 
\begin{eqnarray}
0 && =   - \frac{U'(0)}{2} \phi(0) - \phi'(0) =  - \frac{\mu}{2} \phi(0) - \phi'(0)
\nonumber \\
0 && =  - \frac{U'(L)}{2} \phi(L) - \phi'(L) =   \frac{\mu}{2} \phi(L) - \phi'(L) 
\label{jnphisaw}
\end{eqnarray}
As a consequence, the solution $\phi(x)$ can be written in the two regions as
\begin{eqnarray}
\phi(x) && = \phi(0)  \bigg(  \cos[kx ] - \frac{\mu}{2k}  \sin[kx ]    \bigg)  \ \ {\rm for }  \ \ 0 \leq x \leq \rho L
\nonumber \\
\phi(x) && = \phi(L) \bigg( \cos[k (L-x) ] - \frac{\mu}{2k}  \sin[k (L-x) ]  \bigg)  \ \ {\rm for }  \ \ \rho L \leq x \leq  L
\label{phin}
\end{eqnarray}
in terms of the wavevector $k \equiv \sqrt{ \lambda - \frac{ \mu^2 }{4  } } $ of Eq. \ref{wavevector}
and in terms of the two constants $\phi(0) $ and $\phi(L) $ that can be determined as follows.
The delta function $\delta(x-\rho L) $ in Eq. \ref{hamiltoniensaw}
leads to the following continuity condition for $\phi(x)$ at $x=\rho L \pm \epsilon$
\begin{eqnarray}
0 && =\phi(\rho L-\epsilon) - \phi(\rho L+\epsilon) 
= \phi(0)  \bigg(  \cos[k \rho L ] - \frac{\mu}{2k}  \sin[k \rho L ]    \bigg)
- \phi(L) \bigg( \cos[k (1-\rho )L ] - \frac{\mu}{2k}  \sin[k (1-\rho ) L]  \bigg)
\label{phinconti}
\end{eqnarray}
and to the following discontinuity condition for $\phi'(x)$ at $x=\rho L \pm \epsilon$
\begin{eqnarray}
0  && = \phi'(\rho L-\epsilon) - \phi'(\rho L+\epsilon) + \mu \phi(\rho L ) 
 = \phi(0)  \bigg(  -k \sin[k \rho L ] - \frac{\mu}{2}  \cos[k \rho L ]    \bigg)
- \phi(L) \bigg( k \sin[k (1-\rho) L ] + \frac{\mu}{2}  \cos[k (1-\rho) L]  \bigg)
\nonumber \\
&& \ \ \ \ \ \ \ \ \ \ \ \ \ \ \ \ \ \ \ \ \ \ + \mu  \phi(0)  \bigg(  \cos[k \rho L ] - \frac{\mu}{2k}  \sin[k \rho L ]    \bigg)
\nonumber \\
&& = \phi(0)  \bigg(   \frac{\mu}{2}  \cos[k \rho L ] - [ k +\frac{\mu^2}{2k} ]  \sin[k \rho L ]   \bigg)
- \phi(L) \bigg( \frac{\mu}{2}  \cos[k (1-\rho) L] + k \sin[k (1-\rho) L ]   \bigg)
\label{phindisc}
\end{eqnarray}
Eqs \ref{phinconti}
and \ref{phindisc}
 give two different expressions for the ratio $\frac{\phi(L)}{\phi(0)}$ 
\begin{eqnarray}
\frac{\phi(L)}{\phi(0)}  && 
= \frac{  \bigg(  k \cos[k \rho L ] - \frac{\mu}{2}  \sin[k \rho L ]    \bigg) }
{ \bigg( k \cos[k (1-\rho )L ] - \frac{\mu}{2}  \sin[k (1-\rho ) L]  \bigg) }
\nonumber \\
\frac{\phi(L)}{\phi(0)} && 
=  \frac{  \bigg(   \frac{\mu}{2}  \cos[k \rho L ] - [ k +\frac{\mu^2}{2k} ]   \sin[k \rho L ]   \bigg) }
{ \bigg( \frac{\mu}{2}  \cos[k (1-\rho) L] + k \sin[k (1-\rho) L ]   \bigg) }
\label{ratioapam}
\end{eqnarray}
The consistency between these two expressions
\begin{eqnarray}
0 &&  
=   \bigg( k \cos[k \rho L ] - \frac{\mu}{2}  \sin[k \rho L ]    \bigg) 
\bigg( \frac{\mu}{2}  \cos[k (1-\rho) L] + k \sin[k (1-\rho) L ]   \bigg)
  \nonumber \\
&&  - \bigg(   \frac{\mu}{2}  \cos[k \rho L ] - [ k +\frac{\mu^2}{2k} ]   \sin[k \rho L ]   \bigg) 
\bigg( k \cos[k (1-\rho )L ] - \frac{\mu}{2}  \sin[k (1-\rho ) L]  \bigg)   
  \nonumber \\
&&  
=  \left( \frac{\mu^2}{4} + k^2 \right) 
\left[ \cos[k \rho L ] \sin[k (1-\rho) L ]   
+ \sin[k \rho L ]   \cos[k (1-\rho) L] 
-  \frac{\mu}{k}  \sin[k \rho L ]  \sin[k (1-\rho) L ] \right]
  \label{consistency}
\end{eqnarray}
determines the possible values of the wavevector $k$ of Eq. \ref{wavevector},
i.e. of the eigenvalues $\lambda= k^2 + \frac{ \mu^2 }{4  }$.
So besides the vanishing eigenvalue $\lambda=0$ corresponding to $k = i \frac{\mu}{2}$,
the other possible wavevectors $k$ for the excited states are the solutions of
\begin{eqnarray}
0 && =  \cos[k \rho L ] \sin[k (1-\rho) L ]   
+ \sin[k \rho L ]   \cos[k (1-\rho) L] 
-  \frac{\mu}{k}  \sin[k \rho L ]  \sin[k (1-\rho) L ]
\nonumber \\
&& =   \sin(kL) -  \frac{ \mu }{k} \sin [k \rho L ] \sin [k (1- \rho) L ]  
\label{kexcited}
\end{eqnarray}
Let us first focus on the special case $\rho=\frac{1}{2} $ in the next subsection.


\subsection{ Symmetric case $\rho=\frac{1}{2}$: decomposition of the spectrum into the symmetric and antisymmetric sectors }

For $\rho=\frac{1}{2} $ , Eq. \ref{kexcited} can be factorized into
\begin{eqnarray}
0 =   \sin \left( \frac{k  L}{2} \right) \left[ 2 \cos \left( \frac{k  L}{2} \right) -  \frac{\mu }{k} \sin \left( \frac{k  L}{2} \right) \right]
\label{kexcitedrhodemi}
\end{eqnarray}
This factorization is a consequence of the mirror symmetry $x \to L-x$ 
with respect to the middle-point $x=\frac{L}{2}$
of the Hamiltonian $H$ with the quantum potential of Eq. \ref{vfromusaw},
that allows to decompose the spectrum into its symmetric and antisymmetric sectors as follows.


\subsubsection{ Symmetric sector of the spectrum }

When Eq. \ref{kexcitedrhodemi} is satisfied via the vanishing of the first factor
\begin{eqnarray}
 \sin \left( \frac{k  L}{2} \right) =0
\label{group1eq}
\end{eqnarray}
then the ratio of Eq. \ref{ratioapam} reduces to unity 
\begin{eqnarray}
\frac{\phi(L)}{\phi(0)}  =1
\label{ratioapamgroup1}
\end{eqnarray}
so that the wave-function $\phi(x)$ of Eq. \ref{phin} is symmetric with respect to the middle-point $x=\frac{L}{2}$
\begin{eqnarray}
\phi(x)  = \phi(L-x) 
\label{phinsym}
\end{eqnarray}
In this symmetric sector, the possible non-vanishing wavevectors $ k^{[sym]}_n $ satisfying Eq. \ref{group1eq}
and the corresponding eigenvalues $\lambda^{[sym]}_n  $ read
\begin{eqnarray}
 k^{[sym]}_n && = \frac{ 2 \pi n}{L} \ \ \ \ \ \ \ \ \ \ \ \ \ \ \ \ \ \ \ \ \ \ \ \ \ \ \ \ \ \ \ \ \ \ \ \
 {\rm with } \ \  n=1,2,..,+\infty
\nonumber \\
\lambda^{[sym]}_n && =  \frac{ \mu^2 }{4  } +  [k^{[sym]}_n]^2 
= \frac{ \mu^2 }{4  } + \frac{ 4 \pi^2 n^2}{L^2} \ \ \ \ \ {\rm with } \ \ n=1,2,..,+\infty
\label{group1}
\end{eqnarray}

So one can compute the spectral Kemeny time associated to this symmetric sector
 \begin{eqnarray}
\tau^{Spectral[sym]}_{*} 
\equiv  \sum_{n=1}^{+\infty} \frac{1}{\lambda^{[sym]}_n} 
= \frac{L^2}{4 \pi^2} \sum_{n=1}^{+\infty} \frac{1}{ n^2 + \left( \frac{ \mu L}{ 4 \pi  } \right)^2 }
= \frac{2}{\mu^2} \left[  \frac{ \frac{\mu L}{4 }  }{ \tanh \left( \frac{\mu L}{4 } \right) } -1 \right]
\label{tauspectralsym}
\end{eqnarray}
with its asymptotic linear behavior for large $L$
 \begin{eqnarray}
\tau^{Spectral[sym]}_{*} 
\opsimeq_{L \to + \infty} 
 \frac{2}{\mu^2} \left[ \left( \frac{\mu L}{4 } \right) {\rm sgn} (\mu) \right] = \frac{ L}{2 \vert \mu \vert }
\label{tauspectralsyml}
\end{eqnarray}

Note that the condition of Eq. \ref{group1eq} corresponds to Eq. \ref{kexcitedrhozero}
with the replacement $L \to \frac{L}{2}$:
indeed, the mirror symmetry of Eq. \ref{phinsym} for the eigenvector
means that one could consider only the first half $x \in [0,\frac{L}{2}]$ of the system
with a reflecting boundary condition at $\frac{L}{2} $.
Physically, this means that the symmetric sector of the spectrum describes only the relaxation within each valley, while the relaxation between the two valleys involves the antisymmetric sector
of the spectrum described in the next subsection.


\subsubsection{ Antisymmetric sector of the spectrum }

When Eq. \ref{kexcitedrhodemi} is satisfied via the vanishing of the second factor
\begin{eqnarray}
2 \cos \left( \frac{k  L}{2} \right) -  \frac{\mu }{k} \sin \left( \frac{k  L}{2} \right) =0
\label{group2eq}
\end{eqnarray}
the ratio of the second Eq. \ref{ratioapam} reduces to
\begin{eqnarray}
\frac{\phi(L)}{\phi(0)}  =-1
\label{ratioapamgroup2}
\end{eqnarray}
so that the wave-function $\phi(x)$ of Eq. \ref{phin} is antisymmetric with respect to the middle-point $\frac{L}{2}$
\begin{eqnarray}
\phi(x)  = -\phi(L-x) 
\label{phinantisym}
\end{eqnarray}
and in particular vanishes at the middle-point $\phi(\frac{L}{2})=0$.
In this antisymmetric sector, the possible wavevectors $ k^{[anti]}_m $ satisfying Eq. \ref{group2eq}
are the solutions of
\begin{eqnarray}
\tan \left( \frac{k^{[anti]}_m  L}{2} \right) = \frac{2 }{\mu}  k^{[anti]}_m
\label{group2}
\end{eqnarray}
and will give the eigenvalues
\begin{eqnarray}
\lambda^{[anti]}_m && =  \frac{ \mu^2 }{4  } +  [k^{[anti]}_m]^2 
\label{group2m}
\end{eqnarray}

Note that the condition of Eq. \ref{group2} corresponds to Eq. \ref{abslinear}
with the replacement $L \to \frac{L}{2}$:
indeed, the antisymmetry of Eq. \ref{phinantisym} for the eigenvector
means that one could consider only the first half $x \in [0,\frac{L}{2}]$ of the system
with an absorbing boundary condition at $\frac{L}{2} $.


\subsubsection{ Lowest eigenvalue $\lambda^{[anti]}_{m=1} \equiv \lambda_1 $ for $\mu>0$ where the maximum $U(x=\frac{L}{2})  =  \mu \frac{L}{2} $
separates two valleys}

It is now interesting to make the link with the main text:
for $\mu>0$ and $\rho=\frac{1}{2} $, where the maximum $U(x=\frac{L}{2})  =  \mu \frac{L}{2} $
separates the two valleys with minima at the two boundaries $U(x=0)=0=U(x=L)$, 
the Kemeny time of Eq. \ref{kemeny1dsawdemi} grows exponentially with the system size $L$.
The lowest eigenvalue $ \lambda_1 \equiv \lambda^{[anti]}_{m=1} $ in the antisymmetric sector of Eq. \ref{group2m}
 is then expected to be exponentially small in $L$
 and thus smaller than the constant $\frac{ \mu^2 }{4  } $, 
 so that the corresponding wavevector $ k^{[anti]}_{m=1} $ is purely imaginary
\begin{eqnarray}
k^{[anti]}_{1} && = i \kappa
\nonumber \\
\kappa && \equiv \sqrt{ \frac{ \mu^2 }{4  } - \lambda_1 } = \frac{\mu}{2} - \frac{\lambda_1}{\mu} +O(\lambda_1^2)
\label{group2k1}
\end{eqnarray}
and Eq. \ref{group2} becomes
\begin{eqnarray}
0 && = \tanh \left( \frac{\kappa  L}{2} \right) - \frac{2  \kappa }{\mu}  
\nonumber \\
&&= \tanh \left( \frac{\mu L}{4} - \frac{L}{2\mu}\lambda_1  \right) - 1 + \frac{2   }{\mu^2}  \lambda_1  +O(\lambda_1^2)
\nonumber \\
&& = \left[\tanh \left( \frac{\mu L}{4}   \right) -1 \right]
 +\lambda_1 \frac{2   }{\mu^2} \left[  1  - \frac{L \mu }{4 \cosh^2 \left( \frac{\mu L}{4}   \right) } \right]  +O(\lambda_1^2)
\label{group2h}
\end{eqnarray}
The lowest eigenvalue $\lambda_1$ thus reads at leading order 
\begin{eqnarray}
\lambda_1  
=\frac{\mu^2} {2} \left( \frac{1- \frac{\sinh \left( \frac{\mu L}{4}   \right) }{\cosh \left( \frac{\mu L}{4}   \right) } }
{   1  - \frac{L \mu }{4  \cosh^2 \left( \frac{\mu L}{4}   \right) }  } \right) 
 =\frac{\mu^2} {2} \left( \frac{ e^{- \frac{\mu L}{4} }}
{   \cosh \left( \frac{\mu L}{4}   \right)  - \frac{L \mu }{4  \cosh \left( \frac{\mu L}{4}   \right) }  } \right)
 \opsimeq_{L \to + \infty} \mu^2   e^{- \frac{\mu L}{2} }
\label{group2hgap}
\end{eqnarray}
i.e. the relaxation time $\tau_1$ associated to this lowest eigenvalue $\lambda_1$
\begin{eqnarray}
\tau_1 \equiv \frac{1}{\lambda_1 } \opsimeq_{L \to + \infty} \frac{1}{\mu^2}   e^{ \frac{\mu L}{2} }
\label{group2hgapinverse}
\end{eqnarray}
has exactly the same behavior for large $L$ as the Kemeny time of Eqs \ref{kemeny1dsawdemi}
discussed in the main text.


\subsection{ Lowest eigenvalue $\lambda_1 $ for $\mu>0$ and $\rho \ne (0,1/2,1)$ with the maximum $U(x=\rho L)  $ between two valleys}

Let us now generalize the previous analysis concerning 
the lowest eigenvalue $\lambda_1 $ for $\mu>0$ when $\rho \ne (0,1/2,1)$.
Plugging the analog of Eq. \ref{group2k1}
\begin{eqnarray}
k && = i \kappa
\nonumber \\
\kappa && \equiv \sqrt{ \frac{ \mu^2 }{4  } - \lambda_1 } = \frac{\mu}{2} - \frac{\lambda_1}{\mu} +O(\lambda_1^2)
\label{kappa}
\end{eqnarray}
into Eq. \ref{kexcited} yields
\begin{small}
\begin{eqnarray}
0 && =   \sinh[ \kappa L] -  \frac{ \mu }{ \kappa}  \sinh [\kappa \rho L ]  \sinh [\kappa (1- \rho) L ]  
\label{kexcitedimaginary}
 \\
&& =   \sinh[ \frac{\mu}{2} L] -  \frac{ 1 }{ 2}  \sinh [\frac{\mu}{2} \rho L ]  \sinh [\frac{\mu}{2} (1- \rho) L ]
\nonumber \\
&& - \frac{\lambda_1}{\mu} \left[
 L \cosh[ \frac{\mu}{2} L] +  \frac{ 4  }{ \mu}  \sinh [\frac{\mu}{2} \rho L ]  \sinh [\frac{\mu}{2} (1- \rho) L ]  
 -  2 \rho L  \cosh [\frac{\mu}{2} \rho L ]  \sinh [\frac{\mu}{2} (1- \rho) L ]  
 -  2 (1- \rho) L  \sinh [\frac{\mu}{2} \rho L ]  \cosh [\frac{\mu}{2} (1- \rho) L ]  
 \right]  +O(\lambda_1^2)
\nonumber 
\\
&& = \left(  \cosh[ \frac{\mu}{2} (1-2\rho) L] -  e^{- \frac{\mu}{2} L} \right)
 - \frac{\lambda_1}{\mu} 
 \left[  \left(L + \frac{2}{\mu} \right)\cosh[ \frac{\mu}{2} L] - L \sinh[ \frac{\mu}{2} L]
 + L (1-2 \rho) \sinh[ \frac{\mu}{2} (1-2 \rho) L] -  \frac{2}{\mu}\cosh[ \frac{\mu}{2} (1-2 \rho) L]
  \right]  +O(\lambda_1^2)
 \nonumber 
\end{eqnarray}
\end{small}

So the lowest eigenvalue $\lambda_1$ reads at leading order for $\mu>0$ and $\rho \ne (0,1/2,1)$
\begin{eqnarray}
\lambda_1 && 
= \mu \ \frac{ \cosh[ \frac{\mu}{2} (1-2\rho) L] -  e^{- \frac{\mu}{2} L}}
 {  \frac{2}{\mu}\cosh[ \frac{\mu}{2} L] + L e^{- \frac{\mu}{2} L}
 + L (1-2 \rho) \sinh[ \frac{\mu}{2} (1-2 \rho ) L] -  \frac{2}{\mu}\cosh[ \frac{\mu}{2} (1-2\rho) L]
}
\nonumber \\
&& \opsimeq_{L \to + \infty} \frac{\mu^2}{2} e^{ - L \frac{\mu}{2} \left[ 1-\vert 1-2 \rho \vert \right] }
\label{lambda1imaginary}
\end{eqnarray}
i.e. the relaxation time $\tau_1$ associated to this lowest eigenvalue $\lambda_1$
\begin{eqnarray}
\tau_1 \equiv \frac{1}{\lambda_1 } \opsimeq_{L \to + \infty} \frac{2}{\mu^2} e^{  L \frac{\mu}{2} \left[ 1-\vert 1-2 \rho \vert \right] }
= \begin{cases}
\frac{2}{\mu^2} e^{  L \mu \rho }  = \frac{2}{\mu^2} e^{B_1(L)} & \text{for $0<\rho<\frac{1}{2}$ } \\
 \frac{2}{\mu^2} e^{  L \mu ( 1- \rho)   } = \frac{2}{\mu^2} e^{B_2(L)}  & \text{for $\frac{1}{2}<\rho<1$}
\end{cases}
\label{tau1imaginary}
\end{eqnarray}
has exactly the same behavior for large $L$ as the Kemeny time of Eqs \ref{kemeny1dsawmuposrhosmall}
and \ref{kemeny1dsawmuposrhobig}
discussed in the main text.



\section{ Kemeny times for diffusion processes on infinite intervals $(x_R-x_L) \to + \infty$ }

\label{app_infinite}

In this Appendix, we focus on diffusion processes on infinite intervals $(x_R-x_L) \to + \infty$:
even if the normalized steady state $P_*(x)$ exists on the infinite interval,
the Kemeny time $\tau_*$ can diverge as illustrated by the two following simple examples.

\subsection{ Divergence of the Kemeny time $\tau_*$ for the linear valley $U(x)=\mu x$ with $\mu>0$ and $D(x)=1$ on $[0,+\infty[ $}

Let us consider the limit $L \to + \infty$ of the case of Eq. \ref{URsawtooth} with $\rho=1$ 
corresponding to the linear potential $U(x)=\mu x$ with $\mu>0$ and $D(x)=1$:
the normalized steady state exists for $L \to + \infty$
\begin{eqnarray}
  p_*(x)  = \mu \frac{e^{- \mu x} \theta(0 \leq x \leq L)}{1-e^{-\mu L} }  \opsimeq_{L \to + \infty}  \mu e^{ - \mu x }  \ \ \ \ \ {\rm for } \ \ x \in [0,+\infty[
 \label{steadyeqsawinf}
\end{eqnarray}
but the corresponding Kemeny time computed in Eqs \ref{kemeny1dsawrhozero} \ref{kemeny1dsawrhozeroscaling}
diverges
\begin{eqnarray}
\tau_* = \frac{2}{ \mu^2 } \left[ \frac{ \frac{\mu L}{2} }{\tanh \left(\frac{\mu L}{2} \right)}  -1\right] 
\opsimeq_{L \to + \infty}  \frac{L }{  \vert \mu \vert } \opsimeq_{L \to + \infty} + \infty
\label{kemeny1dsawrhozeroscalingdv}
\end{eqnarray}

Since the quantum potential of Eq. \ref{vfromu} associated to the diffusion coefficient $D(x)=1$
and to the force $F(x)=- \mu$ 
\begin{eqnarray}
V(x)  = \frac{ F^2(x) }{4  } + \frac{F'(x)}{2} = \frac{ \mu^2 }{4  } \ \ {\rm for } \ \ x \in ]0,+\infty[
\label{vfromusawcte}
\end{eqnarray}
is simply constant for $ x \in ]0,+\infty[$, it is clear that 
there will be a continuum of excited states with eigenvalues $\lambda \in ] \frac{ \mu^2 }{4  },+ \infty[$,
as can be also seen from the large $L$-limit as the exact spectrum for finite $L$ given in Eq. \ref{eigenexcitedrhozerotext}.

More generally, the Kemeny time will diverge whenever there is a continuum of eigenvalues 
\begin{eqnarray}
\tau_*^{Spectral[Continuum]} = +\infty
\label{kemeny1dcontinuum}
\end{eqnarray}


\subsection{ Divergence of the Kemeny time $\tau_*$ for the quadratic valley $U(x)=\omega \frac{x^2}{2}$ 
and $D(x)=1$ on $]-\infty,+\infty[$ }

The Ornstein-Uhlenbeck process corresponding to the linear restoring force towards the origin $x=0$
with the parameter $\omega>0$
\begin{eqnarray}
F(x)=-  \omega x
\label{FOU}
\end{eqnarray} 
 and to the diffusion coefficient $D(x)=1$
 leads to the quadratic potential in Eq. \ref{UR}
\begin{eqnarray}
U(x)= \omega \frac{ x^2}{2}
\label{muOU}
\end{eqnarray}
The corresponding Gaussian steady state
\begin{eqnarray}
p_*(x) =  \frac{e^{-U(x)}}{Z}  = \sqrt{  \frac{\omega }{ 2 \pi } }  e^{- \omega  \frac{ x^2}{2}} 
\label{gaussOUeq}
\end{eqnarray}
can be plugged into the real-space Kemeny time of Eq. \ref{kemeny1d}
to obtain that it diverges
\begin{eqnarray}
\tau^{Space}_*(x) =   + \infty
\label{kemeny1dOU}
\end{eqnarray}

Since the quantum potential of Eq. \ref{vfromu} is also quadratic
\begin{eqnarray}
V(x)  = \frac{ F^2(x) }{4  } + \frac{F'(x)}{2} = \frac{ \omega^2 x^2 }{4  } - \frac{\omega}{2}
\label{vfromuOU}
\end{eqnarray}
the quantum Hamiltonian of Eq. \ref{hamiltonien} corresponds to the quantum harmonic oscillator 
with its well-known linear spectrum
\begin{eqnarray}
\lambda_n = \left(n+ \frac{1}{2} \right) \omega - \frac{\omega}{2} = n \omega \ \ {\rm with } \ n=0,1,..,+\infty
\label{eigenOU}
\end{eqnarray}
So the spectrum is fully discrete 
but the spectral Kemeny time nevertheless diverges
 \begin{eqnarray}
\tau_{*}^{Spectral} =  \sum_{n=1}^{+\infty} \frac{1}{\lambda_n} = \frac{1}{\omega} \sum_{n=1}^{+\infty} \frac{1}{n} =+\infty
\label{tauspectralinftyOU}
\end{eqnarray}
in agreement with the real-space Kemeny time of Eq. \ref{kemeny1dOU}.

After these two examples where the Kemeny time diverges,
it is thus interesting to discuss what conditions will produce a finite Kemeny time $\tau_*$,
 by considering the real-space Kemeny time and the spectral Kemeny time
 in the two next subsections respectively.


\subsection{ Analysis of the convergence of the real-space Kemeny time $\tau_{*}^{Space}  $ at $x_R=+\infty$  }

For $x_R=+\infty$, the real-space Kemeny time of Eq. \ref{kemeny1d} reads
\begin{eqnarray}
\tau^{Space}_*   =\int_{x_L}^{+\infty} dy \frac{1}{D(y) p_*(y) }
 \left[ \int_{-\infty}^{y} dz p_*(z) \right] \left[  \int_{y}^{+\infty} dx p_*(x) \right]
\label{kemeny1dcumulative}
\end{eqnarray}
so one needs to analyze the convergence of the integral in $y$ when $y \to + \infty$

Let us assume that the normalized steady state displays the following asymptotic decay for $x \to + \infty$
\begin{eqnarray}
p^*(x)  \oppropto_{x \to + \infty}     \frac{ e^{- K x^{\sigma} } }{ x^{1+\nu}} \ \ {\rm where } \ \ \sigma>0
\ \ {\rm or } \ \ \sigma=0 ; \nu>0
 \label{cumulativederi}
\end{eqnarray}
The leading behavior of the corresponding cumulative distribution reads
\begin{eqnarray}
\int_{y}^{+\infty} dz p_*(z)  \oppropto_{y \to + \infty}     \frac{ e^{- K y^{\sigma} } }{ y^{\sigma+\nu}} 
 \label{cumulativeinfty}
\end{eqnarray}
Then the convergence of the Kemeny time of Eq. \ref{kemeny1dcumulative} 
is determined by the convergence at $(+\infty)$ of the integral
 \begin{eqnarray}
 \int^{+\infty} dy \frac{ \frac{ e^{- K y^{\sigma} } }{ y^{\sigma+\nu}}}{ D(y) \frac{ e^{- K y^{\sigma} } }{ y^{1+\nu}}}
 =  \int^{+\infty} dy \frac{ y^{1-\sigma}   }{ D(y)    }  
\label{kemeny1dcumulativeinfinity}
\end{eqnarray}
i.e. it depends on the behavior of the diffusion coefficient $D(y)$ for $y \to + \infty$.
If the diffusion coefficient behaves asymptotically as the power of exponent $\eta $
 \begin{eqnarray}
 D(y) \oppropto_{y \to +\infty} y^{\eta}
 \label{Deta}
\end{eqnarray}
then the integral of Eq. \ref{kemeny1dcumulativeinfinity} will converge for
 \begin{eqnarray}
 \sigma>2 - \eta
 \label{eta2eta}
\end{eqnarray}
with the following important cases:

(i) When the diffusion coefficient is constant $D(y)=1$ corresponding to $\eta=0$,
then the real-space Kemeny time is finite $\tau^{Space}_*<+\infty$ only for exponents
 \begin{eqnarray}
 \sigma>2
 \label{eta2}
\end{eqnarray}
i.e. only if the asymptotic decay of the steady state of Eq. \ref{cumulativederi} is more rapid than Gaussian.
In particular, the Ornstein-Uhlenbeck considered in the previous subsection
corresponding to the Gaussian steady state of exponent $\sigma=2$ 
is just on the border where the Kemeny time diverges.

(ii) when the diffusion coefficient is asymptotically linear $D(y) \propto y$ corresponding to $\eta=1$,
then the real-space Kemeny time is finite $\tau^{Space}_*<+\infty$ only for exponents $\sigma >1$.

(iii) when the diffusion coefficient is asymptotically quadratic $D(y) \propto y^2$ corresponding to $\eta=2$,
then the real-space Kemeny time is finite $\tau^{Space}_*<+\infty$ for any strictly positive exponent $\sigma >0$.

(iv) if $\sigma=0$, i.e. if the steady state of Eq. \ref{cumulativederi}
behaves asymptotically as the power-law $ \frac{ 1 }{ x^{1+\nu}}  $ of exponent $\nu>0$,
then real-space Kemeny time is finite $\tau^{Space}_*<+\infty$ only 
for exponents $\eta>2$ in the diffusion coefficient of Eq. \ref{Deta}.


\subsection{ Analysis of the convergence of the spectral Kemeny time $\tau_{*}^{Spectral} $  at $n=+\infty$}

The convergence of the spectral Kemeny time $\tau^{Spectral}_*$ 
requires that the spectrum is fully discrete 
and that the $n^{th}$ eigenvalue $\lambda_n$ grows more rapidly than linearly for large $n$.

To make the link with the real-space analysis of the previous subsection, 
let us now assume that the diffusion coefficient is constant $D(x)=1$ as considered for Eq. \ref{eta2}
and that the potential $U(x)$ of Eq. \ref{steadyeq} defined on $]-\infty,+\infty[$
displays the following asymptotic behaviors 
\begin{eqnarray}
U(x)  \opsimeq_{x \to \pm \infty}  K \vert x \vert^{\sigma}
 \label{URq}
\end{eqnarray}
where the exponent $\sigma$ was introduced in Eq. \ref{cumulativederi}.
The asymptotic behavior of the force of Eq. \ref{UR}
\begin{eqnarray}
F(x) = - U'(x)  \opsimeq_{x \to \pm  \infty} - K \sigma \ {\rm sgn}(x) \vert x \vert ^{\sigma-1 }
 \label{URFq}
\end{eqnarray}
yields the asymptotic power-law behavior of the quantum potential of Eq. \ref{vfromu}
\begin{eqnarray}
V(x)  = \frac{ F^2(x) }{4  } + \frac{F'(x)}{2} 
\opsimeq_{x \to + \infty}  \frac{K^2 \sigma^2}{4}  x^{2(\sigma-1) } \ \ \ 
\label{vfromuq}
\end{eqnarray}
For large $n$, the scaling of the eigenvalues $\lambda_n$ can be 
obtained via the standard semi-classical formula for high-energy levels
\begin{eqnarray}
\lambda_n \oppropto_{n \to + \infty} n^{  q(\sigma) }  \ \ { \rm with } \ \ q(\sigma) = \frac{ 2 (\sigma-1) }{ \sigma}
\label{vsemiclass}
\end{eqnarray}
So the the spectral Kemeny time $\tau^{Spectral}_*$ will
converge for exponents $q(\sigma) >1 $, i.e. for $\sigma>2$
in agreement with the criterion of Eq. \ref{eta2} as it should for consistency.
In particular, the Ornstein-Uhlenbeck corresponding to 
$\sigma=2$ and to the linear behavior $q(\sigma=2)=1 $
(see Eq. \ref{tauspectralinftyOU})
is just on the border where the spectral Kemeny time diverges.

It is also interesting to mention that the case of finite intervals $(x_R-x_L)<+\infty$
 can be recovered in the limit of an infinite exponent $\sigma \to + \infty$
in the potential of Eq. \ref{URq}, with the corresponding exponent 
\begin{eqnarray}
q(\sigma \to + \infty) = 2
\label{q2finiteinterval}
\end{eqnarray}
for the large-n behavior of the eigenvalues of Eq. \ref{vsemiclass}.
So the semi-classical formula of Eq. \ref{vsemiclass}  
explains the leading quadratic behavior $ \lambda_n \propto n^2$ 
found in all examples on finite intervals with explicit eigenvalues.


\subsection{ Example: finite Kemeny time for the diffusion $D(x)=1$ in the quartic potential $U(x)= \frac{x^4}{2}$ on $]-\infty,+\infty[$}
 
As an example of the condition of Eq. \ref{eta2} with $D(x)=1$, 
let us consider the quartic potential on $]-\infty,+\infty[$ corresponding to the exponent $\sigma=4$
\begin{eqnarray}
U(x) = \frac{x^4}{2}
\label{quartic}
\end{eqnarray}
The steady state of Eq. \ref{steadyeq} 
\begin{eqnarray}
  p_*(x)  && = \frac{ e^{ - \frac{x^4}{2} } }{Z}
  \nonumber \\
  Z && = \int_{-\infty}^{+\infty} dx e^{ - \frac{x^4}{2} } 
  = 2^{-\frac{3}{4}} \int_{-\infty}^{+\infty} dy y^{\frac{1}{4}-1} e^{ - y }
  = 2^{-\frac{3}{4}}  \Gamma \left( \frac{1}{4} \right)
 \label{steadyeqquartic}
\end{eqnarray}
allows to compute the real-space Kemeny time via Eq. \ref{kemeny1d}
 \begin{eqnarray}
\tau_*  =\int_{-\infty}^{+\infty} dy \frac{1}{ p_*(y) }
 \left[\int_{-\infty}^{y} dz p_*(z) \right]
\left[ \int_{y}^{+\infty} dx p_*(x) \right]
= \frac{3 \pi  \Gamma \left(\frac{5}{4}\right)}{ 4  \sqrt{2} \Gamma \left(\frac{3}{4}\right)} 
\label{kemeny1dquartic}
\end{eqnarray}

For the quartic potential of Eq. \ref{quartic} with $D(x)=1$, 
the force of Eq. \ref{UR} is cubic
\begin{eqnarray}
F(x) = - U'(x)  = - 2 x^3
 \label{Fcubic}
\end{eqnarray}
so the quantum potential of Eq. \ref{vfromu} is a combination of sextic and quadratic contributions
\begin{eqnarray}
V(x)  = \frac{ F^2(x) }{4  } + \frac{F'(x)}{2} 
=  x^6 - 3 x^2
\label{vfromuqsextic}
\end{eqnarray}
while the scaling of the eigenvalues $\lambda_n$ for large $n$ given by the semiclassical formula of Eq. \ref{vsemiclass}
involves the exponent
\begin{eqnarray}
 q(\sigma=4) = \frac{ 3 }{ 2}
\label{qfor4}
\end{eqnarray}



\section{Jump-diffusion and jump-drift processes produced by resetting  }

\label{app_JumpDiff}

In this Appendix, we focus on jump-diffusion and jump-drift processes produced by stochastic resetting (see the review \cite{review_reset} and references therein).

\subsection{ Diffusion with diffusion coefficient $D(x)$ and force $F(x)$ ; resetting with rate $\gamma(x)$ towards $\Pi(y )$}

The one-dimensional jump-diffusion dynamics is defined as follows: 
when at position $x$,
the particle diffuses with the diffusion coefficient $D(x)$ in the force $F(x)$ as in section \ref{sec_DiffInterval},
but with the resetting rate $\gamma(x)$ per unit time, it can also make a non-local jump toward
 some new position $y$ chosen with the normalized probability distribution $\Pi(y ) $.

The propagator $p_t ( x \vert x_0)$ 
satisfies the Forward dynamics with respect to the final position $x$
\begin{eqnarray}
 \partial_t p_t(x \vert x_0)  && =    
  \partial_{x}  \left[ - F(x)  + D (x)  \partial_{x} \right] p_t(x \vert x_0) 
  + \Pi(x )  \int d y  \  \gamma( y) p_t(y \vert x_0) -  \gamma( x)p_t(x \vert x_0) 
\label{jumpdiff}
\end{eqnarray}
and the backward dynamics with respect to the initial position $x_0$
\begin{eqnarray}
 \partial_t p_t(x \vert x_0)  && =    
 \left[  F(x_0)  +   \partial_{x_0} D (x_0) \right]    \partial_{x_0}  p_t(x \vert x_0) 
  + \left[ \int d y_0  \  p_t(x \vert y_0) \Pi(y_0) -   p_t(x \vert x_0) \right] \gamma( x_0) 
\label{jumpdiffback}
\end{eqnarray}


\subsection{ Steady state $p_*(x ) $ with its two types of steady currents }

The steady version of the forward Eq. \ref{jumpdiff} satisfied by the steady state $p_*(x)$
\begin{eqnarray}
0 && =  - \frac{ d }{ d x}   \left[  F(x) p_*(x )  - D (x)   \frac{ d  p_*(x)}{ d x}    \right]
 + \Pi(x) \int d y  \gamma( y)  p_*(y)-  \gamma( x)   p_*(x)
 \nonumber \\
&& \equiv  - \frac{ d j_*(x)}{ d x}    +  \int d y  J_*(x , y) 
\label{jumpdiffst}
\end{eqnarray}
involves two types of steady currents $j_*(x) $ and $J_*(x , y)  $:

(i) the local diffusive current $ j_*(x )  $ at position $x$ that involves the diffusion coefficient $D(x)$ and the force $F(x)$
\begin{eqnarray}
 j_*(x )  \equiv  F(x) p_*(x )  - D (x)   \frac{ d  p_*(x)}{ d x}  
\label{jslocal}
\end{eqnarray}

(ii) the non-local current from $y$ to $x$ produced by the jumps
 that involve the rates $\gamma(.)$ and the probability distribution $\Pi(.)$
\begin{eqnarray}
 J_*(x , y)  \equiv \Pi(x ) \gamma( y)  p_*(y) - \Pi(y ) \gamma( x)  p_*(x) = -J_*(y,x)
\label{jsnonlocal}
\end{eqnarray}

Since the backward Eq. for the MFTP $\tau(x, x_0) $ reads
\begin{eqnarray}
 -1+\frac{\delta_{x,x_0} }{p_*(x)}  && =    
 \left[  F(x_0)  +   \partial_{x_0} D (x_0) \right]    \partial_{x_0}  \tau(x, x_0) 
  + \left[ \int d y_0  \  \tau(x, y_0)  \Pi(y_0 ) -   \tau(x, x_0)  \right] \gamma( x_0) 
\label{jumpdiffbacktau}
\end{eqnarray}
it is convenient to introduce the difference
\begin{eqnarray}
g(x,x_0) \equiv \tau(x, x_0)- \int d y_0  \  \tau(x, y_0)  \Pi(y_0 ) 
\label{taudifference}
\end{eqnarray}
that satisfies the differential equation
\begin{eqnarray}
 -1+\frac{\delta_{x,x_0} }{p_*(x)}   =    \partial_{x_0} \left[ D (x_0)   \partial_{x_0}  g(x, x_0) \right]
 +  F(x_0)   \partial_{x_0}  g(x, x_0) -  \gamma( x_0) g(x, x_0)
\label{eqtaudifference}
\end{eqnarray}
For arbitrary parameters $[D(x);F(x),\gamma(x)]$, the general solution of this second-order
differential equation cannot be written explicitly, so we will focus 
on two soluble cases in the two following subsections: 

(i)  the case where the three parameters $[D(x);F(x),\gamma(x)]$ are space-independent

(ii) the case without diffusion $D(x)=0$ while the force $F(x)$ and the resetting rate $\gamma(x)$
are space-dependent.


\subsection{ Diffusion $D(x)=1$ and force $F(x)=\mu$ on the ring $[0,L]$ with resetting at rate $\gamma(x)=\gamma$ towards $\Pi(y)$    }

For the diffusion with diffusion coefficient $D(x)=1$ and force $F(x)=\mu$ on the ring $[0,L]$ with resetting at constant rate $\gamma(x)=\gamma$ towards $\Pi(y)$, the spectral Kemeny time has been already computed 
in Eq.  \ref{KemenyResetDiff} of the main text. 
The goal of the present section is
to give an alternative derivation via the real-space expression of the Kemeny time.

The difference of Eq. \ref{taudifference}
\begin{eqnarray}
g(x,x_0) \equiv \tau(x, x_0)- \int_0^L d y_0  \  \tau(x, y_0)  \Pi(y_0 ) 
\label{taudifferencereset}
\end{eqnarray}
satisfies the differential equation of Eq. \ref{eqtaudifference}
\begin{eqnarray}
 -1+ \frac{\delta_{x,x_0} }{p^{[Reset]}_*(x)}    =       \partial^2_{x_0}  g(x, x_0) 
 + \mu \partial_{x_0}  g(x, x_0) -  \gamma g(x, x_0)
\label{eqtaudifferenceuni}
\end{eqnarray}
The general solution of the homogeneous equation is a linear combination of the two
exponential solutions $e^{ x_0 \kappa_{\pm} }$ where the two coefficients $ \kappa_{\pm}$ are solutions of the quadratic equation
\begin{eqnarray}
0 = \kappa^2+ \mu \kappa - \gamma
\label{quadratic}
\end{eqnarray}
and read
\begin{eqnarray}
\kappa_{\pm} = - \frac{\mu}{2} \pm \sqrt{ \frac{\mu^2}{4}+\gamma}
\equiv - \frac{\mu}{2} \pm \kappa
\label{quadraticsolu}
\end{eqnarray}
with the notation
\begin{eqnarray}
\kappa \equiv \sqrt{ \frac{\mu^2}{4}+\gamma}
\label{kappagamma}
\end{eqnarray}

In the two regions $x_0 <x$ and $x_0>x$, 
the general solution of the inhomogeneous equation of Eq. \ref{eqtaudifferenceuni}
reads
\begin{eqnarray}
0 \leq x_0 \leq x : \ \  g(x, x_0) 
&& = \frac{1}{\gamma} + e^{- \frac{\mu}{2} x_0 } 
\left[ A_-(x) \cosh\left( x_0 \kappa \right) 
+ B_-(x) \sinh\left( x_0 \kappa \right)
\right]
\nonumber \\
x \leq x_0 \leq L : \ \  g(x, x_0) 
&& =  \frac{1}{\gamma} + e^{- \frac{\mu}{2} (x_0-L) } 
\left[ A_+(x) \cosh\left((x_0-L)\kappa \right) 
+ B_+(x) \sinh\left((x_0-L)\kappa  \right)
\right]
\label{gabove}
\end{eqnarray}
with their corresponding derivatives with respect to $x_0$
\begin{eqnarray}
0 \leq x_0 \leq x : \ \  \partial_{x_0}g(x, x_0) 
&& = - \frac{\mu}{2} e^{- \frac{\mu}{2} x_0 } 
\left[ A_-(x) \cosh\left( x_0 \kappa \right) 
+ B_-(x) \sinh\left( x_0 \kappa \right)
\right]
\label{gabovederi}
 \\
&& + \kappa e^{- \frac{\mu}{2} x_0 } 
\left[ A_-(x) \sinh\left( x_0 \kappa \right) 
+ B_-(x) \cosh\left( x_0 \kappa \right)
\right]
\nonumber \\
x \leq x_0 \leq L : \ \  \partial_{x_0}g(x, x_0) 
&& =   - \frac{\mu}{2} e^{- \frac{\mu}{2} (x_0-L) } 
\left[ A_+(x) \cosh\left((x_0-L)\kappa \right) 
+ B_+(x) \sinh\left((x_0-L)\kappa  \right)
\right]
\nonumber \\
&&+\kappa e^{- \frac{\mu}{2} (x_0-L) } 
\left[ A_+(x) \sinh\left((x_0-L)\kappa \right) 
+ B_+(x) \cosh\left((x_0-L)\kappa  \right)
\right]
\nonumber
\end{eqnarray}

The four functions $[A_{\pm}(x);B_{\pm}(x)]$ are determined by the 
following boundary conditions:

(i) The periodicity between $x_0 =0 $ and $x_0=L$ for the function $g(x,x_0)$ 
and its derivative $\partial_{x_0}  g(x, x_0) $ 
\begin{eqnarray}
0 && = g(x, x_0=0) - g(x, x_0=L) =   A_-(x) -A_+(x)
\nonumber \\
0 &&    =      \partial_{x_0}  g(x, x_0) \vert_{x_0=0} -  \partial_{x_0}  g(x, x_0) \vert_{x_0=L}
=  \frac{\mu}{2}   \left[ A_+(x) -A_-(x)  \right] 
 + \kappa  \left[ B_-(x) - B_+(x) \right]
\label{CLABperio}
\end{eqnarray}
yields that the two functions $A_{\pm}(x)$ coincide, as well as the two functions $B_{\pm}(x)$
\begin{eqnarray}
   A_-(x) && = A_+(x) \equiv A(x)
\nonumber \\
 B_-(x) && = B_+(x) \equiv B(x)
\label{CLABperiocoincide}
\end{eqnarray}
so that from now on we will use the simplified notations $A(x)$ and $B(x)$.

(ii) At $x_0=x$, the continuity of the function $g(x,x_0)$ 
and the discontinuity of its derivative $\partial_{x_0}  g(x, x_0) $ at $x_0=x$
required by Eq. \ref{eqtaudifferenceuni}
\begin{eqnarray}
 0  && = \left[ g(x, x_0=x^+) - g(x, x_0=x^-) \right]  e^{ \frac{\mu}{2} x } 
\nonumber
 \\
&&=
 -  A(x) \left[ e^{ \frac{\mu}{2} L }\cosh\left((L-x)\kappa \right) 
- \cosh\left( x \kappa \right) \right]
+ B(x) \left[ e^{\frac{\mu}{2} L }\sinh\left((L-x)\kappa  \right)
+ \sinh\left( x \kappa \right)\right]
 \nonumber \\&& 
 \equiv - A(x) C(x)+B(x) S(x)
\nonumber \\
 \frac{e^{ \frac{\mu}{2} x }  }{p^{[Reset]}_*(x)}   &&   =   \left[     \partial_{x_0}  g(x, x_0) \vert_{x_0=x^+} -  \partial_{x_0}  g(x, x_0) \vert_{x_0=x^-} \right]  e^{ \frac{\mu}{2} x } 
\nonumber \\ 
&& = 
- \frac{\mu}{2}  A(x) \left[ e^{ \frac{\mu}{2} L}   \cosh\left((L-x)\kappa \right)
 -\cosh\left( x \kappa \right)   \right]
 + \frac{\mu}{2} B(x) \left[ e^{ \frac{\mu}{2} L}  \sinh\left((L-x)\kappa  \right)
 +\sinh\left( x \kappa \right)
 \right]
\nonumber \\
&& 
- \kappa A(x) 
\left[e^{ \frac{\mu}{2}L }   \sinh\left((L-x)\kappa \right)  + \sinh\left( x \kappa \right) 
\right]
+\kappa B(x) 
\left[ e^{ \frac{\mu}{2}L }   \cosh\left((L-x)\kappa  \right)
- \cosh\left( x \kappa \right)\right]
\nonumber \\
&& \equiv  - \frac{\mu}{2} \left[ A(x) C(x)-B(x) S(x)\right]
- \kappa\left[ A(x) S(x)-B(x) C(x)\right]
 \label{eqtaudifferenceunidisc}
\end{eqnarray}

where we have introduced the two functions
\begin{eqnarray}
C(x) && \equiv e^{ \frac{\mu}{2}L }   \cosh\left((L-x)\kappa  \right)
- \cosh\left( x \kappa \right)
\nonumber \\
S(x) && \equiv e^{ \frac{\mu}{2}L }   \sinh\left((L-x)\kappa \right)  + \sinh\left( x \kappa \right)
\label{CSX}
\end{eqnarray}
satisfying
\begin{eqnarray}
C^2(x) - S^2(x)&& = e^{ \mu L } \left[  \cosh^2\left((L-x)\kappa  \right)
-  \sinh^2\left((L-x)\kappa \right) \right]
+ \left[ \cosh^2\left( x \kappa \right) - \sinh^2\left( x \kappa \right)\right]
\nonumber \\
&& - 2 e^{ \frac{\mu}{2}L } \left[  \cosh\left((L-x)\kappa  \right)
 \cosh\left( x \kappa \right)+  \sinh\left((L-x)\kappa \right)   \sinh\left( x \kappa \right) \right]
 \nonumber \\
 && =e^{ \mu L } +1 - 2 e^{ \frac{\mu}{2}L } \cosh\left( L\kappa  \right)
 =  -2  e^{ \frac{\mu}{2}L } \left[ \cosh\left( L\kappa  \right) - \cosh\left( L \frac{\mu}{2}  \right) \right]
\label{CSX2}
\end{eqnarray}

The solution of the system \ref{eqtaudifferenceunidisc} for the two functions $A(x)$ and $B(x)$
\begin{eqnarray}
 0  && =  A(x) C(x)-B(x) S(x)
\nonumber \\
 \frac{e^{ \frac{\mu}{2} x }  }{\kappa p^{[Reset]}_*(x)}   &&   =  -A(x) S(x)+B(x) C(x)
 \label{systemABSC}
\end{eqnarray}
thus reads
\begin{eqnarray}
A(x) && = -  \frac{e^{ \frac{\mu}{2} x }  S(x)}{p^{[Reset]}_*(x) \kappa \left[ S^2(x) - C^2(x)\right]}    
= - \frac{e^{ \frac{\mu}{2} x } 
\left[   \sinh\left((L-x)\kappa \right)  + e^{- \frac{\mu}{2}L } \sinh\left( x \kappa \right)\right] }
{p^{[Reset]}_*(x)   2 \kappa   
 \left[ \cosh\left( L\kappa  \right) - \cosh\left( L \frac{\mu}{2}  \right) \right] }
\nonumber \\
B(x) && =  -  \frac{e^{ \frac{\mu}{2} x } C(x) }{p^{[Reset]}_*(x) \kappa \left[ S^2(x) - C^2(x)\right]} 
= - \frac{e^{ \frac{\mu}{2} x } 
\left[    \cosh\left((L-x)\kappa  \right)- e^{ - \frac{\mu}{2}L } \cosh\left( x \kappa \right)\right]  }
{p^{[Reset]}_*(x) 2 \kappa
 \left[ \cosh\left( L\kappa  \right) - \cosh\left( L \frac{\mu}{2}  \right) \right]} 
 \label{solABx}
\end{eqnarray}

For $x_0=x$ with the vanishing MFTP $\tau(x, x_0=x) =0$, Eq. \ref{taudifference} reads
\begin{eqnarray}
g(x,x) = - \int_0^L d y_0  \  \tau(x, y_0)  \Pi(y_0 ) 
\label{taudifferencexx}  
\end{eqnarray}
so that Eq. \ref{taudifference} allows to compute the MFTP $\tau(x, x_0) $ 
from the function $g(.,.)$ via
\begin{eqnarray}
\tau(x, x_0) = g(x,x_0) - g(x,x)
\label{taufromg}
\end{eqnarray}
The solution found for $g(.,.)$ in Eq. \ref{gabove}
then yields in the two regions
\begin{eqnarray}
x_0 \in [0,x] : \ \ \tau(x, x_0)  && = g(x,x_0) - g(x,x)  
\nonumber \\ &&
=   A(x) \left[ e^{- \frac{\mu}{2} x_0 }  \cosh\left( x_0 \kappa \right) 
- e^{- \frac{\mu}{2} x }  \cosh\left( x \kappa \right)\right]
\nonumber \\ 
&& + B(x) \left[e^{- \frac{\mu}{2} x_0 } \sinh\left( x_0 \kappa \right)
-e^{- \frac{\mu}{2} x } \sinh\left( x \kappa \right)
\right]
\nonumber \\
x_0 \in [x,L] : \ \ \tau(x, x_0)  && = g(x,x_0) - g(x,x) 
 \nonumber \\ &&= 
 A(x) \left[ e^{- \frac{\mu}{2} (x_0-L) } \cosh\left((L-x_0)\kappa \right) 
 - e^{- \frac{\mu}{2} (x-L) } \cosh\left((L-x)\kappa \right) \right]
 \nonumber \\ &&
 - B(x) \left[ e^{- \frac{\mu}{2} (x_0-L) } \sinh\left((L-x_0)\kappa  \right)
- e^{- \frac{\mu}{2} (x-L) } \sinh\left((L-x)\kappa  \right)
\right]
\label{taureset}  
\end{eqnarray}

Let us now compute more explicitly the MFTP $\tau (x, x_0=0) $ as a function of the final point $x$
when starting at $x_0=0$ using Eq. \ref{solABx}
\begin{eqnarray}
&& \tau(x, x_0=0)  
  =  - A(x) \left[  e^{- \frac{\mu}{2} x }  \cosh\left( x \kappa \right) - 1 \right]
 - B(x) e^{- \frac{\mu}{2} x } \sinh\left( x \kappa \right)
 \nonumber \\
&&=  \frac{ 
 \left[   \sinh\left((L-x)\kappa \right)  + e^{ - \frac{\mu}{2} L } \sinh\left( x \kappa \right)\right]\left[    \cosh\left( x \kappa \right) - e^{ \frac{\mu}{2} x } \right]
  +\left[    \cosh\left((L-x)\kappa  \right)- e^{ - \frac{\mu}{2} L } \cosh\left( x \kappa \right)\right]  
   \sinh\left( x \kappa \right) }
{p^{[Reset]}_*(x)   2 \kappa   
 \left[ \cosh\left( L\kappa  \right) - \cosh\left( L \frac{\mu}{2}  \right) \right] }
  \nonumber \\
&&=  \frac{ \sinh\left( L \kappa \right)
- e^{ \frac{\mu}{2} x }  \sinh\left((L-x)\kappa \right) 
 - e^{- \frac{\mu}{2}(L -x) } \sinh\left( x \kappa \right)    }
{p^{[Reset]}_*(x)   2 \kappa   
 \left[ \cosh\left( L\kappa  \right) - \cosh\left( L \frac{\mu}{2}  \right) \right] }
 \label{tauresetzero}  
\end{eqnarray}

in order to compute
the real-space Kemeny time 
 \begin{eqnarray}
&& \tau^{[Reset]Space}_{*}  = \int_0^L dx p^{[Reset]}_{*}(x) \tau (x, x_0=0) 
 = \int_0^L dx 
  \frac{ \sinh\left( L \kappa \right)
- e^{ \frac{\mu}{2} x }  \sinh\left((L-x)\kappa \right) 
 - e^{- \frac{\mu}{2}(L -x) } \sinh\left( x \kappa \right)    }
{   2 \kappa   
 \left[ \cosh\left( L\kappa  \right) - \cosh\left( L \frac{\mu}{2}  \right) \right] }
 \nonumber \\
 && =
  \frac{ L \sinh\left( L \kappa \right)   }
{   2 \kappa   
 \left[ \cosh\left( L\kappa  \right) - \cosh\left( L \frac{\mu}{2}  \right) \right] }
- \int_0^L dx 
  \frac{  e^{ \left( \frac{\mu}{2} -\kappa \right) x } 
 \left[ e^{L\kappa } -e^{- \frac{\mu}{2} L }  \right]
 + e^{ \left( \frac{\mu}{2} +\kappa \right) x } 
 \left[ e^{- \frac{\mu}{2}L } -e^{- L\kappa } \right]  
      }
{   4 \kappa   
 \left[ \cosh\left( L\kappa  \right) - \cosh\left( L \frac{\mu}{2}  \right) \right] }
 \nonumber \\
&& =
  \frac{ L \sinh\left( L \kappa \right)   }
{   2 \kappa   
 \left[ \cosh\left( L\kappa  \right) - \cosh\left( L \frac{\mu}{2}  \right) \right] }
-     \frac{ 1 } { \kappa^2 - \frac{\mu^2}{4} } 
 \label{taukemenyresetspacekappa}
\end{eqnarray}
Using the value $\kappa = \sqrt{ \frac{\mu^2}{4}+\gamma} $ of Eq. \ref{kappagamma},
the final result
\begin{eqnarray}
\tau^{[Reset]Space}_{*}
&& =
  \frac{ L \sinh\left( L \sqrt{ \frac{\mu^2}{4}+\gamma} \right)   }
{   2 \sqrt{ \frac{\mu^2}{4}+\gamma}   
 \left[ \cosh\left( L \sqrt{ \frac{\mu^2}{4}+\gamma}  \right) - \cosh\left( L \frac{\mu}{2}  \right) \right] }
-     \frac{ 1 } { \gamma } 
 \nonumber \\&&
 =  \frac{ L \sinh\left( L \sqrt{ \frac{\mu^2}{4}+\gamma} \right)   }
{   2 \sqrt{\mu^2+4 \gamma} 
 \sinh \left( L \frac{\sqrt{\mu^2+4 \gamma}+ \mu}{4}\right)
 \sinh \left( L \frac{\sqrt{\mu^2+4 \gamma}- \mu}{4}\right) }
-     \frac{ 1 } { \gamma } 
 \label{taukemenyresetspace}
\end{eqnarray}
is in agreement with the spectral Kemeny time given in Eq.  \ref{KemenyResetDiff} of the main text.


\subsection{ Jump-drift process with force $F(x)>0$ on the ring $[0,L]$ with resetting 
at rate $\gamma(x) $ towards $\Pi( y)=\delta(y)  $    }

Let us now consider the following 
jump-drift process without diffusion $D(x)=0$ 
with the space-dependent strictly positive force $F(x) >0 $
on the ring periodic $[0,L]$,
while the resetting occurs with the space-dependent rate $\gamma(x)$ towards 
the origin $\Pi(y)=\delta(y)$.


\subsubsection{ Steady state $p_*(x)$  }

Since there is no diffusion $D(x)=0$,
the local steady current $ j_*(x )  $ of Eq. \ref{jslocal} is directly proportional to the steady state
\begin{eqnarray}
 j_*(x )  =  F(x) p_*(x ) 
\label{jslocaljump}
\end{eqnarray}
As a consequence, the steady state Eq. \ref{jumpdiffst} 
can be rewritten as the following first-order
differential equation for the local current $_*(x)$
\begin{eqnarray}
  \frac{ d  j_*(x )}{ d x}   +  \frac{\gamma( x)}{F(x) }   j_*(x) 
 =  \delta(x)  \int_0^L d y  \frac{\gamma( y)}{F(y) }   j_*(y) 
\label{jumpdiffstjump}
\end{eqnarray}
The current 
\begin{eqnarray}
 j_*(x )  =  j_*(0^+ )  e^{- \int_0^x dy \frac{\gamma (y)} {F(y) } } \ \ \ {\rm for } \ \ x \in ]0,L[
\label{jsi}
\end{eqnarray}
thus decays from its value $j_*(x=0^+ ) $ along the ring up to 
its value at $x=L^-$ that coincides with $x=0^-$
\begin{eqnarray}
 j_*(0^- ) = j_*(L^- )  =  j_*(0^+ )  e^{- \int_0^L dy \frac{\gamma (y)} {F(y) } } 
\label{jsizero}
\end{eqnarray}
while the discontinuity at the origin as determined by Eq. \ref{jumpdiffstjump}
\begin{eqnarray}
   j_*(0^+) -    j_*(0^- )  =  \int_0^L d y  \frac{\gamma( y)}{F(y) }   j_*(y)  
\label{jzeroring}
\end{eqnarray}
is satisfied for any $ j_*(0^+) $ using Eqs \ref{jsi} and \ref{jsizero}.
It is the normalization of the steady state $p_*(x)$ obtained from the current $ j_*(x ) $ via Eq. \ref{jslocaljump} 
\begin{eqnarray}
 p_*(x )  =  \frac{ j_*(x )}{F(x)} = \frac{j(0^+)}{F(x)} e^{- \displaystyle\int_0^x dy \frac{\gamma (y)} {F(y) } }
 \ \ \ {\rm for } \ \ x \in ]0,L[
\label{steadyresetsi}
\end{eqnarray}
that determines the current $ j_*(0^+ ) $ at the origin
\begin{eqnarray}
1= \int_0^L dx p_*(x) = \int_0^L dx \frac{ j_*(x )  }{  F(x) }  
=   j_*(0^+ ) \int_0^L dx \frac{1  }{  F(x) }   e^{- \int_0^x dy \frac{\gamma (y)} {F(y) } }
\label{normasteadyjumpjump}
\end{eqnarray}


\subsubsection{ Mean-First-Passage-Time $\tau(x,x_0)$ }

The difference of Eq. \ref{taudifference}
\begin{eqnarray}
g(x,x_0) \equiv \tau(x, x_0)-   \tau(x, 0)  
\label{taudifferenceorigin}
\end{eqnarray}
satisfies the first-order differential equation of Eq. \ref{eqtaudifference}
\begin{eqnarray}
 -1+\frac{\delta_{x,x_0} }{p_*(x)}   =     F(x_0)   \partial_{x_0}  g(x, x_0) -  \gamma( x_0) g(x, x_0)
\label{eqtaudifferenceorigin}
\end{eqnarray}
The change of function
\begin{eqnarray}
 g(x, x_0) = h(x,x_0) e^{\int_0^{x_0} dy \frac{\gamma (y)} {F(y) }}
\label{tauh}
\end{eqnarray}
yields that the new function $h(x,x_0)$
satisfies
\begin{eqnarray}
  \partial_{x_0} h(x, x_0)  && = \left[ -1
  +\frac{\delta_{x,x_0} }{p_*(x)}  \right] \frac{1}{ F(x_0) } e^{-\int_0^{x_0} dy \frac{\gamma (y)} {F(y) }} 
  =  \left[ -1  +\frac{\delta_{x,x_0} }{p_*(x)}  \right] \frac{p_*(x_0) }{j_*(0^+) }
 \nonumber \\
 && 
 = \frac{1 }{j_*(0^+) } \left[ - p_*(x_0)  +\delta_{x,x_0}   \right] 
\label{hreset}
\end{eqnarray}
where we have used the steady state $p_*(x_0)$ of Eq. \ref{steadyresetsi} 
and the current $j_*(0^+)$ to rewrite the right hand side.

Since $h(x,x_0)$ of Eq. \ref{tauh} vanishes at $x_0 =0$ as the function $g(x,x_0)$ of Eq. \ref{taudifferenceorigin},
the integration of Eq. \ref{hreset} yields in the two regions
\begin{eqnarray}
x_0 \in [0,x[  : \ \  h(x, x_0)   && 
= - \frac{1 }{j_*(0^+) } \int_0^{x_0} d y   p_*(y) 
\nonumber \\
x_0 \in ]x,L[  : \ \  h(x, x_0)   
&& =  \frac{1 }{j_*(0^+) } \left[ - \int_0^{x_0} d y   p_*(y) +1 \right] 
=   \frac{1 }{j_*(0^+) } \int_{x_0}^L d y   p_*(y)
\label{hresetinteg}
\end{eqnarray}
where we have used the normalization of the steady state for the second region.
Putting everything together, one obtains that 
the MFTP $\tau(x, x_0) $ 
can be rewritten in terms of the force $F(x)$ and of the steady state $p_*(x)$ of Eq. \ref{steadyresetsi}
\begin{eqnarray}
x_0 \in [0,x[  : \ \  \tau(x, x_0) && 
=  \tau(x, 0) 
- \frac{1 }{j_*(0^+) } e^{\int_0^{x_0} dy \frac{\gamma (y)} {F(y) }}\int_0^{x_0} d y   p_*(y) 
=
\tau(x, 0) 
- \frac{1 }{ F(x_0) p_*(x_0) } \int_0^{x_0} d y   p_*(y)
\nonumber \\
x_0 \in ]x,L]  : \ \  \tau(x, x_0) && =  \tau(x, 0)   
 +  \frac{1 }{j_*(0^+) } e^{\int_0^{x_0} dy \frac{\gamma (y)} {F(y) }} \int_{x_0}^L d y   p_*(y)
  =  \tau(x, 0)   
 +  \frac{1 }{ F(x_0) p_*(x_0) }  \int_{x_0}^L d y   p_*(y)
\label{tauresetinteg}
\end{eqnarray}
In the the present model without diffusion $D(x)=0$,
 the particle can only moves along the force $F(x)>0$ or jump at rate $\gamma(x)$ towards the origin, 
so the MFTP $\tau(x, x_0) $ vanishes only for $x_0=x^-$
but not for $x_0=x^+$ 
\begin{eqnarray}
0 && = \tau(x, x_0=x^-)  = \tau(x, 0) 
- \frac{1 }{ F(x) p_*(x) } \int_0^{x} d y   p_*(y)
\nonumber \\
0 && \ne \tau(x, x_0=x^+)  =  \tau(x, 0)   
 +  \frac{1 }{ F(x) p_*(x) }  \int_{x}^L d y   p_*(y)
\label{taudirected}
\end{eqnarray}
So the first line yields the MFTP $ \tau(x, 0) $ at position $x$ when starting at the origin $x_0=0$
\begin{eqnarray}
  \tau(x, 0)   = \frac{1 }{ F(x) p_*(x) } \int_0^{x} d y   p_*(y) 
\label{tauxzero}
\end{eqnarray}
that can be plugged into Eq. \ref{tauresetinteg}
to obtain the final solution for the MFTP in the two regions
\begin{eqnarray}
x_0 \in [0,x[  : \ \  \tau(x, x_0) && 
= \frac{1 }{ F(x) p_*(x) } \int_0^{x} d y   p_*(y) 
- \frac{1 }{ F(x_0) p_*(x_0) } \int_0^{x_0} d y   p_*(y)
\nonumber \\
x_0 \in ]x,L]  : \ \  \tau(x, x_0) && 
  =  \frac{1 }{ F(x) p_*(x) } \int_0^{x} d y   p_*(y)  
 +  \frac{1 }{ F(x_0) p_*(x_0) }  \int_{x_0}^L d y   p_*(y)
\label{tauresetintegfinal}
\end{eqnarray}


\subsubsection { Real-space Kemeny time }

The real-space Kemeny time of Eq. \ref{taukemeny} can be evaluated for the initial position 
at $x_0=0$ using the MFTP $ \tau(x,0) $ of Eq. \ref{tauxzero} 
 \begin{eqnarray}
\tau^{Space}_{*} = \int_0^L dx p_*(x)  \tau(x,0) 
=\int_0^L dx  \frac{1}{F(x)} \int_0^{x} d y   p_*(y) 
= \int_0^L d y   p_*(y) \int_y^L dx  \frac{1}{F(x)} 
\label{taukemenyreset}
\end{eqnarray}
where the steady state is given in Eqs \ref{steadyresetsi} \ref{normasteadyjumpjump}.


\subsubsection { Simplest example with the constant force $F(x)=1$ and the constant resetting rate $\gamma(x)=\gamma$ }

The steady state $p_*(x)$ of Eq. \ref{steadyresetsi} then reduces to 
\begin{eqnarray}
 p_*(x )   = \frac{\gamma}{1- e^{-\gamma L}}  e^{- \gamma x }
  \ \ \ {\rm for } \ \ x \in ]0,L[
\label{steadyresetsiuni}
\end{eqnarray}
and the real-space Kemeny time of Eq. \ref{taukemenyreset} becomes
 \begin{eqnarray}
\tau^{Space}_{*} 
=\frac{\gamma}{1- e^{-\gamma L}}    \int_0^L d y   (L-y)  e^{- \gamma y }
=  \frac{L}{1- e^{-\gamma L}} - \frac{1}{\gamma}
\label{taukemenyresetuni}
\end{eqnarray}

In the limit $L \to + \infty$ where one recovers the Sisyphus process on the half-line $x \in [0,+ \infty[$
with its exponential steady state (see \cite{c_reset} and references therein)
\begin{eqnarray}
 p_*(x )   = \gamma  e^{- \gamma x }
  \ \ \ {\rm for } \ \ x \in ]0,+\infty[
\label{steadyresetuniinfty}
\end{eqnarray}
the real-space Kemeny time of Eq. \ref{taukemenyresetuni} diverges
 \begin{eqnarray}
\tau^{Space}_{*} = + \infty
\label{taukemenyresetuniDV}
\end{eqnarray}


\section{ Computation of the spectral Kemeny times involving the roots of unity  }

\label{app_roots}

In this purely technical Appendix, we describe how to compute the various spectral Kemeny times
of the main text that involve the roots of unity.

\subsection{ Computation of some spectral Kemeny times involving directly the roots of unity } 

The characteristic polynomial 
\begin{eqnarray}
P(z) = z^N-1 = \prod_{n=0}^{N-1} (z-\omega_n)   
\label{polynomialrootsunity}
\end{eqnarray}
associated to the $N$ roots of unity
\begin{eqnarray}
\omega_n = e^{i n \frac{2 \pi}{N} } \ \ { \rm for } \ n=0,1,..,N-1
\label{rootsunity}
\end{eqnarray}
can be factorized into
\begin{eqnarray}
P(z)  = z^N-1  = (z-1) Q(z)
\label{polynomialrootsunityfactorQ}
\end{eqnarray}
where the polynomial $Q(z)$ only involves the $(N-1)$ roots $\omega_n$ with $n=1,..,N-1$ that are different from unity $1=\omega_0$
\begin{eqnarray}
Q(z) =  \prod_{n=1}^{N-1} (z-\omega_n)   = \frac{ z^N-1  }{ z-1 } = \sum_{k=0}^{N-1} z^k
\label{polynomialrootsunityfactor}
\end{eqnarray}
Its logarithmic derivative allows to compute the sum of the inverses $\frac{1}{z-\omega_n } $ over $n=1,..,N-1$
\begin{eqnarray}
\sum_{n=1}^{N-1} \frac{1}{z-\omega_n}  = \frac{Q'(z) }{Q(z) }   
=   \frac{ N z^{N-1}  }{ z^N-1 } - \frac{ 1 }{ z-1 }
= \frac{ \displaystyle \sum_{k=1}^{N-1} k z^{k-1} }{ \displaystyle  \sum_{k=0}^{N-1} z^k}
\label{polynomialderilog}
\end{eqnarray}

\subsubsection{ Computation of the spectral Kemeny time of Eq. \ref{tauspectralDir}}

For $z=1$, Eq. \ref{polynomialderilog} reduces to
\begin{eqnarray}
\sum_{n=1}^{N-1} \frac{1}{1-\omega_n}  
= \frac{ \displaystyle \sum_{k=1}^{N-1} k  }{ \displaystyle  \sum_{k=0}^{N-1} 1}
 = \frac{\frac{N(N-1)}{2}}{N} = \frac{N-1}{2}
\label{polynomialderilog1}
\end{eqnarray}
that allows to compute the spectral Kemeny time of Eq. \ref{tauspectralDir} of the main text.

\subsubsection{ Computation of the spectral Kemeny time of Eq. \ref{KemenyResetDirTrap}}

For $z \ne 1$, Eq. \ref{polynomialderilog} 
\begin{eqnarray}
\sum_{n=1}^{N-1} \frac{1}{z-\omega_n}  
=    \frac{ N   }{ z (1-z^{-N}) } - \frac{ 1 }{ z-1 }
\label{polynomialderilogz}
\end{eqnarray}
allows to compute the spectral Kemeny time of Eq. \ref{KemenyResetDirTrap} of the main text
with $z=\frac{\gamma}{2}+1$.


\subsection{ Computation of some spectral Kemeny times involving the cosine function and the roots of unity} 

\subsubsection{ Computation of the spectral Kemeny time of Eq. \ref{tauspectralcos}}

Using $\omega_n $ of Eq. \ref{rootsunity}, one can rewrite
\begin{eqnarray}
 \frac{1}{2-2 \cos\left( n  \frac{2 \pi}{N}\right)} 
= \frac{1}{2- \omega_n -\frac{1}{\omega_n } } 
= \frac{1}{(1- \omega_n)(1 -\frac{1}{\omega_n }) } 
= \frac{- \omega_n}{(1- \omega_n)^2 } 
= \frac{(1- \omega_n)-1}{(1- \omega_n)^2 } 
=  \frac{1}{(1- \omega_n) } -  \frac{1}{(1- \omega_n)^2 } 
\label{cos1fraction}
\end{eqnarray}

The derivation of Eq. \ref{polynomialderilog} with respect to $z$
\begin{eqnarray}
- \sum_{n=1}^{N-1} \frac{1}{(z-\omega_n)^2} 
= \frac{ \displaystyle \sum_{k=2}^{N-1} k (k-1) z^{k-2} }{ \displaystyle  \sum_{k=0}^{N-1} z^k}
- \left( \frac{ \displaystyle \sum_{k=1}^{N-1} k z^{k-1} }{ \displaystyle  \sum_{k=0}^{N-1} z^k} \right)^2
\label{polynomialderilogderi}
\end{eqnarray}
yields for $z=1$
\begin{eqnarray}
- \sum_{n=1}^{N-1} \frac{1}{(1-\omega_n)^2} 
= \frac{ \displaystyle \sum_{k=2}^{N-1} k (k-1)  }{ \displaystyle  \sum_{k=0}^{N-1} 1}
- \left( \frac{ \displaystyle \sum_{k=1}^{N-1} k  }{ \displaystyle  \sum_{k=0}^{N-1} } \right)^2
= \frac{\frac{N(N-1)(N-2)}{3}}{N} -\frac{(N-1)^2}{4}
= \frac{(N-1)(N-5)}{12}
\label{polynomialderilogderi1}
\end{eqnarray}

So the sum of Eq. \ref{cos1fraction}
over $n=1,..,N-1$ can be obtained from Eqs \ref{polynomialderilog1} and \ref{polynomialderilogderi1}
\begin{eqnarray}
\sum_{n=1}^{N-1}  \frac{1}{2-2 \cos\left( n  \frac{2 \pi}{N}\right)} 
= \sum_{n=1}^{N-1}  \frac{1}{(1- \omega_n) } - \sum_{n=1}^{N-1}  \frac{1}{(1- \omega_n)^2 } 
=  \frac{N-1}{2} + \frac{(N-1)(N-5)}{12}
= \frac{N^2-1}{12}
\label{sumcos1}
\end{eqnarray}

For $N=2 M$, Eq. \ref{sumcos1} reads
\begin{eqnarray}
 \frac{ 4 M^2-1}{12} && = \sum_{n=1}^{2M-1}  \frac{1}{2-2 \cos\left( n  \frac{ \pi}{M}\right)} 
= \sum_{n=1}^{M-1}  \frac{1}{2-2 \cos\left( n  \frac{ \pi}{M}\right)} 
+  \frac{1}{2-2 \cos\left(  \pi \right)} 
+ \sum_{n=M+1}^{2M-1}  \frac{1}{2-2 \cos\left( n  \frac{ \pi}{M}\right)} 
\nonumber \\
&& =  \sum_{n=1}^{M-1}  \frac{1}{2-2 \cos\left( n  \frac{ \pi}{M}\right)} 
+  \frac{1}{4} 
+ \sum_{m=1}^{M-1}  \frac{1}{2-2 \cos\left( (2M-m)  \frac{ \pi}{M}\right)} 
= 2 \sum_{n=1}^{M-1}  \frac{1}{2-2 \cos\left( n  \frac{ \pi}{M}\right)} 
+  \frac{1}{4} 
\label{sumcoseven}
\end{eqnarray}
so that one obtains
\begin{eqnarray}
\sum_{n=1}^{M-1}  \frac{1}{2-2 \cos\left( n  \frac{ \pi}{M}\right)}  
= \frac{1}{2} \left[ \frac{ 4 M^2-1}{12} - \frac{1}{4} \right]
=  \frac{  M^2-1}{6}  
\label{sumcos2}
\end{eqnarray}
that allows to compute the spectral Kemeny time of Eq. \ref{tauspectralcos}
of the main text.


\subsubsection{ Computation of the spectral Kemeny time of Eq. \ref{tauspectralcosmu} }

Using $\omega_n $ of Eq. \ref{rootsunity}, one can rewrite
\begin{eqnarray}
 \frac{1}{\left(y+\frac{1}{y}\right)-2 \cos\left( n  \frac{2 \pi}{N}\right)} 
&& = \frac{1}{y+\frac{1}{y}- \omega_n -\frac{1}{\omega_n } } 
 = \frac{1}{(y- \omega_n)(1 -\frac{1}{y \omega_n }) } 
=   \frac{-\omega_n }{(y- \omega_n)(\frac{1}{y}- \omega_n ) } 
\nonumber \\
&& = \frac{1}{y^2-1} \left[ \frac{y^2}{y- \omega_n} - \frac{1}{ \frac{1}{y}- \omega_n} \right]
\label{cos2fraction}
\end{eqnarray}
So the sum of Eq. \ref{cos2fraction}
over $n=1,..,N-1$ can be obtained from Eq. \ref{polynomialderilog} for the two values $z=y$ and $z=\frac{1}{y}$
\begin{eqnarray}
\sum_{n=1}^{N-1}  \frac{1}{\left(y+\frac{1}{y}\right)-2 \cos\left( n  \frac{2 \pi}{N}\right)} 
&& = \frac{1}{y^2-1} \left[ y^2 \sum_{n=1}^{N-1} \frac{1}{y- \omega_n} -\sum_{n=1}^{N-1}  \frac{1}{ \frac{1}{y}- \omega_n} \right]
\nonumber \\
&& = \frac{1}{y^2-1} \left[ y^2  \frac{ N y^{N-1}  }{ y^N-1 } - \frac{ y^2 }{ y-1 }
 -  \frac{ N y^{1-N}  }{ y^{-N}-1 } + \frac{ 1 }{ \frac{1}{y}-1 }  \right]
 \nonumber \\
&& = \frac{y}{y^2-1} \left[   \frac{ N   (y^{N} +1) }{ y^N-1 }  - \frac{ y+1}{ y-1 }  \right]
\label{cos2sum}
\end{eqnarray}

For $N=2 M$, Eq. \ref{cos2sum} reads
\begin{eqnarray}
 \frac{y}{y^2-1} \left[   \frac{ 2 M   (y^{2 M} +1) }{ y^{2M}-1 }  - \frac{ y+1}{ y-1 }  \right]
 && =\sum_{n=1}^{2M-1}  \frac{1}{\left(y+\frac{1}{y}\right)-2 \cos\left( n  \frac{ \pi}{M}\right)} 
\nonumber \\
&& = \sum_{n=1}^{M-1}  \frac{1}{\left(y+\frac{1}{y}\right)-2 \cos\left( n  \frac{ \pi}{M}\right)}
+   \frac{1}{\left(y+\frac{1}{y}\right)-2 \cos (\pi) }
+ \sum_{n=M+1}^{2M-1}  \frac{1}{\left(y+\frac{1}{y}\right)-2 \cos\left( n  \frac{ \pi}{M}\right)}
\nonumber \\
&& = \sum_{n=1}^{M-1}  \frac{1}{\left(y+\frac{1}{y}\right)-2 \cos\left( n  \frac{ \pi}{M}\right)}
+   \frac{1}{y+\frac{1}{y}+2  }
+ \sum_{m=1}^{M-1}  \frac{1}{\left(y+\frac{1}{y}\right)-2 \cos\left( (2M-m)  \frac{ \pi}{M}\right)}
\nonumber \\
&& =2  \sum_{n=1}^{M-1}  \frac{1}{\left(y+\frac{1}{y}\right)-2 \cos\left( n  \frac{ \pi}{M}\right)}
+   \frac{y}{(y+1)^2 }
\label{cos2sumeven}
\end{eqnarray}
so that one obtains
\begin{eqnarray}
  \sum_{n=1}^{M-1}  \frac{1}{\left(y+\frac{1}{y}\right)-2 \cos\left( n  \frac{ \pi}{M}\right)}
&& =
\frac{1}{2} \left(  \frac{y}{y^2-1} \left[   \frac{ 2 M   (y^{2 M} +1) }{ y^{2M}-1 }  - \frac{ y+1}{ y-1 }  \right]
 -   \frac{y}{(y+1)^2 } \right)
 \nonumber \\
 && =  \frac{y}{ (y^2-1)} \left[   \frac{  M   (y^{2 M} +1) }{ y^{2M}-1 }
-    \frac{  (y^2+1)}{ y^2-1} \right]
\label{cos2sumevenres}
\end{eqnarray}

For $y=e^{ \frac{\mu}{2} }$ and the label change $M \to N$, the sum of Eq. \ref{cos2sumevenres}
\begin{eqnarray}
S_{\mu}(N) \equiv  \sum_{n=1}^{N-1}  \frac{1}{2 \cosh \left( \frac{\mu}{2} \right)-2 \cos\left( n  \frac{ \pi}{N}\right)}
  =  \frac{ 1}
{ 2 \sinh \left( \frac{\mu}{2} \right) }  
 \left[  \frac{ N }{\tanh \left( N\frac{\mu}{2} \right)}
  - \frac{ 1 }{ \tanh \left( \frac{\mu}{2} \right) }
  \right]
\label{cos2sumhyperbolic}
\end{eqnarray}
allows to compute the spectral Kemeny time of Eq. \ref{tauspectralcosmu}.

Since the leading behavior for large $N$ of the sum $S_{\mu}(N) $ of Eq. \ref{cos2sumhyperbolic} is linear for $\mu \ne 0$
\begin{eqnarray}
S_{\mu}(N) \opsimeq_{N \to +\infty}  \frac{ N}{ 2 \sinh \left( \frac{\vert \mu \vert }{2} \right) }  
\label{cos2sumhyperboliclargeN}
\end{eqnarray}
while it is quadratic for $\mu=0$ (Eq. \ref{sumcos2})
\begin{eqnarray}
S_{\mu=0}(N) = \frac{N^2-1}{6} \opsimeq_{N \to +\infty}   \frac{  N^2}{6}  
\label{cos2sumhyperboliclargeNmuzero}
\end{eqnarray}
one can analyze the crossover between these two scaling behaviors for large $N$
via the introduction of the appropriate rescaled variable 
\begin{eqnarray}
 {\hat N} \equiv N\frac{ \vert \mu \vert }{2}  
\label{hatNmu}
\end{eqnarray}
that is considered fixed in the double limit $N \to + \infty$ and $\mu \to 0$.
Then one obtains that the ratio of Eq. \ref{cos2sumhyperbolic} and \ref{sumcos2}
\begin{eqnarray}
\frac{S_{\mu}(N)}{ S_{\mu=0}(N)}  
\displaystyle \opsimeq_{\substack{N \to + \infty \\ \mu \to 0 \\  {\hat N} = N\frac{\vert \mu \vert}{2} \text{fixed}}}
  \frac{3}{   {\hat N}^2} \left[ \frac{ {\hat N} }{\tanh {\hat N}}  -1\right] \equiv {\cal T} ({\hat N}) 
\label{cos2sumhyperboliccrossover}
\end{eqnarray}
involves the same scaling function ${\cal T}(.) $ encountered in Eq. \ref{KemenyResetdiffuniCrossoverCalT}
of the main text
that governs the analog crossover the Kemeny time of Eq. \ref{kemeny1dsawrhozeroCrossover}
 for the corresponding continuous-space diffusion process.


\subsubsection{ Translation to obtain the spectral Kemeny time of Eq. \ref{KemenyResetcos}  }

In Eq. \ref{KemenyResetcos} of
the main text, one needs the sum of Eq. \ref{cos2sumhyperbolic}
for the values
\begin{eqnarray}
 \cosh \left( \frac{\mu}{2} \right) && = 1 + \frac{\gamma}{2}
\nonumber \\
\sinh \left( \frac{\mu}{2} \right) && = \sqrt{ \cosh^2 \left( \frac{\mu}{2} \right) -1 } = \sqrt{ \gamma \left( 1 + \frac{\gamma}{4} \right) }
\label{eqyg}
\end{eqnarray}
 leading to
\begin{eqnarray}
 \frac{\mu}{2}  = 2 \ln \left( \sqrt{ \frac{\gamma}{4} }+ \sqrt{ \frac{\gamma}{4} +1}\right)
\label{agamma}
\end{eqnarray}
So Eq. \ref{cos2sumhyperbolic} becomes the sum
\begin{eqnarray}
\Sigma_{\gamma}(N) \equiv  \sum_{n=1}^{N-1}  \frac{1}{ (\gamma + 2) -2 \cos\left( n  \frac{ \pi}{N}\right)}
 && =  \frac{1}{  \sqrt{ \gamma ( 4 + \gamma ) } }
 \left[ 
   \frac{  N   }{\tanh \left[ 2 N \ln \left( \sqrt{ \frac{\gamma}{4} }+ \sqrt{ \frac{\gamma}{4} +1}\right)\right] }
-    \frac{  \gamma +2 }{ \sqrt{ \gamma ( 4 + \gamma ) }} \right]
\nonumber \\
 && =  \frac{1}{  \sqrt{ \gamma ( 4 + \gamma ) } }
 \left[ N
   \frac{ \left( \sqrt{ \frac{\gamma}{4} }+ \sqrt{ \frac{\gamma}{4} +1}\right)^{4N} -1   }
   { \left( \sqrt{ \frac{\gamma}{4} }+ \sqrt{ \frac{\gamma}{4} +1}\right)^{4N} +1}
-    \frac{  \gamma +2 }{ \sqrt{ \gamma ( 4 + \gamma ) }} \right]
\label{cos2sumhyperbolicres}
\end{eqnarray}
and allows to compute the Kemeny time of Eq. \ref{KemenyResetcos} of the main text.
 
Similarly, the crossover of Eq. \ref{cos2sumhyperboliccrossover}
translates into
\begin{eqnarray}
\frac{\Sigma_{\gamma}(N)}{ \Sigma_{\gamma=0}(N)}  
\displaystyle \opsimeq_{\substack{N \to + \infty \\ \gamma \to 0 \\  {\hat N} =\sqrt{\gamma} N \text{fixed}}}
  \frac{3}{   {\hat N}^2} \left[ \frac{ {\hat N} }{\tanh {\hat N}}  -1\right] 
  \equiv   {\cal T} \left( {\hat N} = \sqrt{\gamma} N \right)
\label{cos2sumhyperboliccrossovertranslation}
\end{eqnarray}
that involves the new scaling variable
\begin{eqnarray}
 {\hat N} \equiv \sqrt{\gamma} N
\label{hatNgamma}
\end{eqnarray}
and the same scaling function ${\cal T}(.) $ encountered in Eq. \ref{KemenyResetdiffuniCrossoverCalT}
of the main text
that governs the analog crossover the Kemeny time of Eq. \ref{KemenyResetdiffuniCrossover}
 for the corresponding continuous-space diffusion process.



\end{document}